\documentclass[
preprint,
 amsmath,amssymb,
 aps,
 prd,
]{revtex4-2}

\usepackage{url}
\usepackage{subcaption}
\usepackage[tracking=true]{microtype}

\usepackage{xcolor}

\usepackage{graphicx}
\usepackage{dcolumn}
\usepackage{bm}
\usepackage{hyperref}

\begin{document}


\title{\boldmath  Left-Right Symmetry Breaking and Gravitational Waves : A Tale of Two Phase Transitions}



\author{Z. A. Borboruah}
\email{zafri123@iitb.ac.in}
\affiliation{
 Indian Institute of Technology Bombay, Mumbai 400076, India
}

\author{U. A. Yajnik}
\email{yajnik@iitb.ac.in}
\affiliation{
 Indian Institute of Technology Gandhinagar, Gandhinagar 382055, India
}%
\affiliation{
 Indian Institute of Technology Bombay, Mumbai 400076, India
}%

%

\date{\today}

\begin{abstract}
We study possible ways gravitational waves (GW) are sourced in a theory with minimal left-right symmetry breaking. Generically first order phase transitions (FOPT) lead to gravitational waves sourced by bubble dynamics, while second order phase transitions (SOPT) do not. However, due the presence of two degenerate fields, we obtain domain walls in the putative SOPT case, giving rise to GW via disintegrating domain walls, testable at experiments such as IPTA, DECIGO, and LISA. On the other hand, for the case of FOPT, we get the usual signal from spontaneously created bubbles, but there also arises a  late forming domain wall structure separating the two
types of vacua. The disintegration of these walls provides an additional source of GW. Thus the parameter range signalling FOPT case gives rise to two distinct peaks in the spectrum of GW. This is verifiable for the low symmetry breaking scales $10^4 - 10^6$ GeV, but a high scale such as $\sim10^{10}$ GeV  remains beyond the reach of currently planned experiments. Finally, we point out that a version of  the left-right symmetric model which separates the scale of parity breaking from that of gauge symmetry breaking is also subject to domain wall formation and amenable to GW observations.
\end{abstract}

\maketitle


\section{Introduction}
\label{sec:intro}
The detection of gravitational waves (GWs) \cite{Abbott2016ObservationMergerb} has opened a new window into the unknown of the Universe. Originating from the linearized theory of general relativity \cite{Einstein1916NaherungsweiseGravitation}, GWs are a new independent source of information about cosmic events. Also, since gravitation is the weakest force in nature, gravitons are believed to decouple from the rest of the forces and matter as early as the Planck scale itself, meaning they can convey information from the very dawn of the Universe.

The early Universe GW sources include not only astronomical objects, but cosmological events like inflation, phase transitions, etc. Throughout time, GWs from all such events superimpose to create what's called a \emph{stochastic background of GWs}, which is believed to be homogeneous, isotropic, unpolarized with an amplitude following a Gaussian distribution \cite{Caprini2018CosmologicalWaves}. Several current and future experiments are aimed at detecting this background, directly via interferometers \cite{Abbott2017ExploringDetectors,Saulson2017FundamentalsDetectors} or through indirect methods like Pulsar Timing Arrays (PTAs) \cite{Lommen2015PulsarDetection}. This GW background can be a good alternative probe for high and ultra high energy phenomena including those arising from grand unified theories (GUTs) and other particle physics theories beyond the Standard Model (SM).

One such high energy theory is the Left-Right Symmetric Model (LRSM), especially since it naturally accommodates massive neutrinos, along with a gauge charge for the new right handed neutrino states. Based on the gauge group $SU(2)_L\otimes SU(2)_R \otimes U(1)_{B-L}$, this model is fully symmetric between right and left chiral states at high temperatures in the early Universe when the thermal ground state does not break symmetry spontaneously. As such why low energy physics is governed by  $SU(2)_L\times U(1)_Y$ is unknown. An early proposal to rectify this situation was made in  \cite{Chang1984DecouplingModels,Chang1984NewTheories} by dissociating spontaneous breaking of the so called \emph{D-parity} from the breaking of right-handed gauge symmetry, by invoking an extra scalar beyond the basic LRSM particle content. But as we shall discuss later in this paper, this scenario also cannot avoid domain walls \cite{Banerjee:2023hcx} and will give rise to characteristic GW background  of the type discussed here. 

Conventionally LRSM is spontaneously broken down to SM through additional Higgs particles, for example using scalar $SU(2)$ doublets or triplets in the supersymmetric (SUSY) \cite{Babu2016HiggsModels,Basso2015Doubly-chargedSupersymmetry} and Non-SUSY \cite{Mohapatra1975quotNaturalquotSymmetry,Senjanovic1975ExactParity,Branco1985NaturalTheories,Basecq:1985sx,
Gunion1989HiggsModels,Barenboim2002HiggsModel,Kiers2005HiggsViolation,Chakrabortty2014CopositivePotential,Brahmachari1994PotentialModels} contexts. A minimal left-right symmetric model (MLRSM) with scalar triplets has garnered much attention because it can implement the seesaw mechanism and can be embedded in a $SO(10)$ grand unified theory (GUT) nicely \cite{Fritzsch1975UnifiedHadrons,Georgi1975TheTheories,Arbelaez2014LHC-scaleUnification,Deppisch2017SurveyingCase,Shaban1992MinimalMeasurements}. LRSM has been studied extensively in view of the present and future colliders \cite{Chen2013ProbingLHC,Lindner2016Left-rightCollider,Patra2016StringentData,Dev2016DisambiguatingColliders,Rodriguez2002SpontaneousModel,Nemevsek2018Keung-SenjanovicDecays,Dev2016ProbingColliders,Maiezza2010Left-rightLHC} but LHC has detected no evidence of it up to $\mathcal{O}(10 \text{ TeV})$. 
Therefore the GW background presents the exciting possibility of providing the first hints of this theory, especially if originating at significantly higher energies, thus complementing the role of future colliders.

The breaking of discrete symmetry $L\leftrightarrow R$ which is generically exact in left-right symmetric models leads to the formation of domain walls (DW) in the early Universe. Domain walls are massive topological defects created when a discrete symmetry is broken \cite{Rajaraman1982SolitonsOxford,Gelmini1989CosmologyBreaking}. They separate two regions of distinct degenerate vacua of the theory. This is a problem for standard cosmology, as stable domain walls come to dominate the energy density of the Universe at later times \cite{Zeldovich1974CosmologicalSymmetry} and conflict with structure formation and Big Bang Nucleosynthesis (BBN). Therefore, they must be transient \cite{Rai1994GravityProblem,Vilenkin1981GravitationalStrings,Sikivie1982AxionsUniverse,Mohanty1984Matter-antimatterUniverse,Mishra:2009mk,Banerjee:2023hcx}. This can be achieved by making one of the vacua slightly lower in energy so that the domain walls move under the vacuum pressure. Eventually, violent collisions between DW cause their annihilation returning the Universe to a homogeneous state after entropic equilibrium is reached.

While it is usual to identify phase transitions in gauge theories as First Order (FOPT), Second order (SOPT), or crossover, the left-right symmetric models present a peculiar situation, due to the presence of two independent complex scalar fields but with identical or nearly identical values governing their dynamics. Specifically, the putative SOPT also gives rise to domain walls separating regions of either $\Delta_L$ or $\Delta_R$ fields acquiring non-trivial VEVs. The size of such domains is determined by the causal limitation on the length scale discussed by Kibble\cite{Kibble:1976sj}. Thus an apparent case of SOPT where gravitational waves are \textit{not} expected also gives rise to gravitational waves due to the degradation dynamics of such domain walls. 

Likewise, even the putative FOPT is fundamentally modified in that, starting with the fully symmetric high temperature phase,  the spontaneous formation of bubbles of true vacuum could either lead to bubbles of left-like vacuum or right-like vacuum. When these bubbles grow and encounter each other, the collision can result in a larger bubble, provided the interiors contain the same type of vacuum. On the other hand, if one bubble contains left-like while the other contains right-like vacuum then the result is a left-right domain wall. The wall configuration is made up of non-trivial condensates of two or more fields \cite{Cline:2002ia,Sarkar:2007er}. The situation is depicted in Fig. \ref{fig:DegFldFOPT}. The eventual result of such a phase transition will also be a network of domain walls enclosing distinct types of vacua, making it look similar to the modified SOPT discussed in the previous paragraph. However the sizes of domains would be of a different scale. The model with such parameters would therefore result in two well separated peaks in the GW spectrum. We discuss these cases in Sec. \ref{sub:twokinds}.

Our strategy for obtaining the GW spectrum for the latter case is as follows. Gravitational waves from first-order phase transition \cite{Witten1984CosmicPhases,Hogan1983NucleationTransitions,Hogan1986GravitationalTransitions,Turner1990RelicInflation,Kamionkowski1994GravitationalTransitions,Grojean2007GravitationalBeyond,Ellis2019OnSignal} and more specifically from MLRSM \cite{Brdar2019GravitationalBreakingb,Borah:2022wdy} focusing on single field case have been studied extensively which we will discuss in section \ref{section:GWFOPT}. 
On the other hand,  GW resulting from 
$Z_2$  or axionic domain walls have been studied in \cite{Kawasaki2011StudyWalls,Hiramatsu2014OnWalls,IchiSaikawa2017AWalls}.
For obtaining this GW contribution, our approach parallels the methods used in \cite{Kawasaki2011StudyWalls,Hiramatsu2014OnWalls,IchiSaikawa2017AWalls} but with the difference that we consider the late time evolution of domain walls after the Ginzburg temperature has been reached in the context of LRSM. 
Thus for models with FOPT type case, by combining these two types of contributions we determine the double peak spectrum  with symmetry breaking scales $10^4 - 10^6$ GeV and also the generic GUT scale  $\sim10^{10}$GeV and discuss the possibility of their verifiability.

The arrangement of this paper is as follows: in the next section, Sec. \ref{sec:MLRSM} we describe the MLRSM with triplets and the loop corrected temperature dependent effective potential. In a subsection \ref{sub:twokinds} we emphasize the cosmology of the two different classes of models, those that signal SOPT versus those that signal FOPT. In Sec. \ref{section:GWFOPT}, we discuss GWs for the FOPT case in MLRSM, disregarding the contribution from residual domain walls. In Sec. \ref{section:DWLRSM}, we discuss domain wall structure in MLRSM and GW in the relevant SOPT case. Here we consider the effect of adding a bias motivated by phenomenology, to the tree level potential. We outline the numerical methods used to simulate the decaying domain walls and calculation of the spectrum. In Sec.~\ref{sec:doublepeak} we describe the two peak feature of the GW spectra in MLRSM from the relevant FOPT case. In sec. \ref{sec:CMPmodel} we show that the spontaneous parity breaking scenario of \cite{Chang1984NewTheories,Chang1984DecouplingModels} also results in characteristic GW background. Sec. \ref{sec:conclusion} contains a summary and conclusion. We mention the benchmark points for SOPT and FOPT for our analysis in Appendix~\ref{app:BP}. Appendix \ref{app:A} contains some of the additional details of the numerical procedure used and results from simulations.

\section{Minimal Left-Right Symmetric Model}
\label{sec:MLRSM}

The MLRSM~\cite{Mohapatra:1974gc,Pati:1974yy,Senjanovic:1975rk,Senjanovic:1978ev,
Mohapatra:1979ia,Mohapatra:1980yp,Pati:1973uk,Pati:1974vw} is an appealing extension of the SM which naturally accords gauge charges to the  right-handed neutrino states which are 
generically needed to provide masses to neutrinos.  It is based on the gauge group, 
\begin{align}
\mathcal{G}_{LR}\equiv SU(2)_L \times SU(2)_R \times U(1)_{B-L} \times SU(3)_{C}\, 
\label{eq:LRSM}
\end{align}
where the electric charge is defined as 
\begin{align}
Q=T_{3L}+T_{3R}+\frac{B-L}{2}\, .
\end{align}
Under this left-right symmetric gauge group, the usual quarks and leptons transform as
\begin{eqnarray}
&&q_{L}=\begin{pmatrix}u_{L}\\
d_{L}\end{pmatrix}\equiv[2,1,1/3,3]\,, ~ q_{R}=\begin{pmatrix}u_{R}\\
d_{R}\end{pmatrix}\equiv[1,2,1/3,3]\,,\nonumber \\
&&\ell_{L}=\begin{pmatrix}\nu_{L}\\
e_{L}\end{pmatrix}\equiv[2,1,-1,1] \, , ~ \ell_{R}=\begin{pmatrix}\nu_{R}\\
e_{R}\end{pmatrix}\equiv[1,2,-1,1] \, . \nonumber
\end{eqnarray}
The Higgs sector may vary in LRSM. We will focus on the minimal Higgs sector containing the triplets and the bidoublet \cite{Deshpande1991Left-right-symmetricField}, $\Delta_L(1,0,2,0)$, 
$\Delta_R(0,1,2,0)$, $ \phi(1/2,1/2^*,0,0)$. They can be represented in the matrix form,
\begin{equation}
\label{eq:tripletsanddoublet}
     \phi=\left(\begin{array}{cc}
 \phi_{1}^{0} &  \phi_{1}^{+} \\
 \phi_{2}^{-} &  \phi_{2}^{0}
\end{array}\right), \quad \Delta_{L}=\left(\begin{array}{cc}
\frac{1}{\sqrt{2}} \delta_{L}^{+} & \delta_{L}^{++} \\
\delta_{L}^{0} & -\frac{1}{\sqrt{2}} \delta_{L}^{+}
\end{array}\right), \quad \Delta_{R}=\left(\begin{array}{cc}
\frac{1}{\sqrt{2}} \delta_{R}^{+} & \delta_{R}^{++} \\
\delta_{R}^{0} & -\frac{1}{\sqrt{2}} \delta_{R}^{+}
\end{array}\right)
\end{equation}
so that they transform as
\begin{align}
S U(2)_{L} \times S U(2)_{R}&:  \phi \rightarrow U_{L}  \phi U_{R}^{\dagger}, & \Delta_{L} \rightarrow U_{L} \Delta_{L} U_{L}^{\dagger}, \quad \Delta_{R} \rightarrow U_{R} \Delta_{R} U_{R}^{\dagger} \\
U(1)_{B-L}&:  \phi \rightarrow  \phi, & \Delta_{L} \rightarrow e^{i 2 \theta} \Delta_{L}, \quad \Delta_{R} \rightarrow e^{i 2 \theta} \Delta_{R}
\end{align}
for general symmetry transformations: $U_{L} \in S U(2)_{L}, U_{R} \in S U(2)_{R}$, and $e^{i \theta} \in U(1)_{B-L}$.
\subsection{Potential}
The tree-level standard scalar potential of the LRSM with triplets is given by,
\begin{equation}
    V_0 = V_ \phi + V_\Delta + V_{ \phi\Delta}
\end{equation}
where
\textls[1]{
\begin{align}\label{LRSMpot}
\nonumber V_{ \phi}=&-\mu_{1}^{2} \operatorname{Tr}\left[ \phi^{\dagger}  \phi\right]-\mu_{2}^{2}\left(\operatorname{Tr}\left[\tilde{ \phi}  \phi^{\dagger}\right]+\operatorname{Tr}\left[\tilde{ \phi}^{\dagger}  \phi\right]\right)+\lambda_{1} \operatorname{Tr}\left[ \phi^{\dagger}  \phi\right]^{2}+\lambda_{2}\left(\operatorname{Tr}\left[\tilde{ \phi}  \phi^{\dagger}\right]^{2}+\operatorname{Tr}\left[\tilde{ \phi}^{\dagger}  \phi\right]^{2}\right) \\
\nonumber &+\lambda_{3} \operatorname{Tr}\left[\tilde{ \phi}  \phi^{\dagger}\right] \operatorname{Tr}\left[\tilde{ \phi}^{\dagger}  \phi\right]+\lambda_{4} \operatorname{Tr}\left[ \phi^{\dagger}  \phi\right]\left(\operatorname{Tr}\left[\tilde{ \phi}  \phi^{\dagger}\right]+\operatorname{Tr}\left[\tilde{ \phi}^{\dagger}  \phi\right]\right),\\
\nonumber V_{\Delta}=& -\mu_{3}^{2}\left(\operatorname{Tr}\left[\Delta_{L} \Delta_{L}^{\dagger}\right]+\operatorname{Tr}\left[\Delta_{R} \Delta_{R}^{\dagger}\right]\right) + \rho_{1}\left(\operatorname{Tr}\left[\Delta_{L} \Delta_{L}^{\dagger}\right]^{2}+\operatorname{Tr}\left[\Delta_{R} \Delta_{R}^{\dagger}\right]^{2}\right) \\
\nonumber &+\rho_{2}\left(\operatorname{Tr}\left[\Delta_{L} \Delta_{L}\right]\operatorname{Tr}\left[\Delta_{L}^{\dagger} \Delta_{L}^{\dagger}\right]+\operatorname{Tr}\left[\Delta_{R} \Delta_{R}\right] \operatorname{Tr}\left[\Delta_{R}^{\dagger} \Delta_{R}^{\dagger}\right]\right)+\rho_{3} \operatorname{Tr}\left[\Delta_{L} \Delta_{L}^{\dagger}\right] \operatorname{Tr}\left[\Delta_{R} \Delta_{R}^{\dagger}\right]\\
\nonumber &+\rho_{4}\left(\operatorname{Tr}\left[\Delta_{L} \Delta_{L}\right] \operatorname{Tr}\left[\Delta_{R}^{\dagger} \Delta_{R}^{\dagger}\right]+\operatorname{Tr}\left[\Delta_{L}^{\dagger} \Delta_{L}^{\dagger}\right] \operatorname{Tr}\left[\Delta_{R} \Delta_{R}\right]\right), \\
\nonumber V_{ \phi \Delta}=& \alpha_{1} \operatorname{Tr}\left[ \phi^{\dagger}  \phi\right]\left(\operatorname{Tr}\left[\Delta_{L} \Delta_{L}^{\dagger}\right]+\operatorname{Tr}\left[\Delta_{R} \Delta_{R}^{\dagger}\right]\right)+\alpha_{3}\left(\operatorname{Tr}\left[ \phi  \phi^{\dagger} \Delta_{L} \Delta_{L}^{\dagger}\right]+\operatorname{Tr}\left[ \phi^{\dagger}  \phi \Delta_{R} \Delta_{R}^{\dagger}\right]\right) \\
\nonumber &+\alpha_{2}\left(\operatorname{Tr}\left[\Delta_{L} \Delta_{L}^{\dagger}\right] \operatorname{Tr}\left[\tilde{ \phi}  \phi^{\dagger}\right]+\operatorname{Tr}\left[\Delta_{R} \Delta_{R}^{\dagger}\right] \operatorname{Tr}\left[\tilde{ \phi}^{\dagger}  \phi\right]+\text { h.c. }\right) \\
\nonumber &+\beta_{1}\left(\operatorname{Tr}\left[ \phi \Delta_{R}  \phi^{\dagger} \Delta_{L}^{\dagger}\right]+\operatorname{Tr}\left[ \phi^{\dagger} \Delta_{L}  \phi \Delta_{R}^{\dagger}\right]\right)+\beta_{2}\left(\operatorname{Tr}\left[\tilde{ \phi} \Delta_{R}  \phi^{\dagger} \Delta_{L}^{\dagger}\right]+\operatorname{Tr}\left[\tilde{ \phi}^{\dagger} \Delta_{L}  \phi \Delta_{R}^{\dagger}\right]\right) \\
&+\beta_{3}\left(\operatorname{Tr}\left[ \phi \Delta_{R} \tilde{ \phi}^{\dagger} \Delta_{L}^{\dagger}\right]+\operatorname{Tr}\left[ \phi^{\dagger} \Delta_{L} \tilde{ \phi} \Delta_{R}^{\dagger}\right]\right)
\end{align}}
where $\tilde{ \phi}=\sigma_2 \phi^*\sigma_2$. All the couplings are assumed to be real for simplicity. The vacuum expectation values (VEVs) of the Higgs fields that give the correct breakdown of LR symmetry are as follows,
\begin{equation}\label{eq:VEV}
    \langle \phi\rangle=\frac{1}{\sqrt{2}}\left(\begin{array}{cc}
\kappa_{1} & 0 \\
0 & \kappa_{2} e^{i \theta_{2}}
\end{array}\right), \quad\left\langle\Delta_{L}\right\rangle=\frac{1}{\sqrt{2}}\left(\begin{array}{cc}
0 & 0 \\
v_{L} e^{i \theta_{L}} & 0
\end{array}\right), \quad\left\langle\Delta_{R}\right\rangle=\frac{1}{\sqrt{2}}\left(\begin{array}{cc}
0 & 0 \\
v_{R} & 0
\end{array}\right)
\end{equation}
The bidoublet VEVs provide the electroweak scale,
\begin{equation}
    \label{eq:bidoublet}
    \sqrt{\kappa_1^2+\kappa_2^2}=v=246 \text{ GeV}
\end{equation}
and the parameter $\beta=\arctan{\kappa_2/\kappa_1}$ captures the ratio of the VEV's.
The minimization conditions on the potential, with respect to the four VEVs lead to a
relationship between $v_L$ and $v_R$ through a see-saw-like formula,
\begin{equation}
    \label{eq:seesaw}
    \beta_{1} \kappa_{1} \kappa_{2} \cos \left(\theta_{2}-\theta_{L}\right)+\beta_{2} \kappa_{1}^{2} \cos \theta_{L}+\beta_{3} \kappa_{2}^{2} \cos \left(2 \theta_{2}-\theta_{L}\right)=\left(2 \rho_{1}-\rho_{3}\right) v_{L} v_{R}
\end{equation}
It is customary to set $\beta_{1,2,3}=0$ so that the strong condition of this see-saw can be circumvented, $v_L$ can be chosen $0$ while $v_R$ can be assigned phenomenologically desirable large value. Additionally $\theta_2$ and $\theta_L$ may be set to $0$ for simplicity.

We shall use the VEVs from SM and expected phenomenology as the inputs in the minimization conditions, in which case the parameters $\mu^2_{1,2,3}$ get determined as follows.
\begin{equation}
    \label{eq:mu}
    \begin{array}{l}
\mu_{1}^{2}=\lambda_{1}\left(\kappa_{1}^{2}+\kappa_{2}^{2}\right)+2 \kappa_{1} \kappa_{2} \lambda_{4}+\frac{1}{2} v_{R}^{2} \alpha_{1}-\frac{\alpha_{3}}{2} \frac{v_{R}^{2} \kappa_{2}^{2}}{\kappa_{1}^{2}-\kappa_{2}^{2}} \\
\mu_{2}^{2}=\left(2 \lambda_{2}+\lambda_{3}\right) \kappa_{1} \kappa_{2}+\frac{\lambda_{4}}{2}\left(\kappa_{1}^{2}+\kappa_{2}^{2}\right)+\frac{\alpha_{2}}{2} v_{R}^{2}+\frac{\alpha_{3}}{4} \frac{v_{R}^{2} \kappa_{1} \kappa_{2}}{\kappa_{1}^{2}-\kappa_{2}^{2}} \\
\mu_{3}^{2}=\rho_{1} v_{R}^{2}+\frac{1}{2} \alpha_{1}\left(\kappa_{1}^{2}+\kappa_{2}^{2}\right)+2 \alpha_{2} \kappa_{1} \kappa_{2}+\frac{1}{2} \alpha_{3} \kappa_{2}^{2}
\end{array}    
\end{equation}

An important point that has not received due attention in the literature is that the see-saw-like relation equally well implies a possible large value for $v_L$ and $v_R\approx 0$ due to the left-right symmetry of the model. In that case Eq. \eqref{eq:mu} takes the same form with $v_R$ replaced by $v_L$. While low energy phenomenology has preferred the $v_L\approx0$ solution, this will not persist at high temperatures in the early Universe. Indeed we expect gauge symmetry restoration at high temperature \cite{Kirzhnits:1972ut,Weinberg:1974hy,Dolan:1973qd}, a scale $T_c$ above which the thermal ground state has no non-zero VEVs. When the Universe cools it must choose between the left-like vacuum with $v_L\neq 0$ or the right-like vacuum with $v_R\neq 0$. As already introduced in Sec. \ref{sec:intro}, this leads to intricate possibilities for the nature of the phase transition, as we further discuss in Sec. \ref{sub:twokinds}

\subsection{Effective Potential}
\label{sec:effpotential}
\begin{figure}[ht]
 \centering
    \includegraphics[width=\linewidth]{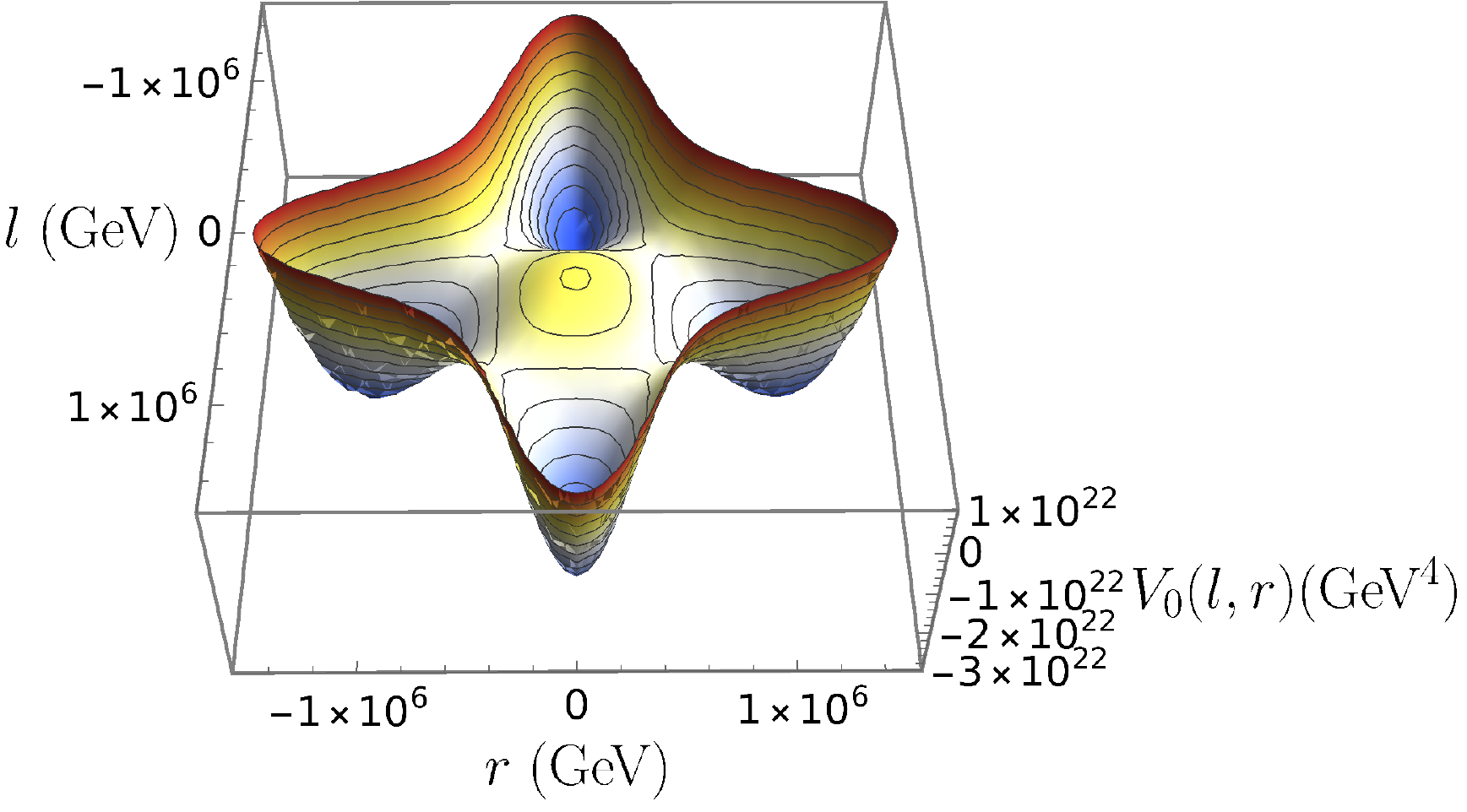}
    \caption{\small 3D view of tree level potential \eqref{eq:tree0} with $(v_L,v_R)=(10^6,0)$ or $(0,10^6)$ GeV. From the contour lines and color gradient it is evident that the potential has degenerate minima at $(l,r)=(10^6,0),(-10^6,0),(0,10^6)$ and $(0,-10^6)$ portraying the inherent $Z_4$ symmetry.}
    \label{fig:mlrsmz4}
\end{figure}

From low energy phenomenology we are interested in scenarios where the $SU(2)_R$ breaking scale is much higher than electroweak symmetry breaking $v_R\gg \kappa_1,\kappa_2$.  It turns out that it is sufficient to focus on large $\delta_R^0$ values in field space, compared to the remaining neutral fields $ \phi_1^0, \phi_2^0$ and $\delta_L^0$ \cite{Brdar2019GravitationalBreakingb}. However, in the context of the early Universe, the alternative possibility of $v_L$ being large while ignoring $v_R$ also arises. Hence we retain the two contributions in the tree level potential from $\delta_R^0$ and $\delta_L^0$
\begin{equation}\label{eq:tree0}
V_0^{total}(r,l) = V_0(r) + V_0(l) + \frac{\rho_3}{4}l^2 r^2
\end{equation}
where $r=\delta_R^0/\sqrt{2}$ and $l=\delta_L^0/\sqrt{2}$, and $V_0$ are both identical functional forms of their arguments,
\begin{equation}
\label{eq:functionalform}
 V_0(r) \approx \frac{\rho_1}{4}\left(r^2-\eta^2\right)^2-\frac{\rho_1}{4}\eta^4
\end{equation}
where $\eta=\sqrt{\mu_3^2/\rho_1}$. In Fig. \ref{fig:mlrsmz4}, we see that the potential \eqref{eq:tree0} has 4 degenerate minima at the expected points in the $r-l$ plane, implying an inherent $Z_4$ symmetry of the system. Indeed due to the see-saw type condition, where $r\rightarrow 0$, $l\rightarrow v_L\neq0$ and vice versa.
As such we expect domain walls, separating regions of left-like vacuum $l\neq0,r=0$ and right-like vacuum $l=0,r\neq0$ with the wall constituting a transition region where one VEV goes to zero and the other turns on.

Let us focus on generalizing the functional form \eqref{eq:functionalform} of the potential to the case of non-zero temperature. The temperature-dependent effective potential is given in \cite{Brdar2019GravitationalBreakingb}, now generalized to both the $l$ and $r$ fields,
\begin{equation}\label{eq:veff}
    V_{\mathrm{eff}}(l,r, T)=V_{0}(l,r)+V_{\mathrm{CW}}(l,r)+V_{\mathrm{FT}}(l,r, T)+V_{\mathrm{D}}(l,r, T)
\end{equation}
where
\begin{align}
    \label{eq:VCW}
    \nonumber
    V_{\mathrm{CW}}(l,r)&=\frac{1}{64 \pi^{2}} {\left[\sum_{i} m_{i}^{4}(l,r)\left(\log \frac{m_{i}^{2}(l,r)}{\mu^{2}}-\frac{3}{2}\right)+6 \sum_{j=L,R} m_{W_{j}}^{4}(l,r)\left(\log \frac{m_{W_{j}}^{2}(l,r)}{\mu^{2}}-\frac{5}{6}\right)\right.} \\
    &\left.+3 \sum_{j=L,R} m_{Z_{j}}^{4}(l,r)\left(\log \frac{m_{Z_{j}}^{2}(l,r)}{\mu^{2}}-\frac{5}{6}\right)-6 \sum_{j=L,R} m_{\nu_{j}}^{4}(l,r)\left(\log \frac{m_{\nu_{j}}^{2}(l,r)}{\mu^{2}}-\frac{3}{2}\right)\right]
\end{align}
is the Coleman-Weinberg correction to the tree potential and,

\begin{align}
    \label{eq:VFT}
    \nonumber
    V_{\mathrm{FT}}(l,r, T)=&\frac{T^{4}}{2 \pi^{2}}{\left[\sum_{i} J_{\mathrm{B}}\left(\frac{m_{i}^{2}(l,r)}{T^{2}}\right)+6\sum_{j=L,R} J_{\mathrm{B}}\left(\frac{m_{W_{j}}^{2}(l,r)}{T^{2}}\right)+3 \sum_{j=L,R} J_{\mathrm{B}}\left(\frac{m_{Z_{j}}^{2}(l,r)}{T^{2}}\right)\right.} \\
    &\left.-6 \sum_{j=L,R} J_{\mathrm{F}}\left(\frac{m_{\nu_{j}}^{2}(l,r)}{T^{2}}\right)\right]
\end{align}
is the temperature correction from finite temperature field theory, where $J_{B/F}$ are integrals for bosons/fermions given by,
\begin{equation}
    \label{eq:JBJF}
    J_{\mathrm{B} / \mathrm{F}}\left(y^{2}\right)=\int_{0}^{\infty} \mathrm{d} x x^{2} \log \left(1 \mp \mathrm{e}^{-\sqrt{x^{2}+y^{2}}}\right)
\end{equation}
The mass terms $m_{i},m_{W_{L/R}},m_{Z_{L/R}},m_{\nu_{L,R}}$ are the field ($l,r$) dependent scalar, gauge and fermion masses given in \cite{Brdar2019GravitationalBreakingb}, now generalized to incorporate both the $l$ and $r$ fields. The $\mu$ in \eqref{eq:VCW} is the renormalization scale which we can conveniently assume to be equal to $\eta$. The last term includes the Daisy diagrams for higher loop corrections,
\begin{equation}
    V_{\mathrm{D}}(l,r, T)=-\frac{T}{12 \pi} \sum_{j}\left[M_{j}^{3}(l,r)-m_{j}^{3}(l,r)\right]
\end{equation}
where $M_j$ are given by eigenvalues of the matrices $\mathcal{M}_j+\Pi_j$ where $\mathcal{M}_j$ are mass matrices of bosons and fermions and $\Pi_j$ are thermal self energies \cite{Brdar2019GravitationalBreakingb}. The summation of $j$ includes all the bosons and fermions of the theory. It can be seen that the temperature dependent effective potential respects the same $Z_4$ symmetry as the tree level potential for temperatures below the critical temperature $T_c$ of phase transition.

In the following two subsection, first we outline the intriguing classes of phase transition scenarios specific to the MLRSM, differentiating them from the generic $Z_2$ symmetry breaking. Then we delineate the parameter ranges that indicate these phase transitions, ensuring they adhere to both theoretical constraints and experimental observations.

\subsection{Two kinds of phase transitions}
\label{sub:twokinds}
In this subsection, we describe the interesting possibilities of two kinds of phase transitions within the MLRSM. As mentioned in Sec.~\ref{sec:intro}, due to the presence of the parity symmetry between the scalar fields $l$ and $r$, one needs to distinguish the phase transitions in our model from the simple $Z_2$ symmetry breaking scenarios. The finite temperature effective potential $V_{\rm eff}$ in Eq. \eqref{eq:veff} demonstrates a phase transition from the parity symmetric $v_R=v_L=0$ state at high temperature to the parity violating $v_R=v_R(T)\neq v_L=0$ or $v_L=v_L(T)\neq v_R=0$ state at low temperature. Numerically $v_R(T)=v_L(T)$ due to the left-right symmetry. This transition occurs at a critical temperature $T_c$ at which the parity symmetric and parity violating minima are degenerate. Generically, phase transitions are of two types: first or second-order, depending on whether there is a potential barrier between the parity-symmetric and parity-violating minima at $T_c$, i.e. they differ in the sense that the transition of the global minimum from the parity symmetric to the parity violating one is discontinuous in case of first-order whereas in second-order, the shift is continuous. It is found that the phase transition is strong first-order for small values of the quartic coupling $\rho_1\lesssim0.3-0.4$. For the left-right case, this classification needs to be modified for two different reasons: finite causal horizon in the early Universe \cite{Kibble:1976sj,Kibble:1980mv}, and the fact that there are two fields $\Delta_L$ and $\Delta_R$ obeying a potential symmetric under $\Delta_L\leftrightarrow\Delta_R$. We may thus identify two possibilities :

\begin{itemize}
\item \textbf{SOPT with two degenerate fields} : The dynamics of this phase transition is the well known case discussed originally in \cite{Kibble:1976sj}, but with a difference due to the presence of two fields. We recapitulate the Kibble mechanism here briefly.  A generic SOPT with a real scalar field and a $Z_2$ symmetric potential as in Eq. \eqref{eq:functionalform} is characterised by fluctuation length $\xi =1/m_r=(\sqrt{2\rho_1} \eta)^{-1}$, the inverse mass of the Higgs degree of freedom $r$. However, the phase transition can be considered to have ended only after the temperature drops below the Ginzburg temperature corresponding to the correlation length $\xi_G$
\begin{equation}
 \xi_G \simeq \frac{1}{2\rho_1 \eta}
\end{equation}
This is determined by the condition that 
\begin{equation}
\xi_G^3 \Delta_c V_{\rm eff} = T_G 
\end{equation}
 which defines $T_G$ implicitly, where $\Delta_c V_{\rm eff}$ is the difference between free energy density before and after the phase transition.
In the early Universe, this length is further restricted by the condition that the correlation length cannot be diverging faster than the speed of light, which then implies that the relevant correlation length is \cite{Kibble:1976sj}
\begin{equation}
 \xi^r_{\rm causal} = \left(\frac{M_{Pl}}{\sqrt{g(T_c)} m_r^2 T_c^2} \right)^{1/3}
\label{eq:correlation_r}
\end{equation}
where $g(T_c)$ is the relativistic degree of freedom at $T_c$. In our model $g(T_c)=134$. But in our case, to this generic $Z_2$ symmetry breaking scenario gets added the feature of the left-right symmetric model that there are two fields $\Delta_L$ and $\Delta_R$ with identical dynamics. For the MLRSM potential in  \eqref{eq:tree0} with 4 degenerate minima, we will show in Sec. \ref{section:DWLRSM} that the properties of SOPT depends on $m_\theta$ defined in \eqref{eq:axionmass} rather than $m_r$ where $\theta=\tan^{-1}\left(\frac{l}{r}\right)$. Thus we 
take the expression for the maximum correlation length to be
\begin{equation}
 \xi_\text{causal} = \left(\frac{M_{Pl}}{\sqrt{g(T_c)} m_\theta^2 T_c^2} \right)^{1/3}
\label{eq:correlation}
\end{equation}
These fields $\Delta_L,\Delta_R$ are complex scalars and it is more appropriate to consider the fate of their modulus values. The energetics dictate that only one of the two moduli have non-zero VEV. The domains are therefore a consequence of regions with $\Delta_L$ having non-trivial VEV giving way to regions with non-trivial VEV of $\Delta_R$ and vice versa.  

These considerations imply that as the correlation length reaches its maximum value, uncorrelated regions with either $v_L\neq 0$ or $v_R \neq 0$ are allowed to develop, with domain walls forming between them. Thus $\xi_\text{causal}$ can be considered to provide the maximum scale of domain sizes just below the temperature $T_G$. We find that  $\xi_\text{causal}$ is larger than the causal horizon size $\sim\tau$ or even Hubble horizon $\sim t$ at temperatures under consideration, where $\tau,t$ are the conformal and cosmic time. So, we take domain sizes limited by the causal horizon as the initial configuration for numerical simulation of domain wall evolution.

\item \textbf{FOPT with two degenerate fields}: The second possibility arises if the theory is governed by a set of parameters that signals an FOPT as a local field theory. It is customary to study such phase transitions by invoking the scenario of spontaneous formation of bubbles of true vacuum by a tunnelling mechanism. The end point of such a mechanism is a homogeneous medium with no domain walls. However due the presence of two fields $\Delta_L$ and $\Delta_R$, the final stages of such a phase transition adds a new subtlety.  Above the critical temperature characterising the FOPT, both the fields have zero VEV. However when the nucleation  of bubbles becomes possible, the latter can be of two types, they can be bubbles with either $v_L\neq0$ or $v_R\neq 0$. Now as in the usual FOPT, the bubbles grow and merge. However, as sketched in Fig. \ref{fig:DegFldFOPT}, after the merger of the same type of vacua, there will arise a residual network of frustrated domain walls characterised either by $v_L\neq0$ or $v_R\neq 0$. The length scale of such domains is not constrained by the causal horizon in contrast to the case of SOPT.
\begin{figure}[ht]
 \centering
    \includegraphics[width=0.8\linewidth]{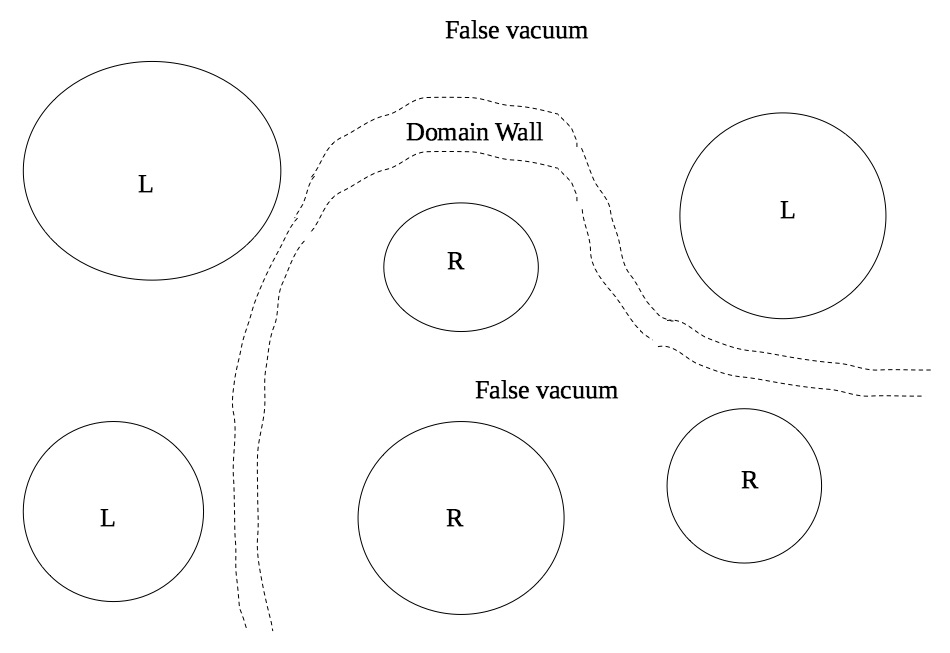}
    \caption{\small Cartoon explaining the emergence of residual domain walls after FOPT in the case of two degenerate fields such as in the left-right symmetric model. The symmetric false vacuum can be destabilized by the spontaneous formation of bubbles of either L type vacuum with $v_L=0$ and $v_R\neq 0$ or R type with $v_R=0$ and $v_L\neq 0$. The domain wall sketched with dotted lines eventually emerges as the bubbles of the same vacua merge to form large homogeneous domains.}  
    \label{fig:DegFldFOPT}
\end{figure}

Using the tunnelling probability formula in terms of the bounce solution, one introduces a time scale: inverse duration of a phase transition ( see Eq. \eqref{eq:inverseduration}) $\beta$.  So long as
\(
\beta \gtrsim H,
\)
a sufficient number of bubbles will be formed so that they can percolate \cite{Essam1980PercolationTheory,Zeldovich1974CosmologicalSymmetry}. However, the percolation can be either among the left-like or the right-like phases. Where the two percolated regions meet, there is a domain wall, until the whole Universe is filled with a frustrated network of domain walls. In the absence of any other scales, it is reasonable to assume that over scales greater than the causal horizon, the next near neighbour domain can randomly be in the same or the other vacuum. This can lead to a frustrated network of scale larger than the causal horizon by a factor $O(1)$ though not much large. For a space filling network of irregular polyhedrons with an average of $S$ faces for each of the polyhedrons, we may assume average enhancement by a factor $\sqrt{S}$ due to merger with an average of $S/2$ near neighbours. The particle physics model, therefore, needs to contain a mechanism for a slow disintegration of these residual walls to not conflict with late cosmology.

The subsequent evolution of these residual domain walls depends on the bias between the two types of approximately degenerate vacua. Finally, the false domains will shrink out of existence according to the bias, producing GWs from the late time evolution of the decay of these domain walls. Thus in addition to the spontaneous bubbles of the usual FOPT, we get a contribution from the evolution of disintegrating DW, of the type similar to the SOPT case. Since the two epochs are well separated in time we can calculate each contribution individually.

\end{itemize}

As we discussed, domain walls are produced in both cases,  whether the parameters of the theory signals SOPT or FOPT. In subsequent sections, we shall develop the theory of these two cases in detail. 

\subsection{Acceptable parameter ranges}
\label{sec:benchmark}
For evaluating the potential given in Eq. \eqref{eq:veff} and to see what type of phase transitions take place, we need to investigate the ground state of the potential by minimizing it, assuming different values of the model parameters. Utilizing the $Z_4$ symmetry, we need to consider only one of the 4 VEV directions for this task, so we choose positive $r$ direction and set $l=0$. LRSM is very constrained in terms of realization and parameters. In~\cite{Deshpande1991Left-right-symmetricField}, it was mentioned that in the triplet model, the only option which yields observable extra gauge and Higgs bosons without fine tuning of the VEV seesaw relation given in Eq.~\eqref{eq:seesaw} is to set the $\beta_i$ Higgs couplings to 0 so that one of the VEVs, $v_L$ or $v_R$ can be made 0. On the other hand, in~\cite{Brdar2019GravitationalBreakingb} and~\cite{Li2021ProspectsModel}, a set of theoretical requirements such as perturbativity limits, unitarity and vacuum stability conditions and correct vacuum criteria along with experimental limits from the LHC are considered. We agree with their analysis and consider the parameter space for our model from these two papers:
\begin{align}
\nonumber v &=246 \mathrm{ GeV}, v_{R} \in\left[10^{4}, 10^{6}\right] \mathrm{GeV}, \tan \beta=\tan 10^{-3} \\
\nonumber \lambda_{1} &=0.13, \lambda_{2}=0, \lambda_{3} \in[0,5], \lambda_{4}=0 \\
\rho_{1} & \in[0,0.5], \rho_{2} \in[0,5], \rho_{3} \in[1,5], \rho_{4}=0 \\
\nonumber \alpha_{1} &=0, \alpha_{2}\in[0,0.5], \alpha_{3} \in[0,5] \\
\nonumber \beta_{1} &=\beta_{2}=\beta_{3}=0
\end{align}

We randomly generate sample points within these limits and check for the global minima of the potential in Eq. \eqref{eq:veff} numerically using Mathematica's \textbf{NMinimize} function as well as using \emph{CosmoTransitions} \cite{Wainwright2012CosmoTransitions:Fieldsb} and match the two results. We look for both strong FOPT and SOPT in CosmoTransitions and Mathematica. A continuous transition VEV from $r=0$ at high temperature to $r=v_R(T)$ at low temperature would imply a second-order or a \emph{cross-over} phase transition whereas a barrier between two minima at the critical temperature would imply a first-order phase transition. Out of all the samples, we take 3 of the SOPT as benchmark points BP1, BP2, BP3 given in Table \ref{tab:BPSOPTFOPT} with energy scales $\eta=10^4,10^5,10^6$ GeV respectively for the GW considerations from domain wall decay. In case of FOPT also, we choose 3 benchmark points named BPF1, BPF2, BPF3 given in Table \ref{tab:BPSOPTFOPT} for calculating the critical bubble profile in CosmoTransitions and using it to calculate the relevant GW parameters in Mathematica as will be described in Sec. \ref{section:GWFOPT}.


\section{Gravitational Waves from First Order Phase Transition}\label{section:GWFOPT}
We begin with studying the contributions to the GW spectrum from a standard FOPT for the relevant benchmark points of our model. As discussed in Sec.~\ref{sub:twokinds}, the bubble formation and mutual collisions will leave behind two-field domain walls of the type discussed in \cite{Cline:2002ia,Sarkar:2007er}. These should disintegrate on their own, giving an additional contribution to the GW. This section is devoted to studying only the FOPT type contribution. After we study the contribution from the disintegrating DW in a subsequent section we shall return to put together the full story of the case of a theory with parameters signalling FOPT. 

For pursuing the FOPT contributions, it will suffice to consider only the positive VEV of field $r$ and ignore $l$. The phase transition begins with the usual spontaneous bubble formation and mutual collisions but towards the end, leaves behind two-field domain walls of the type discussed in \cite{Cline:2002ia,Sarkar:2007er}. These should disintegrate on their own, giving an additional contribution to the GW. Gravitational wave spectrum from FOPT is characterized mainly by two parameters $\alpha$ and $\beta$, corresponding to the strength of the phase transition and the inverse rate of tunnelling. In this section, we will give relevant definitions of these two parameters and speculate the observability of the GW spectrum for typical values for them. First order phase transition (FOPT) takes place through the nucleation of bubbles within which $r$ takes the true VEV, $v_R(T)$. The field $r$ tunnels from a metastable state $v_R(T)=0$ to the stable ground state $v_R(T)\neq 0$ considering enough supercooling. These bubbles are spherically symmetric bounce solutions \cite{Coleman1977FateTheory}. After nucleation, these bubbles of true VEV expand through the plasma in the expanding metastable Universe, and depending on the strength of interaction of the bubble wall with the plasma, either the bubble walls accelerate to attain a terminal velocity before collisions take place (non-runaway scenario in plasma) \cite{Caprini2016ScienceTransitions}, or they keep accelerating without bound to reach approximately the speed of light (runaway scenario) \cite{Caprini2016ScienceTransitions,Lewicki:2022pdb,Ellis:2020nnr,Kierkla:2022odc}. There is a third scenario for super-strong phase transitions, where plasma effects can be ignored and they attain speed of light (runaway in vacuum) \cite{Caprini2016ScienceTransitions}. The nucleation rate of bubbles at temperature $T$ is given by \cite{Brdar2019GravitationalBreakingb,Witten1984CosmicPhases,Hogan1983NucleationTransitions,Linde1983DecayTemperature},
\begin{equation}
    \Gamma(T) \simeq T^{4}\left(\frac{S_{3}(T)}{2 \pi T}\right)^{\frac{3}{2}} \mathrm{e}^{-S_{3}(T) / T}
\end{equation}
where $S_3(T)$ is the three-dimensional Euclidean action evaluated for the bounce solution which is obtained by solving the equation of motion,
\begin{equation}\label{eq:eom}
    \frac{\mathrm{d}^{2} r}{\mathrm{~d} x^{2}}+\frac{2}{x} \frac{\mathrm{d} r}{\mathrm{~d} x}=\frac{\mathrm{d} V_{\mathrm{eff}}(r, T)}{\mathrm{d} r}
\end{equation}
with boundary conditions $dr/dx=0$ at $x=0$ and $r\rightarrow 0$ as $x\rightarrow\infty$ where $x$ denotes the 3D radial coordinate. We use \textit{CosmoTransitions} to solve this equation and to get the critical bubble profile $r(x)$. Once the profile is known we can calculate $S_3(T)$ by,

\begin{equation}\label{eq:action}
 S_3(T)=4\pi\int_0^\infty dx x^2\left[\frac{1}{2}\left(\frac{dr}{dx}\right)^2+V_{\text{eff}}(r(x),T)\right]
\end{equation}

Next, the nucleation temperature $T_n$ is the temperature at which there's at least one bubble created per horizon volume,
\begin{equation}\label{eq:nucleation1}
    \int_{T_{n}}^{T_{c}} \frac{\mathrm{d} T}{T} \frac{\Gamma(T)}{H(T)^{4}} = 1
\end{equation}
where the Hubble parameter is given by,
\begin{equation}
    H(T)^{2}=\frac{\rho_{\mathrm{rad}}(T)+\rho_{\mathrm{vac}}(T)}{3 \mathcal{M}_{\mathrm{Pl}}^{2}}=\frac{1}{3 \mathcal{M}_{\mathrm{Pl}}^{2}}\left(\frac{\pi^{2}}{30} g_{*} T^{4}+\Delta V(T)\right)
\end{equation}
with $g_*=134$ is the relativistic degrees of freedom for our model and $\mathcal{M}_{pl}=2.435\times10^{18}$ GeV is the reduced Planck Mass. Vacuum energy density $\rho_{\mathrm{vac}}(T)$ is calculated from $\Delta V(T):=V_{\mathrm{eff}}(0, T)-V_{\mathrm{eff}}\left(v_{R}(T), T\right)$. The condition \eqref{eq:nucleation1} can be approximated by $\Gamma(T)\sim H(T)^4$ and assuming $\Gamma(T)\sim T^4e^{-S_3(T)/T}$ and radiation domination, i.e $H(T)\sim1.66\sqrt{g_*}T^2/3M_{\text{Pl}}$ with $M_{\text{Pl}}=1.22\times10^{19}$ GeV, we get,
\begin{equation}
 \frac{S_3(T_n)}{T_n}\sim-4\log \frac{1.66\sqrt{g_*}T_n}{M_{\text{Pl}}}\sim 100-130
\end{equation}
for $g_*=134$ and $T_n\sim10^6-10^4\text{ GeV}$. Once we have $T_n$, we can define the parameters \cite{Grojean2007GravitationalBeyond,Espinosa2010EnergyTransitions},
\begin{equation}
    \alpha=\frac{1}{\rho_{\mathrm{rad}}\left(T_{n}\right)}\left(\Delta V\left(T_{n}\right)-\left.\frac{T_{n}}{4} \frac{\partial \Delta V(T)}{\partial T}\right|_{T=T_{n}}\right)
\end{equation}
which characterizes the strength of the FOPT; and the inverse duration of the transition given by,
\begin{equation}
    \beta=\left.H\left(T_{n}\right) T_{n} \cdot \frac{\mathrm{d}\left(S_{3} / T\right)}{\mathrm{d} T}\right|_{T=T_{n}}
    \label{eq:inverseduration}
\end{equation}
For strong FOPT, the typical values for these two parameters for our model are, $\alpha\in[0.001,0.1]$ and $\beta/H_*\in[10^2,10^4]$ \cite{Li2021ProspectsModel} where $H_*=H(T_n)$. Alternatively the condition $v_R(T_c)/T_c>1$ where $v_R(T)$ is the temperature dependent VEV of the field $r$ is also used to characterize a strong FOPT. As mentioned in Sec. \ref{sub:twokinds} this condition is satisfied primarily for a small value of $\rho_1$ \cite{Brdar2019GravitationalBreakingb}. Therefore, in Table \ref{tab:BPSOPTFOPT}, we take 3 benchmark points BPF1, BPF2 and BPF3 with small $\rho_1$ values which signals strong FOPT, which we will use to plot the GW spectrum. Apart from the critical and nucleation temperature it is necessary to estimate the percolation temperature $T_p$, defined as the temperature at which a significant portion of the Hubble volume is attains true vacuum and percolation takes place, and make sure that it is not far from the nucleation temperature. Following \cite{Ellis2019OnSignal}, we calculate the percolation temperature by defining the probability of a point still in false vacuum at temperature $T$ to be $P(T)=e^{-I(T)}$ where $I(T)$ is the volume of true vacuum per unit comoving volume given by \cite{PhysRevD.23.876},
\begin{equation}\label{eq:perc1}
 I(T)=\frac{4\pi}{3}\int_T^{T_c}dT'\frac{\Gamma(T')}{T'^4H(T')}\left(\int_T^{T'}\frac{d\tilde{T}}{H(\tilde{T})}\right)^3
\end{equation}
which under the assumption $H(T)\sim\frac{1}{\sqrt{3}\mathcal{M}_\text{pl}}\frac{\pi^2\sqrt{g_*}}{\sqrt{30}}T^2$ takes the form,

\begin{equation}
 \label{eq:perc2}
 I(T)=\frac{4\pi}{3T^3}\left(\sqrt{3}\mathcal{M}_\text{pl}\frac{\sqrt{30}}{\pi\sqrt{g_*}}\right)^4\int_T^{T_c}dT'\frac{\Gamma(T')}{T'^6}\left(1-\frac{T}{T'}\right)^3
\end{equation}
The percolation temperature is calculated using $I(T_p)=0.34$ \cite{Shante1971AnIT}. As long as $T_p$ is of the same order as   $T_n$ we can assume that bubbles nucleated at $T_n$ will percolate and take $T_n$ to be the proxy for temperature in the calculation of $\alpha$ and $\beta$. The results of our calculations for the 3 benchmark points BPF1, BPF2, BPF3 are shown in Table \ref{tab:BPFOPTres}.

Now, there are three sources of GWs in case of a first-order phase transition; i) bubble wall collision \cite{Kosowsky1992GravitationalBubbles}, ii) sound waves \cite{Hindmarsh2014GravitationalTransition} and iii) magnetohydrodynamic (MHD) turbulence \cite{Caprini2006GravitationalFields}. The spherical symmetry of the nucleated bubble is broken when the bubbles collide, which is the basic requirement for GW production. Thus, the total GW spectrum is given by,
\begin{equation}
    h^{2} \Omega_{\mathrm{GW}} \simeq h^{2} \Omega_{ r}+h^{2} \Omega_{\mathrm{sw}}+h^{2} \Omega_{\mathrm{turb}}
\end{equation}
The first term is the contribution from bubble collisions which can be calculated from the field evolution. The second and third terms are the sound wave. In order to calculate the contributions from each of these three sources, we need to distinguish between the three scenarios mentioned above, namely non-runaway, runaway and runaway in vacuum. To do this we compare our $\alpha$ parameter with the $\alpha_\infty$ parameter \cite{Caprini2016ScienceTransitions} and found that $\alpha$ is always smaller than $\alpha_\infty$, which corresponds to the non-runaway scenario. In this scenario the bubble walls expand for enough time to attain terminal velocity before collisions. Therefore the contribution to GW spectrum from collisions is negligible compared to the contribution from sound waves and turbulences. This is because the energy released in collisions comes from the gradient of the field $r$ which scales as the surface area of the bubble, while the energy from sound waves and turbulences come from the bulk of the plasma which scale as the volume.

\begin{figure}[ht]
    \centering
    \includegraphics[width=\linewidth]{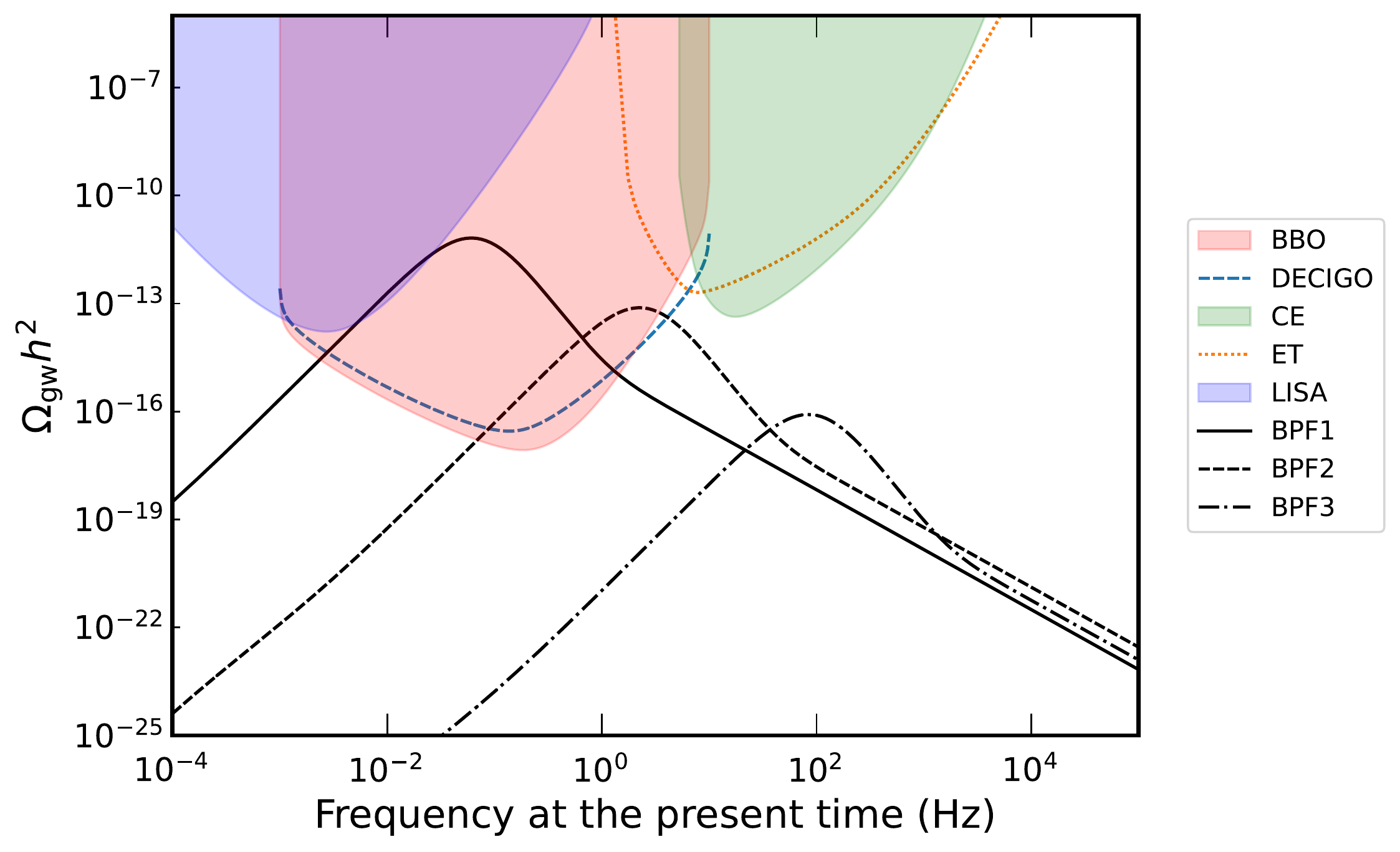}
    \caption{\small Contribution to the GW spectrum from FOPT bubble dynamics alone for BPF1, BPF2, BPF3 with $v_R = 10^4, 10^5, 10^6$ GeV respectively given in Table~\ref{tab:BPSOPTFOPT}, as would be seen today. The coloured regions and lines correspond to \emph{power-law-integrated-sensitivities} \cite{Schmitz2020NewTransitions} of different experiments.}
    \label{fig:spectrumFOPT}
\end{figure}

\begin{figure}[ht]
    \centering
    \includegraphics[width=\linewidth]{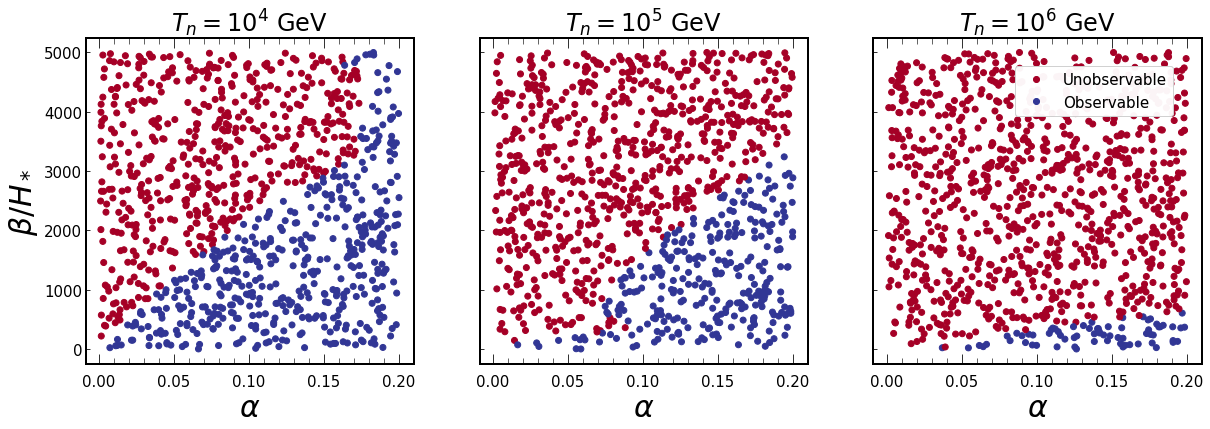}
    \caption{\small Strength of GW peak for different values of $\alpha$ and $\beta/H_*$ for $T_n=10^4,10^5,10^6$ GeV. Red dots signify unobservable peaks in BBO, DECIGO, CE, ET, and LISA whereas blue dots implies observable signal.}
    \label{fig:albeta}
\end{figure}

\subsection{Sound Waves}
The movement of the DWs through the plasma creates pressure waves in the plasma. The contribution of such sound waves to GW is given from numerical fit \cite{Caprini2018CosmologicalWaves,Hindmarsh2015NumericalTransition,Guo_2021},
\begin{equation}\label{eq:Omegasw}
    h^{2} \Omega_{\mathrm{sw}}(f)=2.65 \times 10^{-6}\left(\frac{H_{*}}{\beta}\right)\left(\frac{\kappa_{sw} \alpha}{1+\alpha}\right)^{2}\left(\frac{100}{g_{*}}\right)^{\frac{1}{3}} v_{w} \left(\frac{f}{f_{\mathrm{sw}}}\right)^{3}\left(\frac{7}{4+3\left(f / f_{\mathrm{sw}}\right)^{2}}\right)^{7 / 2}\Upsilon
\end{equation}
where $v_w$ is the wall velocity which we assume as \cite{Lewicki:2021pgr},

\begin{equation}\label{eq:vw}
  v_w=
\begin{cases}
    \sqrt{\frac{\Delta V(Tn)}{\alpha \rho_{\text{rad}}}},& \text{if } \sqrt{\frac{\Delta V(Tn)}{\alpha \rho_{\text{rad}}}}\geq v_J\\
    1,              & \text{otherwise}
\end{cases}
\end{equation}
where $v_J$ is the Jouguet velocity, $v_J=\frac{1}{\sqrt{3}}\frac{1+\sqrt{3\alpha^2+2\alpha}}{1+\alpha}$. The efficiency factor $\kappa_{sw}$ is given by \cite{Schmitz2020NewTransitions,Caprini2016ScienceTransitions},
\begin{equation}\label{eq:kappasw}
  \kappa_{\mathrm{sw}}=
\begin{cases}
    \frac{\alpha}{0.73+0.083 \sqrt{\alpha}+\alpha},& \text{if } v_w\geq v_w^\alpha\\
    \frac{6.9\alpha v_w^{6/5}}{1.36-0.037 \sqrt{\alpha}+\alpha},              & \text{if }v_w<v_w^\alpha
\end{cases}
\end{equation}
where,
\begin{equation}
 v_w^\alpha=\left[\frac{1.36-0.037 \sqrt{\alpha}+\alpha}{6.9(0.73+0.083 \sqrt{\alpha}+\alpha)}\right]^{5/6}
\end{equation}
Lastly the factor $\Upsilon=1-\frac{1}{\sqrt{1+2\tau_{sw}H_*}}$ is a suppression factor dependent on lifetime of sound waves $\tau_{sw}$ \cite{Guo_2021}. It can be parameterized by writing $\tau_{sw}\sim R_*/\overline{U}_f$, where $R_*=(8\pi)^{1/3}v_w/\beta$ and $\overline{U}_f=\sqrt{3\kappa_{\mathrm{sw}}\alpha/4}$ represents mean bubble distance and root-mean-squared fluid velocity \cite{Hindmarsh:2017gnf}.

The peak frequency is given by,
\begin{equation}\label{eq:fsw}
    f_{\mathrm{sw}}=\frac{1.9 \times 10^{-5}}{v_w} \left(\frac{\beta}{H_{*}}\right)\left(\frac{T_{*}}{100 \mathrm{GeV}}\right)\left(\frac{g_{*}}{100}\right)^{\frac{1}{6}} \mathrm{Hz}
\end{equation}

\subsection{MHD Turbulence}
Turbulent motion of the fully ionized plasma also contributes to the GW spectrum, modelled by \cite{Caprini2018CosmologicalWaves,Caprini2009TheTransition,Binetruy2012CosmologicalSources},
\begin{equation}
    h^{2} \Omega_{\mathrm{turb}}(f)=3.35 \times 10^{-4}\left(\frac{H_{*}}{\beta}\right)\left(\frac{\kappa_{\text {turb }} \alpha}{1+\alpha}\right)^{\frac{3}{2}}\left(\frac{100}{g_{*}}\right)^{1 / 3} v_{w} \frac{\left(f / f_{\text {turb }}\right)^{3}}{\left[1+\left(f / f_{\text {turb }}\right)\right]^{\frac{11}{3}}\left(1+8 \pi f / h_{*}\right)}
\end{equation}
where $\kappa_{\mathrm{turb}}=0.05\kappa_{\mathrm{turb}}$ \cite{Caprini2016ScienceTransitions} and,
\begin{equation}
    h_{*}=16.5 \cdot 10^{-6}\left(\frac{T_{n}}{100 \mathrm{GeV}}\right)\left(\frac{g_{*}}{100}\right)^{1 / 6} \mathrm{~Hz}
\end{equation}
The peak frequency is given by,
\begin{equation}
    f_{\text {turb }}=\frac{2.7 \times 10^{-5}}{v_w} \left(\frac{\beta}{H_{*}}\right)\left(\frac{T_{*}}{100 \mathrm{GeV}}\right)\left(\frac{g_{*}}{100}\right)^{\frac{1}{6}}\mathrm{Hz}
\end{equation}

We now summarise the results from all the above-listed contributions. Fig. \ref{fig:spectrumFOPT} shows the GW spectrum from FOPT for the 3 benchmark points BPF1, BPF2, BPF3 with different $v_R$, as given in Table \ref{tab:BPSOPTFOPT} along with the power-law-integrated sensitivity curves (PLISC) of different gravitational wave experiments~\cite{Schmitz2020NewTransitions}. Instead of doing the traditional signal-to-noise ratio (SNR) calculation of each signal against the noise profile of each experiment to infer about the detectability of the signal \cite{Maggiore:1999vm,Allen:1996vm,Allen:1997ad}, we simply use the PLISCs to qualitatively state whether the signal will be detectable or not in the future. A PLIS curve is calculated from the noise profile of the experiment by considering the threshold SNR for the experiment. The PLISCs given here are calculated assuming the threshold to be 1 and observation time to be 1 year for interferometers and 20 years for PTA experiments. Generally signal spectrum $\Omega_\mathrm{signal}$ is said to be detectable in an experiment if the SNR for the signal is larger than the threshold SNR for the experiment. But PLISCs provide a qualitative direct graphical interpretation on the signal detectability without actually calculating the SNR. A signal will be observable if $\Omega_\mathrm{signal}(f)>\Omega_\mathrm{PLIS}(f)$ for some $f$. We see that the peak of BPF1 and BPF2 are within the PLIS curve of BBO and DECIGO whereas BPF3 peak is outside the PLIS curves of any of these experiments. Therefore we conclude that the signals from BPF1 and BPF2 will be observable in BBO and DECIGO assuming a threshold SNR of 1 in both, while BPF3 will not be observed.  In Fig. \ref{fig:albeta}, we show GW peaks for typical values of $\alpha$ and $\beta/H_*$ obtained from our model with different $T_n$, and color code them depending on whether these peaks lie within the PLIS curve of any of these experiments. The red dots correspond to observable peaks whereas the blue dots represent peaks outside the range of BBO, DECIGO, CE, ET, and LISA. We see that for energy $10^6$ GeV it is a lot difficult to get detectable peak, which is verified by us while scanning the model parameter space to find benchmark points.

Note that we did not use the \textit{peak-integrated} sensitivity curves (PISC)s here which also encode the SNR information of the signal, because PISCs are not available for the PTA experiments. This concludes our discussion of the contribution to the GW from FOPT bubble dynamics.


\section{Gravitational Waves from Two-field Second Order Phase Transition}
\label{section:DWLRSM}
As discussed in Sec.~\ref{sec:intro} and~\ref{sub:twokinds}, in left-right symmetric models, a second order phase transition (SOPT) can produce gravitational waves via annihilating domain walls that separates left and right-like domains. Such domain walls also arise at the end of a first-order phase transition (FOPT), where left and right-like bubbles come into contact (see Fig.~\ref{fig:DegFldFOPT}). In this section we will discuss the GW spectrum from disintegrating domain walls in case of an SOPT. First, we will briefly describe the non-trivial axion-like domain wall solutions in MLRSM and then dive into the methods used for numerical simulation. Finally we compare our simulation results with existing formulae for GW spectrum from $Z_2$ and axionic domain walls. 
\subsection{Axion-like domain walls in MLRSM}
First let us look into the mathematical structure of domain walls in MLRSM. Defining $l=v_0\sin\theta,r=v_0\cos\theta$ \cite{PhysRevD.59.103508}, with $v_0$ standing for $\delta_R^0$ or $\delta_L^0$ of Eq. (\ref{eq:tree0}), we can redefine the tree level potential in Eq. \eqref{eq:tree0} as,

\begin{equation}
 \label{eq:axionpot}
 V(v_0,\theta)=\frac{v_0^4}{32}\left(\rho_3-2\rho_1\right)\left(1-\cos4\theta\right)+\frac{v_0^2}{2}\left(\frac{v_0^2\rho_1}{2}-\mu_3^2\right)
\end{equation}
This potential has degenerate minima at the points $(v_0/\sqrt{2},\theta)=(\eta,n\pi/2)$ with $\eta=\mu_3/\sqrt{\rho_1}$ and saddle points at $(2\mu_3/(2\rho_1+\rho_3),n\pi/2+\pi/4)$, where $n=0,1,2,...$ etc. The effective potential respects the $Z_4$ symmetry of the tree level potential as seen in Sec. \ref{sec:effpotential}. It is sufficient to consider only the first quadrant of the $l-r$ plane which corresponds to $\theta\in[0,\pi/2]$, i.e. the Universe cools down to either $\theta=0$ or $\theta=\pi/2$ phase in causally disconnected regions. The regions with $\theta=0$ or $\pi/2$ corresponds to $(l,r)=(0,\eta)$ or $(\eta,0)$ which we call right-like (R-like) or left-like (L-like) domains. 

\begin{figure}[ht]
    \centering
    \includegraphics[width=0.7\linewidth]{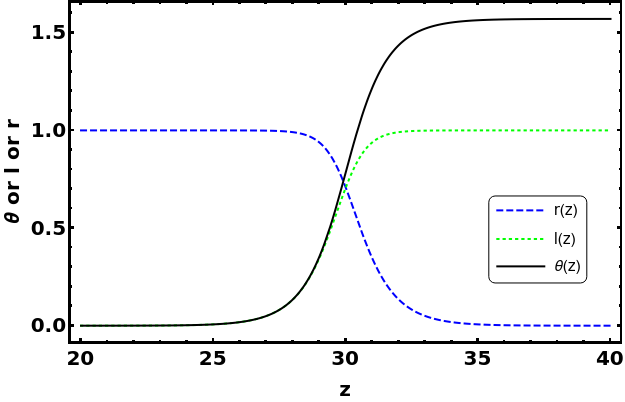}
    \caption{\small The approximate domain wall profile for $\eta=m_\theta=1,z_0=30$. The black solid line shows the angle $\theta(z)$ which changes from $0$ to $\pi/2$ across the wall. The blue and the green lines represents the fields $r$ and $l$ respectively that changes from $0$ to the VEV $\eta$ across the wall.}
    \label{fig:axionkink}
\end{figure}

It is not possible to find a general exact analytical solutions for the above potential but for small $\rho_3-2\rho_1$, we can fix $v_0\sim\eta$ and the above potential is similar to axion-like potential with $N_\text{DW}=4$ \cite{Hiramatsu:2012sc} with a mass term of the ``axion'' like field $\theta$ given by,
\begin{equation}
 \label{eq:axionmass}
 m_\theta=\eta\sqrt{\frac{\rho_3-2\rho_1}{2}}
\end{equation}
Then the Lagrangian takes the form,
\begin{equation}
 \label{eq:thetalag}
    \mathcal{L}=\frac{\eta^2}{2}(\partial_\mu\theta)^2-V(\theta)
\end{equation}
The classical equation of motion for this Lagrangian in Minkowski metric gives an approximate static solitonic solution, a domain wall perpendicular to $z$-axis positioned at $z_0$ \cite{Hiramatsu:2012sc,PhysRevD.59.103508},
\begin{equation}
 \label{eq:axionkink}
 \theta(z)=\tan^{-1}\text{exp}(m_\theta (z-z_0))
\end{equation}
This is also called a ``kink'' solution and each domain wall separates two nearby vacua. The domain wall width can be approximated as the inverse of the mass term,
\begin{equation}
\label{eq:DWWidth}
 \delta_\text{wall}\sim m_\theta^{-1}
\end{equation}
The domain wall has a surface tension, given by,
\begin{equation}
    \label{eq:DWTension}
    \sigma_{\text{wall}} = \int_{-\infty}^{\infty}dz\rho_\theta=\frac{m_\theta\eta^2}{2}
\end{equation}
where $\rho_\theta=\frac{\eta^2}{2}\left| \nabla \theta\right|^2+V(\theta)$ is the static energy density of the field $\theta$. We will assume this approximate profile of the domain wall to be true for ranges of parameters that we consider here. The $l$ and $r$ profiles are then given by $l=\eta\sin\theta=\eta e^{m_\theta(z-z_0)}/\sqrt{1+e^{2m_\theta(z-z_0)}}$ and $r=\eta\cos\theta=\eta/\sqrt{1+e^{2m_\theta(z-z_0)}}$. In Fig. \ref{fig:axionkink}, we show the kink solution for $\eta=m_\theta=1,z_0=30$.

\subsection{The scaling regime of domain wall dynamics}\label{sec:scalingregime}
After production, the domain walls evolve in the expanding Universe according to the Klein-Gordon equations of motion of the scalar fields given in Eq.~\eqref{eomphi}. An important feature of domain wall evolution without bias is the era of scaling when the characteristic length scales like the radius of curvature of domain walls or domain size is comparable to the Hubble radius,
\begin{equation}
    \label{scaling1}
    R\sim H^{-1} \sim t
\end{equation}

This phase of evolution is found both numerically \citep{Press:1989yh,Coulson:1995nv,Garagounis:2002kt,Oliveira:2004he,Avelino:2005pe,Larsson:1996sp} and analytically \citep{Hindmarsh:1996xv,Hindmarsh:2002bq,Avelino:2005kn} in the literature, and it corresponds to ``at least one domain wall per horizon" scenario. While in the scaling regime, the energy density of the domain walls is given by,

\begin{equation}
    \label{scaling2}
    \rho_\text{wall} \sim  \frac{\sigma_\text{wall} S_\text{wall}}{H^{-3}} \sim \sigma_\text{wall} R^2/R^3 \sim \sigma_\text{wall}/t
\end{equation}
with $\sigma_\text{wall}$ given in Eq. \eqref{eq:DWTension} and $S_\text{wall}$ is the wall area within one Hubble horizon. Numerically the scaling regime can be verified by estimating $S_\text{wall}$ from the comoving area within the simulation box averaged over the lattice volume,
\begin{equation}
    \label{scaling3}
    A/V \propto \frac{a(t)S_\text{wall}}{H^{-3}} \propto t^{-1/2} \propto \tau^{-1}
\end{equation}
where $\tau$ is the conformal time. We calculate this quantity from the simulation box using the method described in Appendix ~\ref{sec:AreaDensity}. We define the proportionality constant $\mathcal{A}$ of Eq. \eqref{scaling3} by,
\begin{equation}
 \label{eq:areaparam}
 A/V=\mathcal{A}\,\left(\frac{\tau}{\tau_i}\right)^{-\zeta}
\end{equation}
where $\tau_i$ is the initial conformal time of evolution. For simple models with $Z_2$ symmetric potential, the authors of \cite{Hiramatsu2014OnWalls} have estimated the value of $\mathcal{A}$ to be $\simeq0.8\pm0.1$. They have also shown that for axion-like potential, $\mathcal{A}$ increases a little with $N_\text{DW}$~\cite{Hiramatsu:2012sc}. We will calculate this parameter from our model by simulating domain wall evolution in a 3D lattice in Sec.~\ref{sec:numerical}.

\subsection{Effect of a bias}
An unbiased domain wall is stable and after scaling with the expanding Universe, eventually it dominates the energy density of the Universe, which is in conflict with standard cosmology. As discussed in section \ref{sub:twokinds}, in both the SOPT and FOPT case, \emph{frustrated} domain walls are produced which need to move under the vacuum pressure difference between the true and the false vacua such that they annihilate and vanish after some time. This can be achieved by adding a small parity violating term in the tree potential. Effective field theory suggests the lowest order such terms at dimension six level \cite{Borah:2022wdy,Borah:2013mqa,Ibarra:2018dib,Borah:2019ldn},

\begin{equation}
 \label{eq:bias1}
 V_\text{bias}=\frac{f_L}{\Lambda^2}l^6+\frac{f_R}{\Lambda^2}r^6
\end{equation}
where $\Lambda$ is a cutoff scale, generically the Planck scale $M_{\rm pl}$, but could be the unification or an intermediate scale in a GUT scenario. The dimensionless parameters $f_L,f_R$ are expected to be $\ll 1$. It introduces an energy difference between the $L$ and the $R$ minima $\delta V = \epsilon\,\eta^6/\Lambda^2$ where $\epsilon=(f_L-f_R)$, which in turn exerts a vacuum pressure on the domain wall $p_V\sim \delta V$. The pressure difference should be large enough to make the walls disappear before DWs start dominating the energy density of the Universe. In scaling regime (see Sec. \ref{sec:scalingregime}), the decay time is estimated as the epoch when the pressure due to surface tension $p_T\sim\sigma_\text{wall}/t$ becomes comparable to the volume pressure $p_V$,
\begin{equation}
    \label{eq:decaytime}
    t_{\text{dec}}\sim\frac{m_\theta \eta^2}{2\delta V}
\end{equation}

However, the requirement of percolation of both types of vacuum means that the energy difference cannot be arbitrarily large. The ratio of the percolation probabilities is given by $p_+/p_-\simeq e^{-4\delta V/\rho_1\eta^4}$ \cite{Gelmini1989CosmologyBreaking}, which gives an upper bound on the bias $\delta V\leq-\frac{\rho_1\eta^4}{4}\log\left[(p_+/p_-)_\text{min}\right]$. In a 3D cubic lattice, if the probability of attaining a vacuum $p_\pm$ is greater than a critical value $p_c = 0.311$, then infinite clusters of the vacuum can form \cite{Stauffer1979ScalingClusters}. Assuming $p_+=0.311$ and $p_-=1-0.311=0.689$ we get,
\begin{equation}
    \label{eq:biasupperbound1}
    \delta V\leq 0.2\rho_1\eta^4
\end{equation}

\noindent On the other hand, requirements for the scaling solution gives another bound \cite{Hiramatsu2010GravitationalWalls},
\begin{equation}
    \label{eq:biasscaling}
    \delta V < 0.3\sqrt{g_*}\,\frac{m_\theta \eta^4}{M_\text{pl}}
\end{equation}
where $g_*$ is the number of relativistic degrees of freedom at transition temperature $T_c$ and $M_\text{pl}$ is the Planck mass. However, if domain walls keep scaling for a long time, eventually they dominate the energy density of the Universe. Requiring that the walls decay before they dominate the energy density, we get a lower bound on the bias \cite{Hiramatsu2010GravitationalWalls},

\begin{equation}
    \label{eq:biasdomination}
    \delta V > \frac{8\pi}{3}\left(\frac{m_\theta}{M_\text{pl}}\right)^2\eta^4
\end{equation}


\subsection{Numerical simulation of domain wall evolution: equations of motion}\label{sec:numerical}
In the following subsections, we summarize the methods used to simulate domain wall evolution in MLRSM and calculate the GW spectrum. Detailed overview of the simulation techniques is given in Appendix~\ref{app:A}. We compare the outcome of our simulations with the existing results for domain walls from generic $Z_2$ symmetry breaking~\cite{Kawasaki2011StudyWalls,Hiramatsu2014OnWalls} or string-wall network from Peccei-Quinn type symmetry breaking~\cite{Hiramatsu:2012sc}.

We consider the tree level potential as in Eq. \eqref{eq:tree0} to simulate domain wall decay. Domain walls from LRSM potential with two real fields has not been simulated before as per our knowledge. Note that we are not interested in the formation mechanism of the domain walls but only the late time evolution. We assume that a \textit{frustrated network of domain walls} has already formed at the beginning of our simulation. Also we perform our simulation without bias in order to test the scaling solutions (see Sec. \ref{sec:scalingregime}). Since our simulation time is limited, we need very large biases to make the walls decay within the limited time. But GW detectability dictates small biases. Putting a large bias in the simulation makes the walls deviate from scaling. Therefore we simulate unbiased walls and assume that after scaling till time $t$, domain walls suddenly vanish and calculate the required bias $\delta V$ for this to happen using Eq. \eqref{eq:decaytime} by equating $t$ with $t_\text{dec}$. The Klein-Gordon equations in FLRW metric are given as \cite{Hiramatsu:2012sc,mukhanov2005physical},
\begin{align}
    \label{eomphi}
    \nonumber\ddot{l}+3H\dot{l}-\frac{\nabla^2}{a^2}l+\frac{\partial V}{\partial l}&=0\\
    \ddot{r}+3H\dot{r}-\frac{\nabla^2}{a^2}r+\frac{\partial V}{\partial r}&=0
\end{align}
where dot denotes derivative w.r.t. $t$ and $a=a(t)$ is the scale factor. This equation also describes the evolution of domain walls.

\begin{figure}[ht]
\centering
    \begin{subfigure}{0.24\linewidth}
\includegraphics[width=\linewidth]{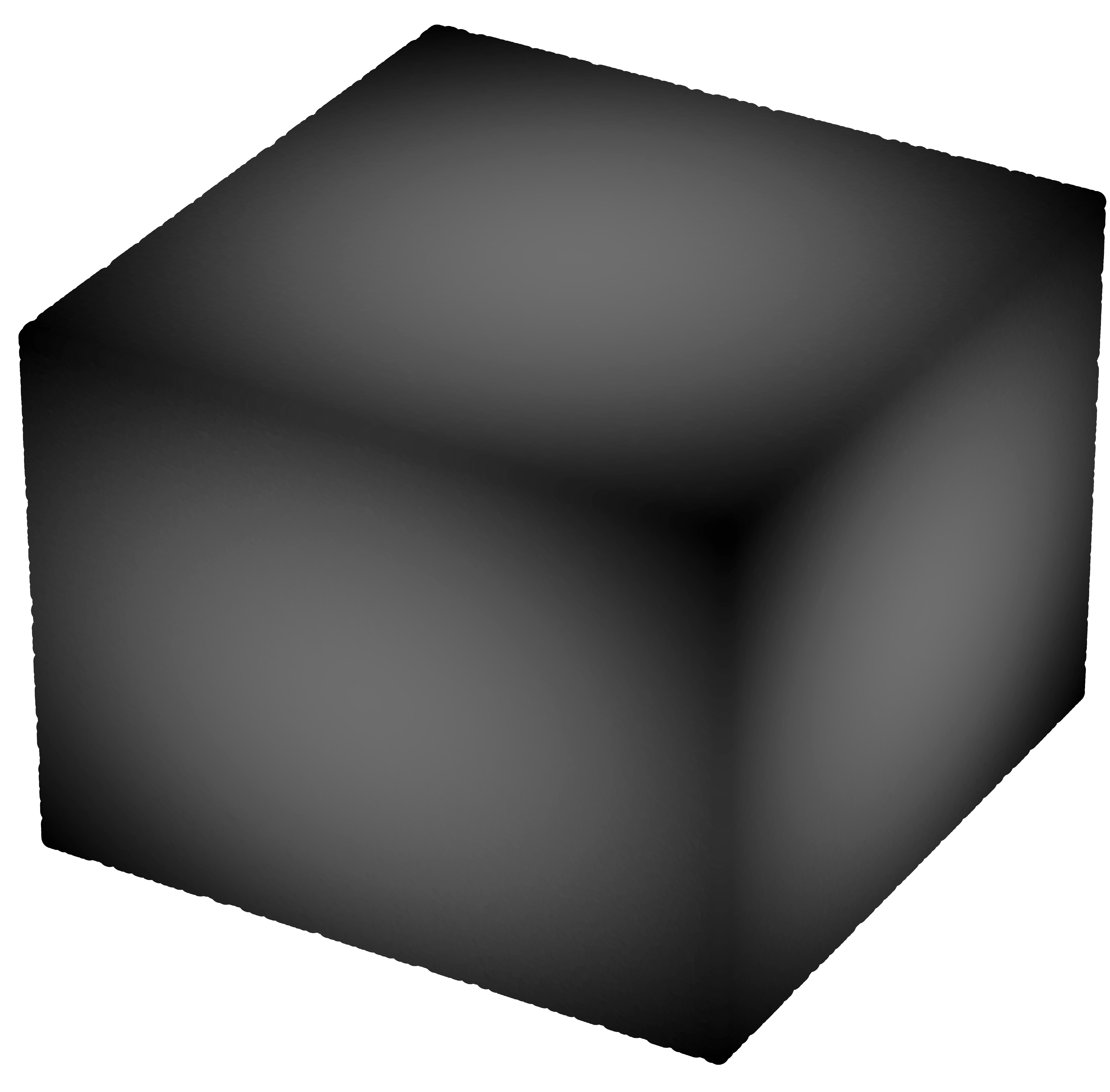}
    \end{subfigure}
    \begin{subfigure}{0.24\linewidth}
\includegraphics[width=\linewidth]{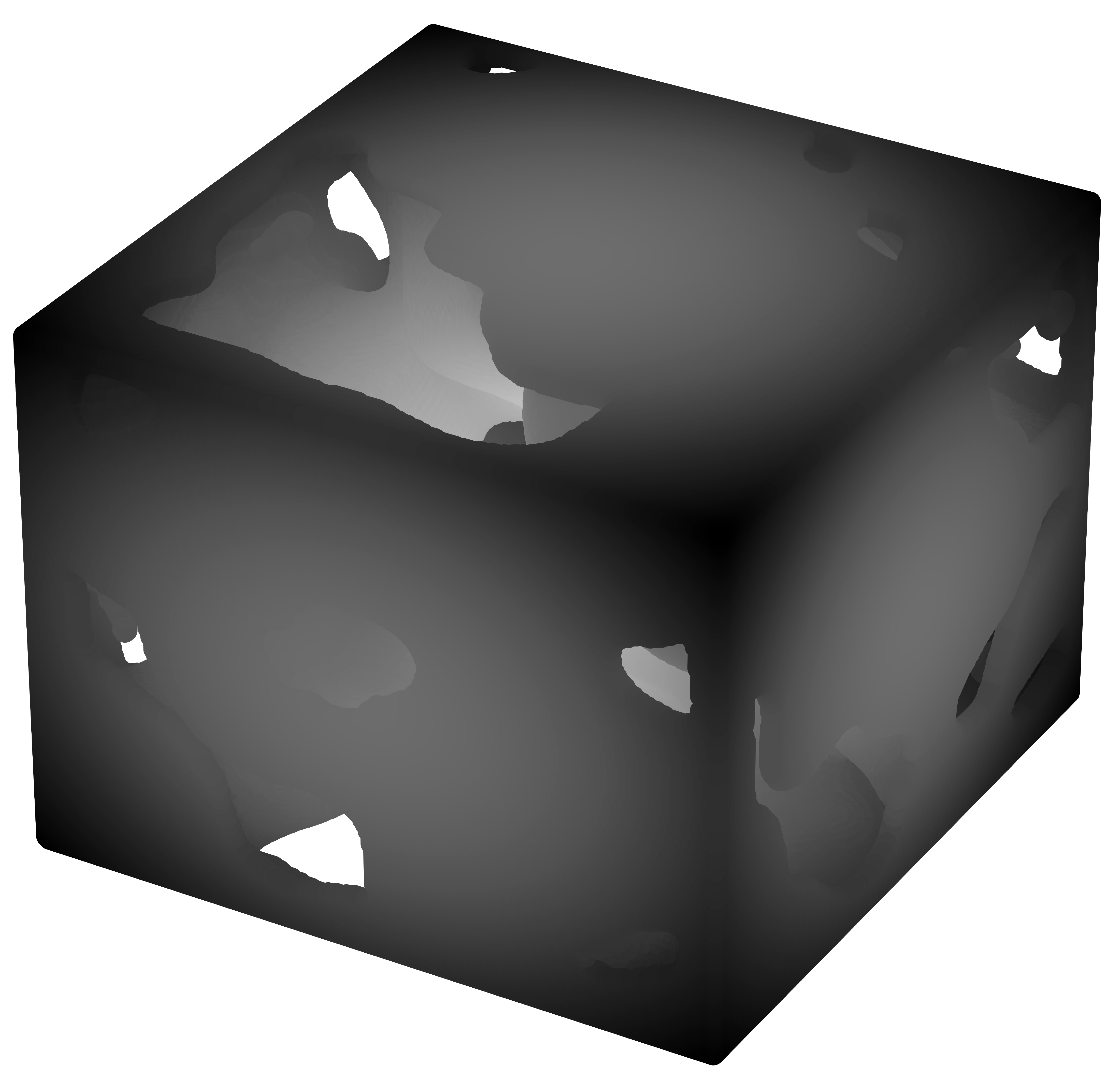}
    \end{subfigure}
    \begin{subfigure}{0.24\linewidth}
\includegraphics[width=\linewidth]{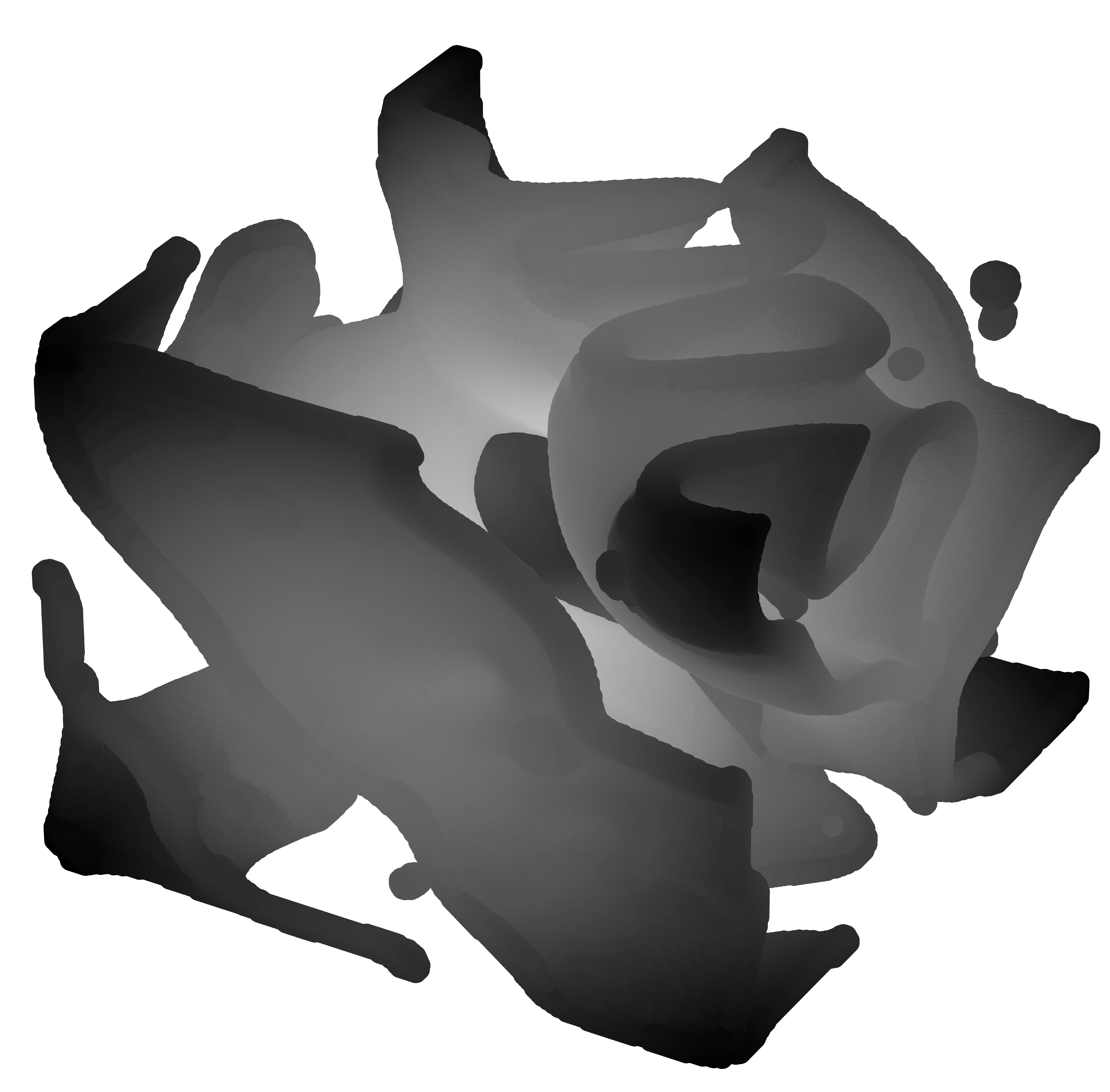}
    \end{subfigure}
    \begin{subfigure}{0.24\linewidth}
\includegraphics[width=\linewidth]{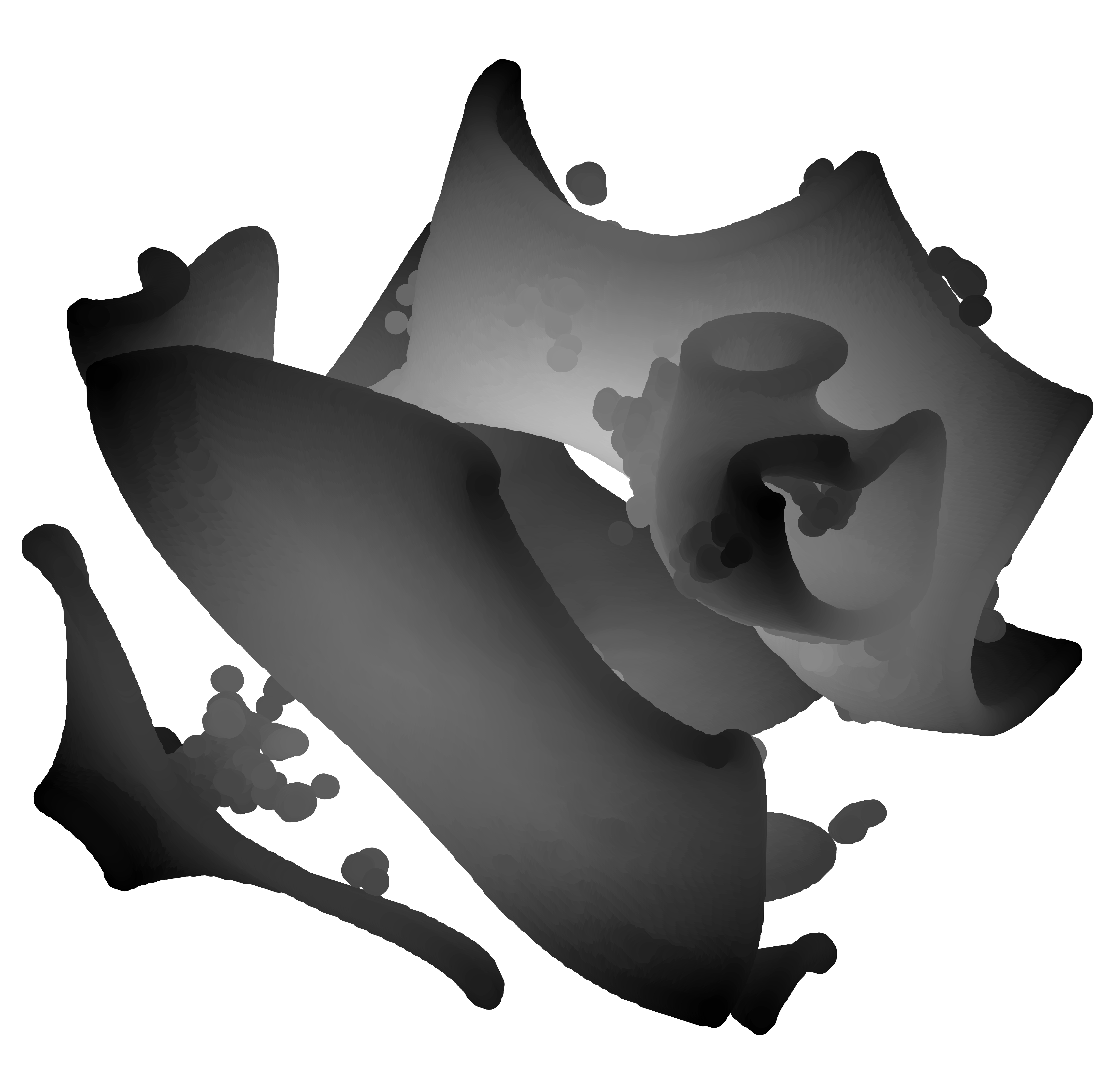}

    \end{subfigure}
\caption{\small Evolution of unbiased domain walls from initial domain sizes comparable to Hubble horizon. We see that wall segments of different sizes emerge with time and small sized walls shrink while larger walls expand in a way that maintains the scaling behavior.}
    \label{fig:3d}
\end{figure}

In Figure \ref{fig:3d}, the evolution of the unbiased domain wall network is shown for $\rho_1=0.1$ and $\rho_3=1.7$ with initial domain sizes of $\mathcal{O}(1)$ (in terms of $\eta^{-1}$). Small domain walls of size $\mathcal{O}(1)$ seems to uniformly cover the whole space within the box of size 50 initially. With time, the Hubble horizon increases within the box and the domains expand with it. It can be seen that still some small domain walls are formed at later time due to random fluctuations of the fields, which shrink and annihilate while large walls expand in order to maintain the scaling solutions. The walls enter the scaling regime discussed in Sec.~\ref{sec:scalingregime} at around conformal time $\tau/\tau_i\sim4$ where $\tau_i=2$ is the initial conformal time of our simulation.

\begin{figure}[ht]
    \centering
    \includegraphics[width=0.8\linewidth]{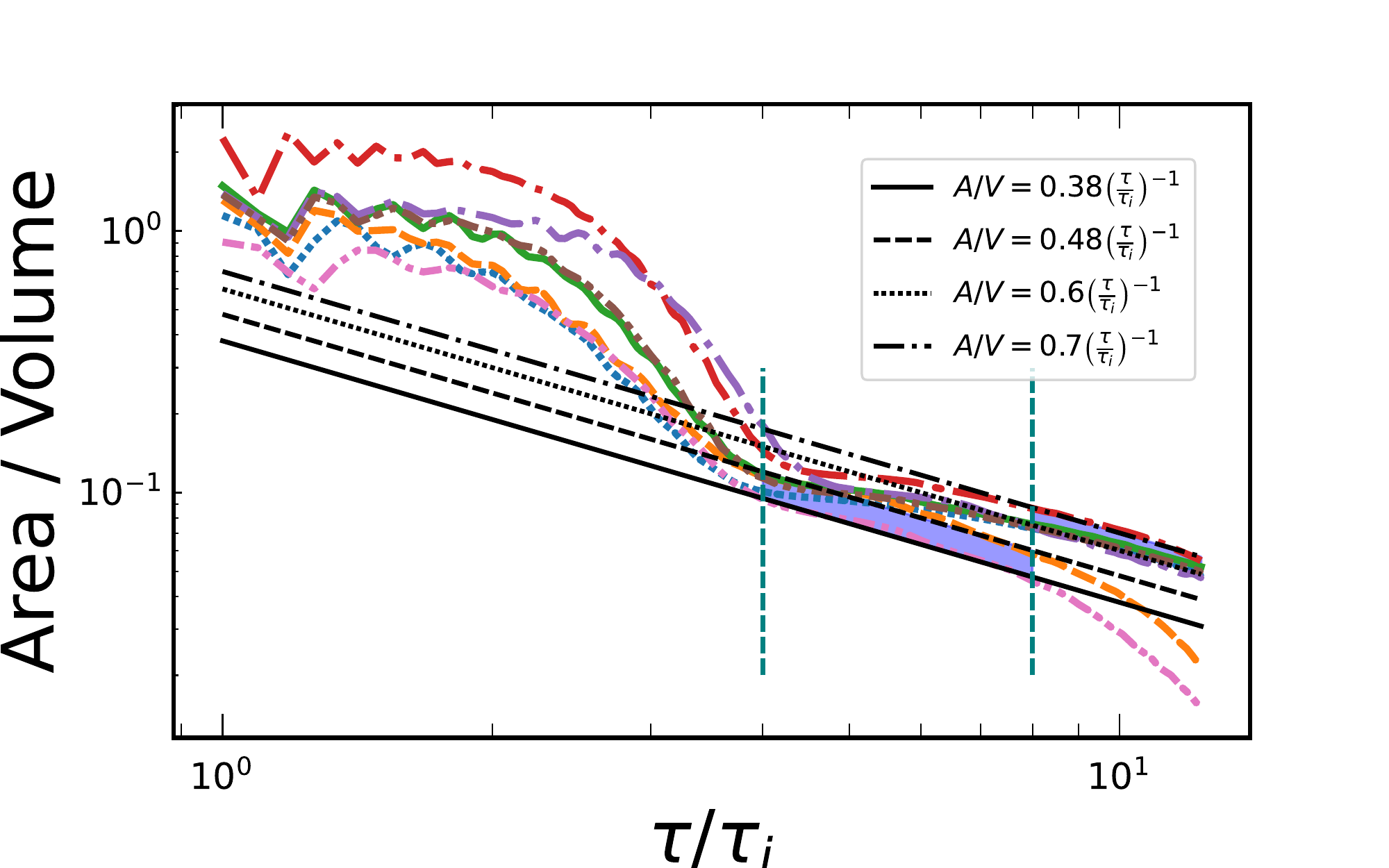}
    \caption{\small Comoving area density with time for several runs. We see that the walls enter scaling regime around $\tau/\tau_i\sim4$ or $8$ for different cases. The black fitted lines represent, $A/V=\mathcal{A}(\tau/\tau_i)^{-1}$ with $\mathcal{A}$ ranging from 0.38 to 0.7. The shaded regions are where we obtain scaling solutions for different initial conditions of the fields $l$ and $r$.}
    \label{fig:areadensity}
\end{figure}

For MLRSM, we calculate $\mathcal{A}$ and $\zeta$ by fitting them in the plot of log of area density vs log of $\tau$.  We calculate the comoving area density $A/V$ of the domain walls within the simulation box at different times and plot in Fig. \ref{fig:areadensity} for several runs. The area density calculation method is described in Appendix~\ref{sec:AreaDensity}. In logarithmic scales, we see that around $\tau/\tau_i\sim4$ or $8$, the domain walls enter scaling regime, where we fit $\mathcal{A}$ and $\zeta$ in Eq. \eqref{eq:areaparam}. Due to small size of lattice we took, we obtain some deviations from exact scaling solutions in many runs. A source of this deviation could be the dependence on initial domain sizes since we observed that bigger initial domain sizes $\mathcal{O}(10)$ results in more deviation from scaling, while domains of initial size comparable to Hubble, $\mathcal{O}(1)$ exhibit exact scaling behaviour or small deviations. A proper study of deviation from scaling for MLRSM potential or other potential, based on initial domain wall size is of general interest and beyond the scope of this article and is left for a future work. In Fig. \ref{fig:areadensity}, the black fitted lines represent Eq. \eqref{eq:areaparam} with $\mathcal{A}$ ranging from $0.38$ to $0.7$ and $\zeta=-1$. The reason of smaller $\mathcal{A}$ compared to $0.8$ could be again dependence on initial conditions or the non trivial structure of the potential of MLRSM. For further analysis, we take $\mathcal{A}\sim0.6$ which is the most common value we obtained from our simulations.

\begin{figure}[ht]
    \centering
    \includegraphics[width=\linewidth]{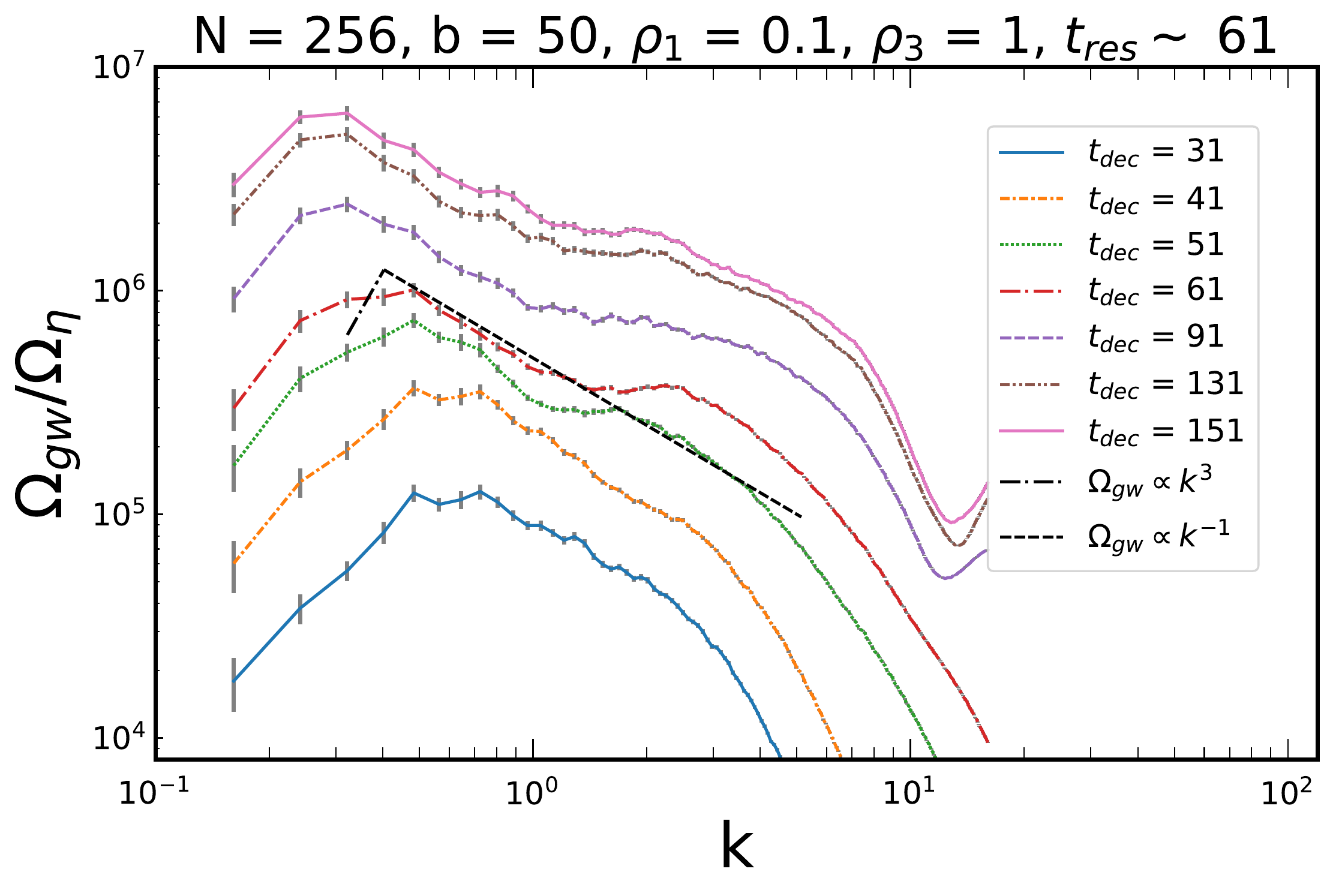}
    \caption{\small Normalized GW spectra as a function of comoving wavenumber for a sample run with initial domain wall size $\sim 1$ with error bars $\sim\sigma/\sqrt{N}$. A straight segment fit with black dashed lines is made to the $t\sim t_{\rm res}\sim 61$ curve. All the physically acceptable curves lie below this as explained in text. The peak is seen to be at the expected location $k\sim k_h\sim\mathcal{O}(0.4-0.5)$, corresponding to the Hubble radius for $t\sim 61$.}
    \label{fig:nobias}
\end{figure}

\begin{figure}[ht]
    \centering
    \includegraphics[width=\linewidth]{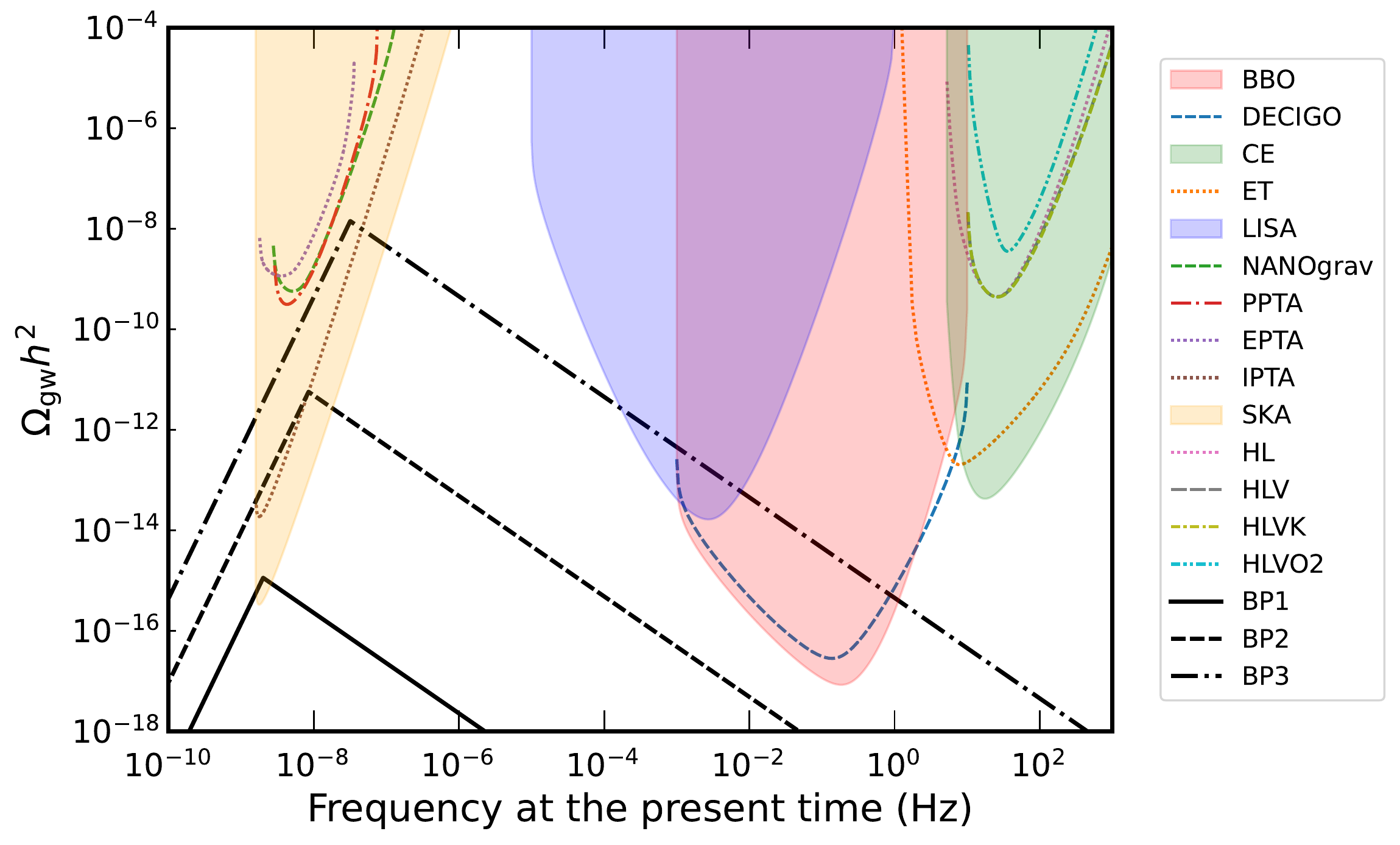}
    \caption{\small \textbf{GW spectra for benchmark points} : The spectra as would be seen today, from disintegrating DWs for BP1, BP2, BP3  with $v_R = 10^4, 10^5, 10^6$ GeV respectively and corresponding scales $\Lambda \sim 10^{16}, 10^{17}, 10^{18}$ GeV as listed in Table \ref{tab:BPSOPTFOPT}.  The coloured regions and lines show the \emph{power-law-integrated-sensitivities} of different experiments.}
    \label{fig:spectrumSOPT}
\end{figure}

\begin{figure}[ht]
    \centering
    \includegraphics[width=\linewidth]{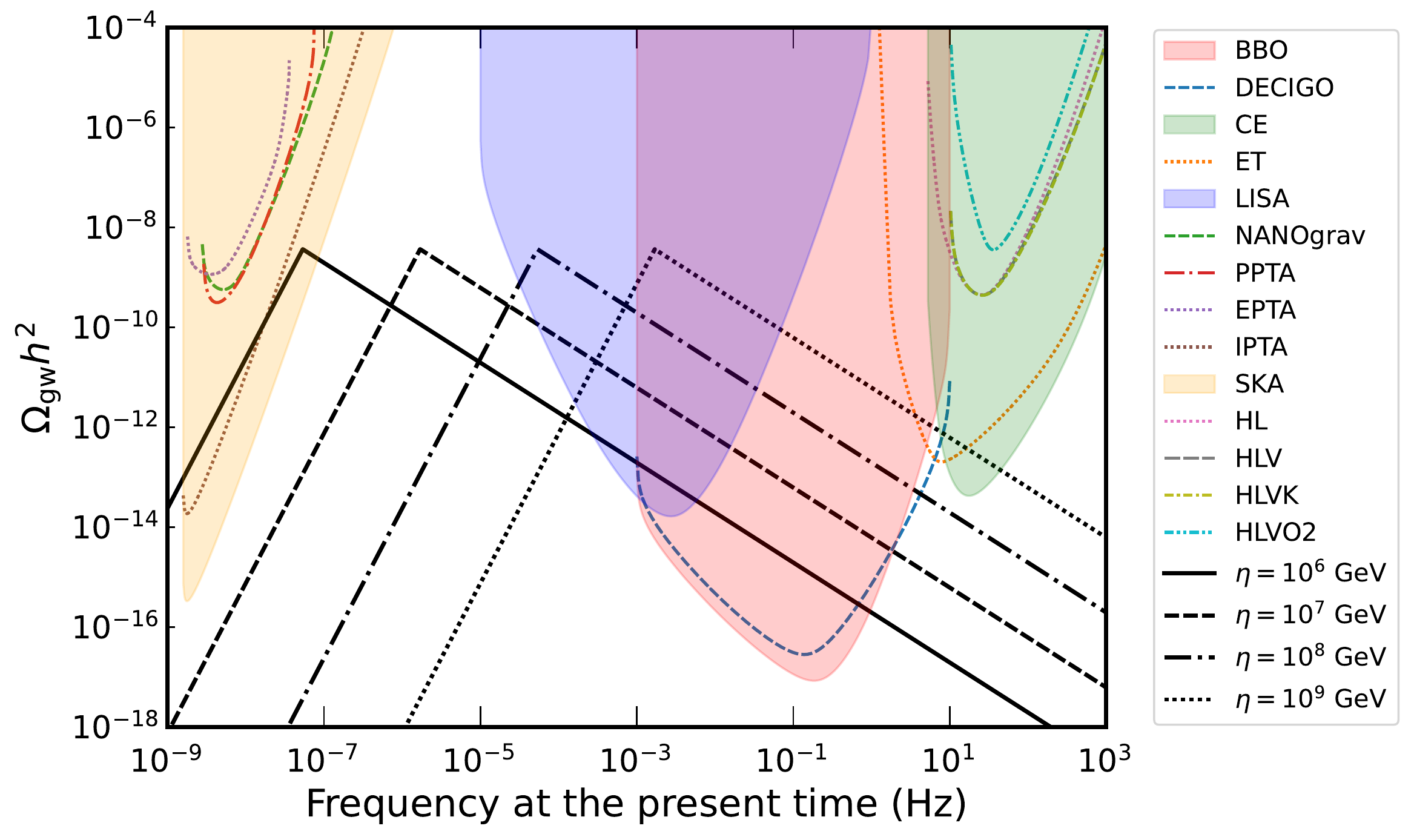}
    \caption{\small \textbf{GW spectra, $M_R$ scale dependence} : Here $\eta=10^6,10^7,10^8$ and $10^9$ GeV and the bias determined by $\epsilon=0.1$ and cutoff scale, $\Lambda=10^{18}$ GeV. Here we have taken $\rho_1=0.4$ and $\rho_3=1.7$. The coloured regions and lines show the \emph{power-law-integrated-sensitivities} of different experiments. }
    \label{fig:spectrumEta}
\end{figure}

\begin{figure}[ht]
    \centering
    \includegraphics[width=\linewidth]{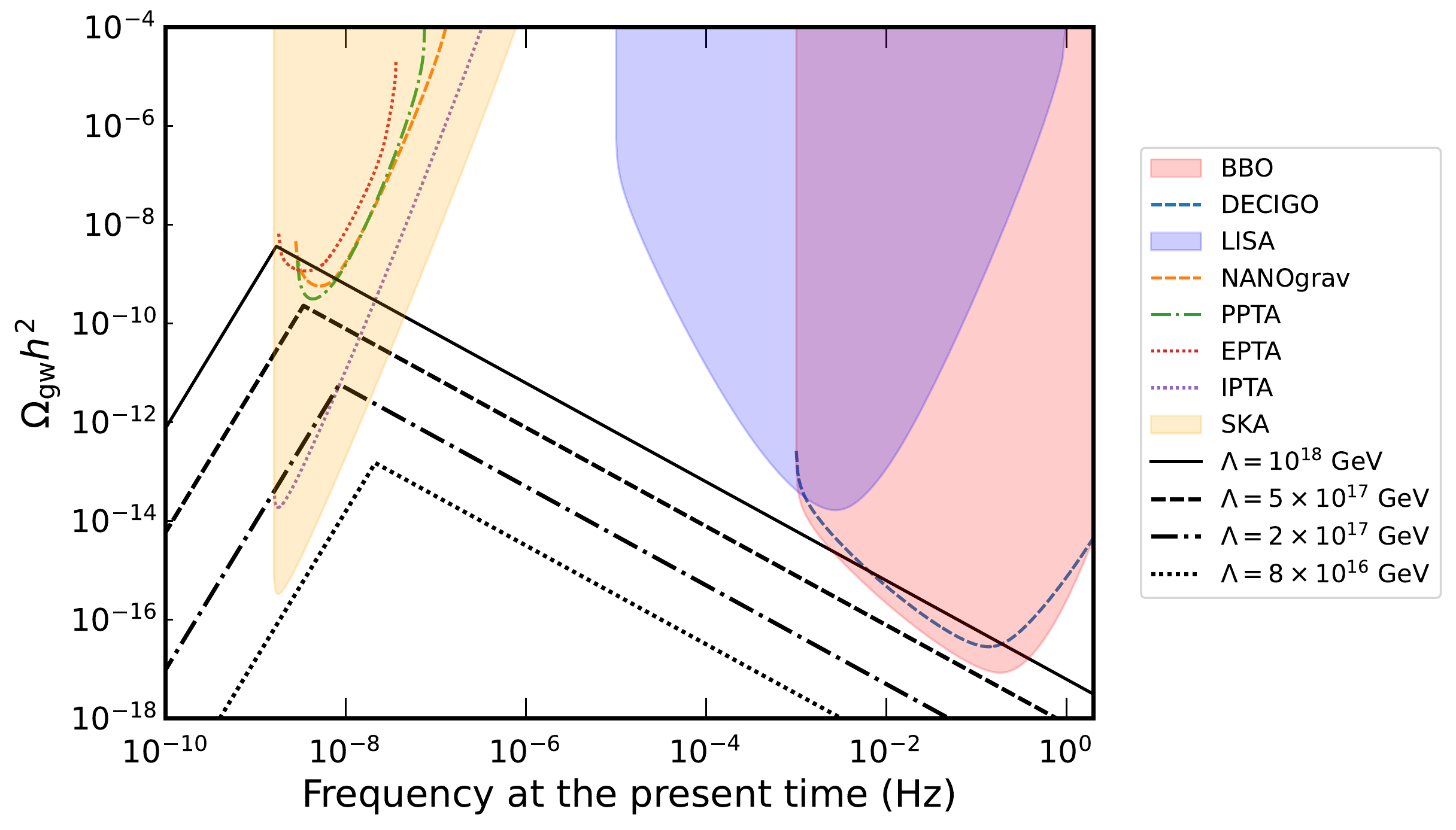}
    \caption{\small \textbf{GW spectra, bias dependence} : Results for potentially observable scale $\eta=10^5$ GeV and bias determined by fixing $\epsilon=0.1$ and varying the cutoff scale $\Lambda$. The smaller the bias, the more the amplitude.  Here we have taken $\rho_1=0.4$ and $\rho_3=1.7$.}
    \label{fig:spectrumBias}
\end{figure}

\begin{figure}[ht]
    \centering
    \includegraphics[width=\linewidth]{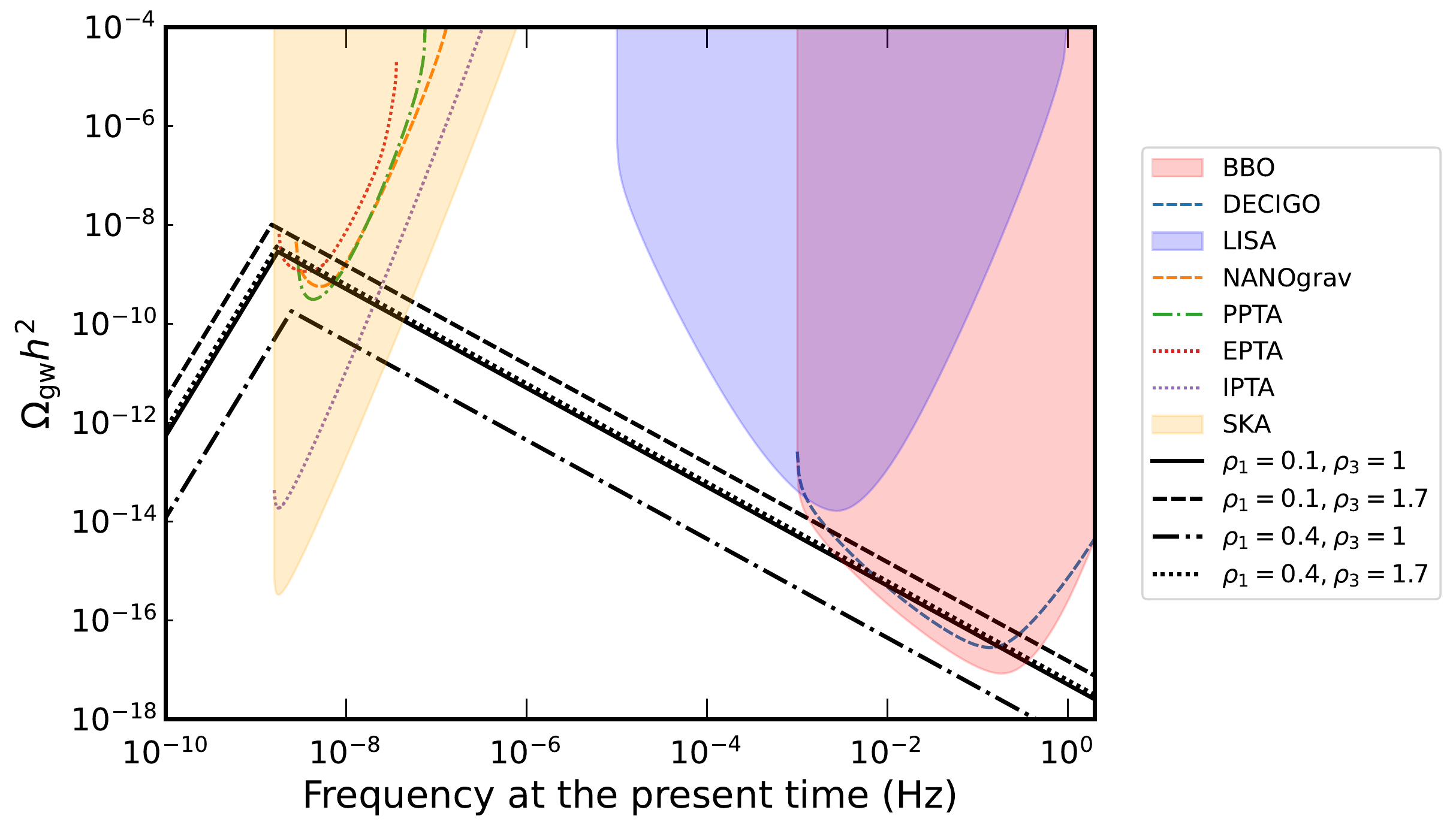}
    \caption{\small \textbf{GW spectra with variation of the quartic parameters $\rho$} : Results for representative $\rho_1$ and $\rho_3$ values. Here we have taken $\eta=10^5$ GeV and bias determined by $\epsilon=0.1$ and cutoff scale $\Lambda=10^{18}$ GeV.}
    \label{fig:spectrumRho}
\end{figure}

\subsection{Numerical simulation of domain wall evolution: calculation of gravitational waves spectrum}
The evolution of domain walls and subsequent annihilation produces gravitational waves that we can calculate from our simulated fields. We use the notation of \cite{Kawasaki2011StudyWalls} and in Appendix \ref{app:A} we summarize the methodology of the calculation,
which utilizes the Green's function method.
The energy density of GW is given by Eq. \eqref{energydensitygw}. Replacing the ensemble average by an average over the volume of the comoving box, we obtain,
\begin{equation}
    \label{gw1}
\rho_{\mathrm{gw}}=\frac{1}{32 \pi G a^{4}} \frac{1}{V} \int \frac{d^{3}\mathbf{k}}{(2 \pi)^{3}} \bar{h}_{i j}^{\prime}(\tau, \mathbf{k})\bar{h}_{ij}^{\prime *}(\tau, \mathbf{k})
\end{equation}

Substituting the form of $\bar{h}_{ij}$ from Eq. \eqref{bessel2} and \eqref{bessel3}, upto first order of $aH$, we obtain,
\begin{align}
\nonumber
\rho_{\mathrm{gw}}= \frac{2 \pi^{2} G}{a^{4} V} \int \frac{d^{3} \mathbf{k}}{(2 \pi)^{3}} \frac{1}{k^{2}} \times \sum_{i j}&\left\{\left|\int_{x_{i}}^{x_{f}} d x^{\prime} \sqrt{x^{\prime}} a\left(x^{\prime}\right) N_{\nu}\left(x^{\prime}\right) T_{i j}^{\mathrm{TT}}\left(\tau^{\prime}, \mathbf{k}\right)\right|^{2}\right.\\
&\left.+\left|\int_{x_{i}}^{x_{f}} d x^{\prime} \sqrt{x^{\prime}} a\left(x^{\prime}\right) J_{\nu}\left(x^{\prime}\right) T_{i j}^{\mathrm{TT}}\left(\tau^{\prime}, \mathbf{k}\right)\right|^{2}\right\}
\end{align}
where we have used approximations for $k\tau\gg1$,
\begin{equation}
    \label{approx1}
    J_\nu(k\tau)\rightarrow\sqrt{\frac{2}{\pi(k\tau)}}\cos\left(k\tau-\frac{\nu\pi}{2}-\frac{\pi}{4}\right),\quad N_\nu(k\tau)\rightarrow \sqrt{\frac{2}{\pi(k\tau)}}\sin\left(k\tau-\frac{\nu\pi}{2}-\frac{\pi}{4}\right)
\end{equation}
and averaged over a period of oscillation of $\bar{h}_{ij}(\tau,\mathbf{k})$ with time. The fraction of energy density of GWs given by Eq. \eqref{omega} becomes,
\begin{align}
\label{numericalomega1}
\nonumber
\Omega_{\mathrm{gw}}(k,t)= \frac{2 G^{2} k}{3 V a(t)^{4} H(t)^{2}} \int d \Omega_{k} & \times \sum_{i j}\left\{\left|\int_{x_{i}}^{x_{f}} d x^{\prime} \sqrt{x^{\prime}} a\left(x^{\prime}\right) N_{\nu}\left(x^{\prime}\right) T_{i j}^{\mathrm{TT}}\left(\tau^{\prime}, \mathbf{k}\right)\right|^{2}\right.\\
&\left.\quad+\left|\int_{x_{i}}^{x_{f}} d x^{\prime} \sqrt{x^{\prime}} a\left(x^{\prime}\right) J_{\nu}\left(x^{\prime}\right) T_{i j}^{\mathrm{TT}}\left(\tau^{\prime}, \mathbf{k}\right)\right|^{2}\right\}
\end{align}
where $\Omega_{k}$ is a unit vector representing the direction of $\mathbf{k}$ and $d \Omega_{k}=d \cos \theta_k d  \phi_k$. The TT part of the stress-energy tensor is computed by applying the projection operator in the momentum space,
\begin{align}
\label{energymomentum1}
\nonumber
T_{i j}^{\mathrm{TT}}(\tau, \mathbf{k}) &=\Lambda_{i j, k l}(\hat{k}) T_{i j}(\tau, \mathbf{k}) \\
&=\Lambda_{i j, k l}(\hat{k})\left\{\partial_{k}  (l,r) \partial_{l}  (l,r)\right\}(\tau, \mathbf{k}) \\
\label{energymomentum2}
\Lambda_{i j, k l}(\hat{k}) &=P_{i k}(\hat{k}) P_{j l}(\hat{k})-\frac{1}{2} P_{i j}(\hat{k}) P_{k l}(\hat{k}) \\
\label{energymomentum3}
P_{i j}(\hat{k}) &=\delta_{i j}-\hat{k}_{i} \hat{k}_{j}
\end{align}
where $\hat{k}=\mathbf{k} /|\mathbf{k}|$, and $\left\{\partial_{k}  (l,r) \partial_{l}  (l,r)\right\}(\tau, \mathbf{k})$ is the Fourier transform of $\partial_{k} l(\tau, \mathbf{x}) \partial_{l} l(\tau, \mathbf{x})$ $+\partial_{k} r(\tau, \mathbf{x}) \partial_{l} r(\tau, \mathbf{x})$.

We present further details about the simulation setup and initial field configurations in Appendix~\ref{app:A}. Here we briefly summarize the interpretations from the GW spectrum. We calculate the GW amplitude using Eq.~\eqref{numericalomega1} from our simulation box, and normalize it by,
\begin{equation}
    \label{omeganormalize}
    \Omega_{\eta}\equiv\frac{\rho^\eta_{gw}}{\rho(t_i)}=\frac{8\pi}{3\beta^2}G^2\eta^4
\end{equation}
where $\rho^\eta_{gw}=G\eta^6$ is the energy density of a source of characteristic scale $\eta$ estimated from Quadrupole approximation, and $\rho(t_i)=3H^2(t_i)/8\pi G=3\beta^2\eta^2/8\pi G$ is the critical energy density at the initial time. We show a representative plot of GW spectra for Hubble sized domains calculated numerically from the simulation box in Fig.~\ref{fig:nobias}. The horizontal axis is the amplitude of the comoving wave vector. It is related to the frequency as in Eq. \eqref{freqwavenumber}. Other plots for different values of $\rho_1$ and $\rho_3$ as well as different initial domain sizes along with careful interpretation of them are given in details in Appendix~\ref{app:results}. Here we list down the key inferences for the GW spectrum,
\begin{enumerate}
 \item The peak of the GW spectrum occurs at frequency corresponding to the Hubble radius at the time of GW production, which is redshifted to the value we see today.
 \item Above the peak frequency, the amplitude of the spectrum has an approximate $f^{-1}$ dependence on frequency $f$.
 \item Below the peak frequency, it is difficult to gauge the frequency dependence of the GW spectrum due to limitations of numerical integration as described in Appendix \ref{app:results}. Therefore an $f^3$ dependence is assumed from causality requirements~\cite{Hindmarsh2014GravitationalTransition}.
\end{enumerate}

Before giving the final GW spectrum, we need to calculate the peak amplitude of the spectrum. In~\cite{Hiramatsu2014OnWalls}, a semi-analytical approach is used to match the magnitude of GW radiation, computed from simulations with that form theoretical expectations. The dominant contribution to the GW spectrum is expected to come from changing Quadrupole moment of the domain walls. With this assumption, the authors of \cite{Hiramatsu2014OnWalls} have introduced an efficiency parameter $\tilde{\epsilon}_\text{gw}$ defined as,
\begin{equation}
 \label{eq:eptilde}
 \tilde{\epsilon}_\text{gw}=\frac{3\pi \,H_\ast^2\, M_\text{pl}^4\,\Omega_\text{gw}|^\ast_{peak}}{8\,\mathcal{A}^2\sigma_\text{wall}^2}
\end{equation}
that relates the calculated amplitude with analytical estimation from Quadrupole moment. Here $H_\ast$ and $\Omega_\text{gw}|^\ast_{peak}$ are the Hubble parameter and the peak GW amplitude at time of decay $t_\text{dec}=t_\ast$. It is shown that $\tilde{\epsilon}_\text{gw}\sim 0.7$ for a $\phi^4$ potential with $Z_2$ symmetry. We will simply assume that in our case also it will not vary much from this value and take $\tilde{\epsilon}_\text{gw}\sim1$.

With these results, finally we compare our interpretation with literature and present the GW spectra from SOPT in MLRSM.

\subsection{Final spectrum as seen today}\label{sec:DWSpec}
In summary, considering the remarks made in the previous subsection regarding the spectrum from simulations, we conclude that our results match with the results from~\cite{Kadota2015GravitationalModel,Hiramatsu2014OnWalls} with $\mathcal{A}\sim0.6$ and $\tilde{\epsilon}_\text{gw}\sim1$ for our model. The GW spectrum that was produced during the decay time of the domain walls, redshifts till today before we see it. We use the same general formulas as given in~\cite{Kadota2015GravitationalModel,Hiramatsu2014OnWalls} for the GW spectrum generated at arbitrary scale with arbitrary bias, as seen today after red-shifting, given by,
\begin{align}
    \label{eq:GWpeakAmp}
    \Omega_{\mathrm{gw}} h^{2}\left(t_{0}\right)_{\text {peak }} &\simeq 5.20 \times 10^{-20} \times \tilde{\epsilon}_{\mathrm{gw}} \mathcal{A}^{4}\left(\frac{10.75}{g_{*}}\right)^{1 / 3}\left(\frac{\sigma_{\text {wall }}}{1 \mathrm{TeV}^{3}}\right)^{4}\left(\frac{1 \mathrm{MeV}^{4}}{\delta V}\right)^{2}\\
    \label{eq:GWpeakFreq}
    f_{\text{peak}} &\simeq 3.99 \times 10^{-9} \mathcal{A}^{-1 / 2}\left(\frac{1 \mathrm{TeV}^{3}}{\sigma_{\text {wall }}}\right)^{1 / 2}\left(\frac{\delta V}{1 \mathrm{MeV}^{4}}\right)^{1 / 2}\mathrm{~Hz}
\end{align}
where $t_0$ denotes the present time and the numerical factors in both the equations account for the red-shifting. Here $g_*$ is the relativistic degrees of freedom at the time of the DW decay, $g_*=134$ in MLRSM. The wall tension $\sigma_\text{wall}$ is given in Eq. \eqref{eq:DWTension} and $\delta V=\epsilon\eta^6/\Lambda^2$, the bias that makes the domain walls annihilate depends on parity breaking term $\epsilon=f_L-f_R$ defined in Eq. \eqref{eq:bias1}. The spectrum is approximately given by,
\begin{equation}
    \label{eq:spectrum}
    \left.\Omega_{\mathrm{GW}} \simeq \Omega_{\mathrm{GW}}\right|_{\text {peak }} \times\left\{\begin{array}{ll}
\left(\frac{f_{\text {peak }}}{f}\right) & \text { for } f>f_{\text {peak }} \\
\left(\frac{f}{f_{\text {peak }}}\right)^{3} & \text { for } f<f_{\text {peak }}
\end{array}\right.
\end{equation}

We plot the GW spectrum for different benchmark points (as given in Table \ref{tab:BPSOPTFOPT}) in Fig. \ref{fig:spectrumSOPT} along with the \textit{power-law-integrated} sensitivity curves (PLISC)s of different GW experiments \cite{Schmitz2020NewTransitions} described in Sec.~\ref{section:GWFOPT}. The 3 benchmark points (BP1, BP2, BP3) in Fig. \ref{fig:spectrumSOPT} stands for $\eta=10^4,10^5,10^6$ GeV respectively. From Fig. \ref{fig:spectrumSOPT} it is clear that these signals are detectable at the PTA experiments assuming a threshold SNR of 1. Fig. \ref{fig:spectrumEta}, shows the GW spectrum for the right handed symmetry breaking scale $M_R\equiv \eta=10^6,10^7,10^8,10^9$ GeV with fixed bias $\epsilon=2$. Fig. \ref{fig:spectrumBias} shows the effect of bias on the spectrum, with the limitation that we can not have an arbitrarily small bias since that will make the domain walls enter the scaling regime and dominate the energy density of the Universe.  Fig. \ref{fig:spectrumRho} shows the spectrum for different configurations of $\rho_1$ and $\rho_3$. Note that the parameter $\rho_1$ is primarily responsible for determining the strength of the phase transition. Especially for bigger values of $\rho_1\gtrsim0.3$, the phase transition seems to be second order or weak first order in nature, as can be seen in our benchmark points.

\section{Gravitational waves from two-field FOPT: Doubly peaked spectrum}
\label{sec:doublepeak}
\begin{figure}[ht]
    \centering
    \includegraphics[width=\linewidth]{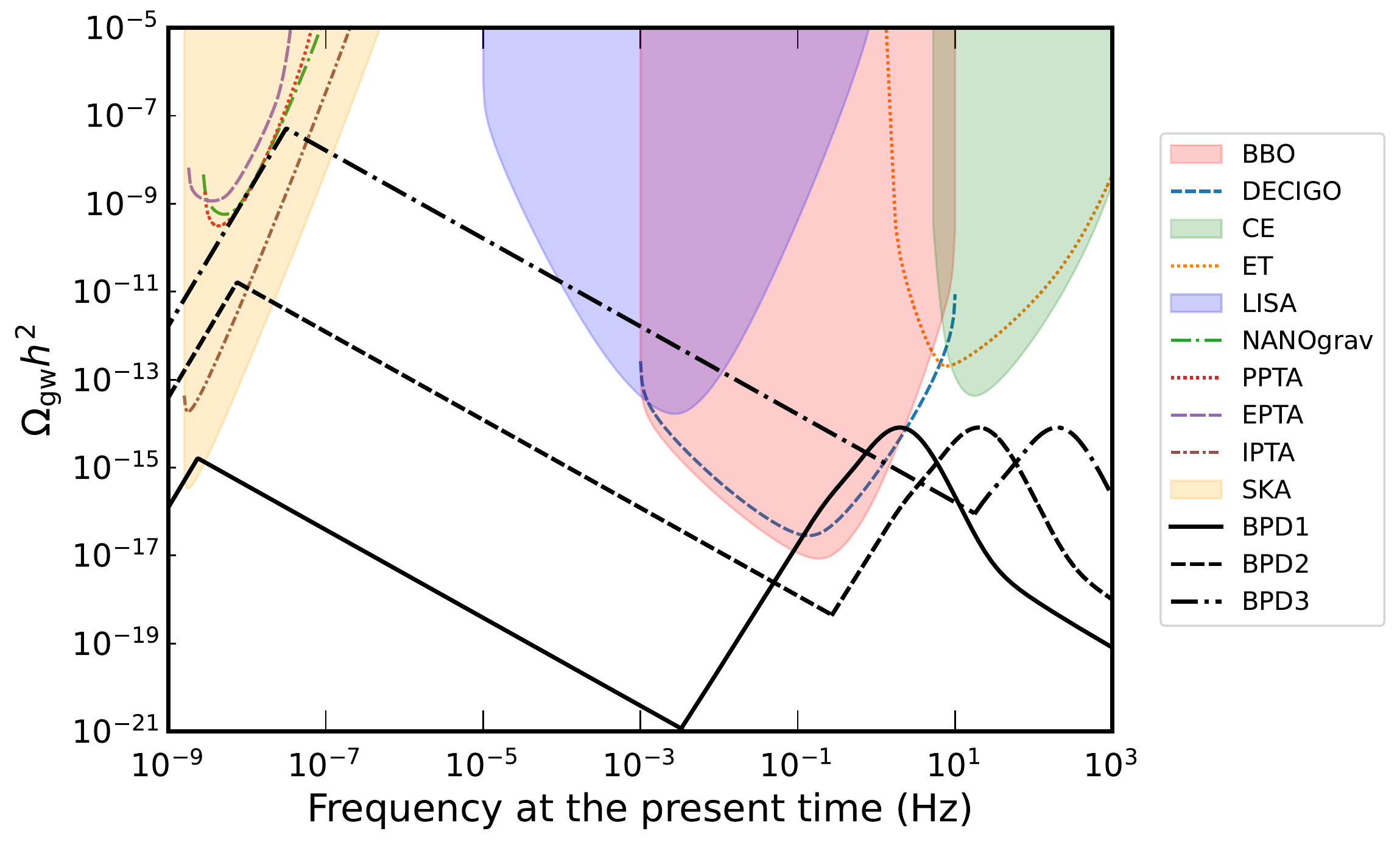}
    \caption{\small Doubly peaked spectra from two-field FOPT followed by domain wall disintegration. Plotted for benchmark points BPD1, BPD2, BPD3 given in Table \ref{tab:doublepeak} corresponding to $M_R$ scale $10^4, 10^5$ and $10^6$ GeV respectively. The peak resulting from FOPT dynamics lies generally within the sensitivity of BBO and DECIGO whereas the peak from disintegration of residual domain walls generically occurs at low frequencies accessible to PTA experiments for appropriate bias.
    \label{fig:doublepeak}}
\end{figure}

\begin{figure}[ht]
    \centering
    \includegraphics[width=\linewidth]{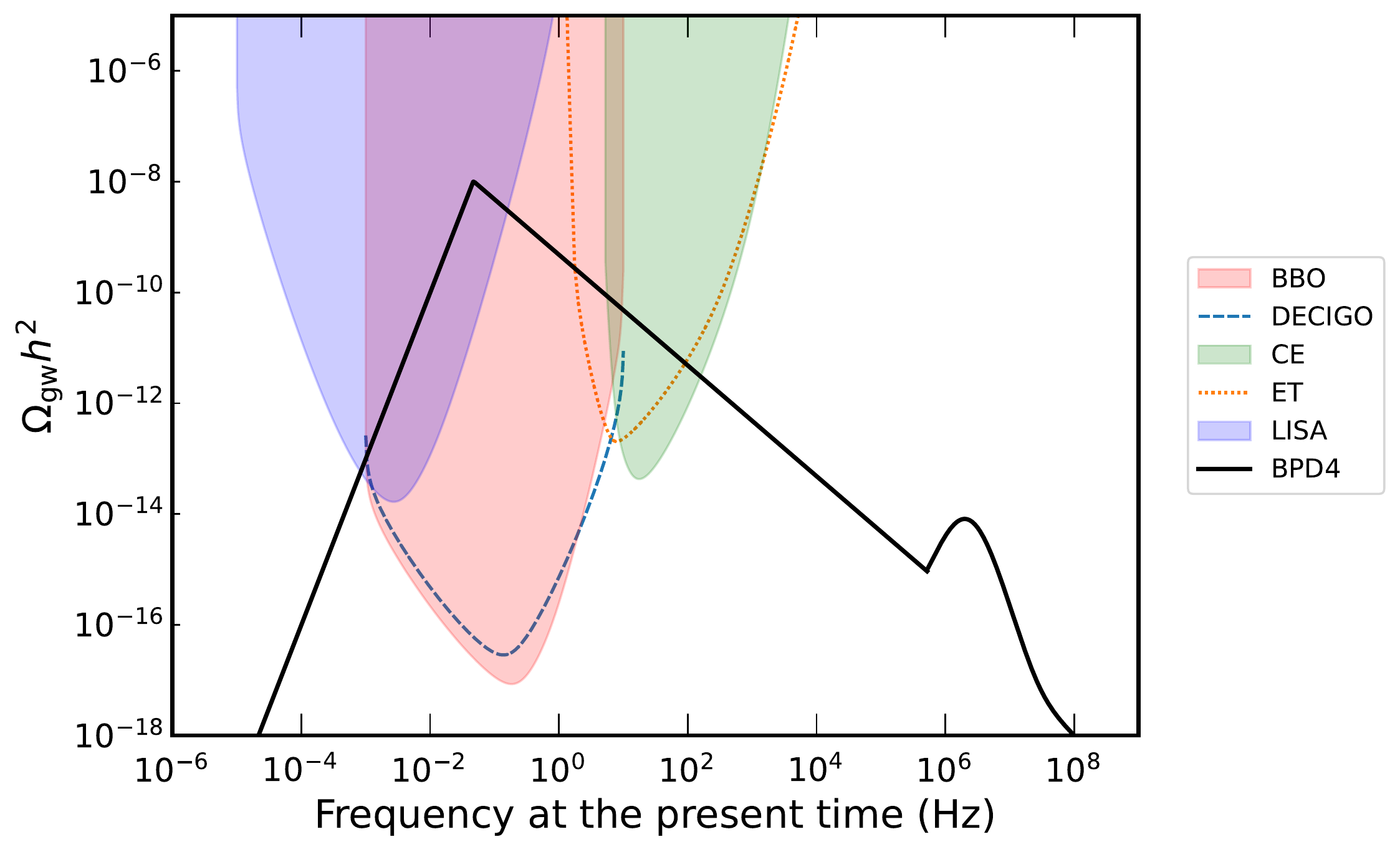}
    \caption{\small The doubly peaked spectrum similar to Fig. \ref{fig:doublepeak}, but for benchmark point BPD4 of Table \ref{tab:doublepeak} corresponding to $M_R$ scale $10^{10}$ GeV. Here the contribution from FOPT dynamics lies beyond the currently planned experiments.
    \label{fig:doublepeak10}}
\end{figure}

As discussed before, the FOPT bubble contributions need to be supplemented by the contribution from the disintegration of the residual DWs. This leads to a two-peaked GW spectrum expected from an FOPT in MLRSM. In Fig. \ref{fig:doublepeak} we plot doubly peaked spectra from typical strong FOPTs at different scales, for the benchmark points (BPDs) 1, 2, 3 given in Table \ref{tab:doublepeak}. The peak on the right corresponds to FOPT while the peak on the left corresponds to the decay of residual domain walls. In Fig. \ref{fig:doublepeak10} we plot the same for a high scale FOPT, $\eta\sim 10^{10}$GeV.
It is seen from Fig. \ref{fig:doublepeak} that a direct verification of FOPT is marginally possible for low scale L-R models, whereas a high scale case such as in Fig. \ref{fig:doublepeak10} is beyond the reach of currently planned experiments. We derive an analytical formula for the frequency, $f_{\rm tr}$ at which the transition from DW spectrum to FOPT spectrum takes place, by evaluating the condition $h^2\Omega_{\rm GW}^{\rm DW}(f_{\rm tr})=h^2\Omega_{\rm GW}^{\rm FOPT}(f_{\rm tr})$, which gives,

\begin{equation}
 \label{eq:ftr}
  f_{\rm tr}=5.6\times10^{-4}\left(\frac{f_{\rm peak}}{10^{-9}}\right)^{\frac{1}{4}}\left(\frac{f_{\rm sw}}{10^{-1}}\right)^{\frac{3}{4}}\left(\frac{\Omega_{\rm gw}h^2(t_0)_{\rm peak}}{10^{-11}}\right)^{\frac{1}{4}}\left(\frac{10^{-10}}{\Omega_{\rm FOPT}h^2}\right)^{\frac{1}{4}}\text{ Hz}
\end{equation}
where $f_{\rm peak}$ is given by Eq.~\eqref{eq:GWpeakFreq}, $f_{\rm sw}$ is given by Eq.~\eqref{eq:fsw}, $\Omega_{\rm gw}h^2(t_0)_{\rm peak}$ is given by Eq.~\eqref{eq:GWpeakAmp} and from Eq.~\eqref{eq:Omegasw},
\begin{equation}
 \Omega_{\rm FOPT}h^2\simeq1.87 \times 10^{-5}\left(\frac{H_{*}}{\beta}\right)\left(\frac{\kappa_{sw} \alpha}{1+\alpha}\right)^{2}\left(\frac{100}{g_{*}}\right)^{\frac{1}{3}} v_{w}\Upsilon
\end{equation}
where we have assumed that sound waves are the only source of the FOPT spectrum and $f/f_{\rm sw}\ll1$ for simplicity.

\begin{figure}[ht]
    \centering
    \includegraphics[width=0.8\linewidth]{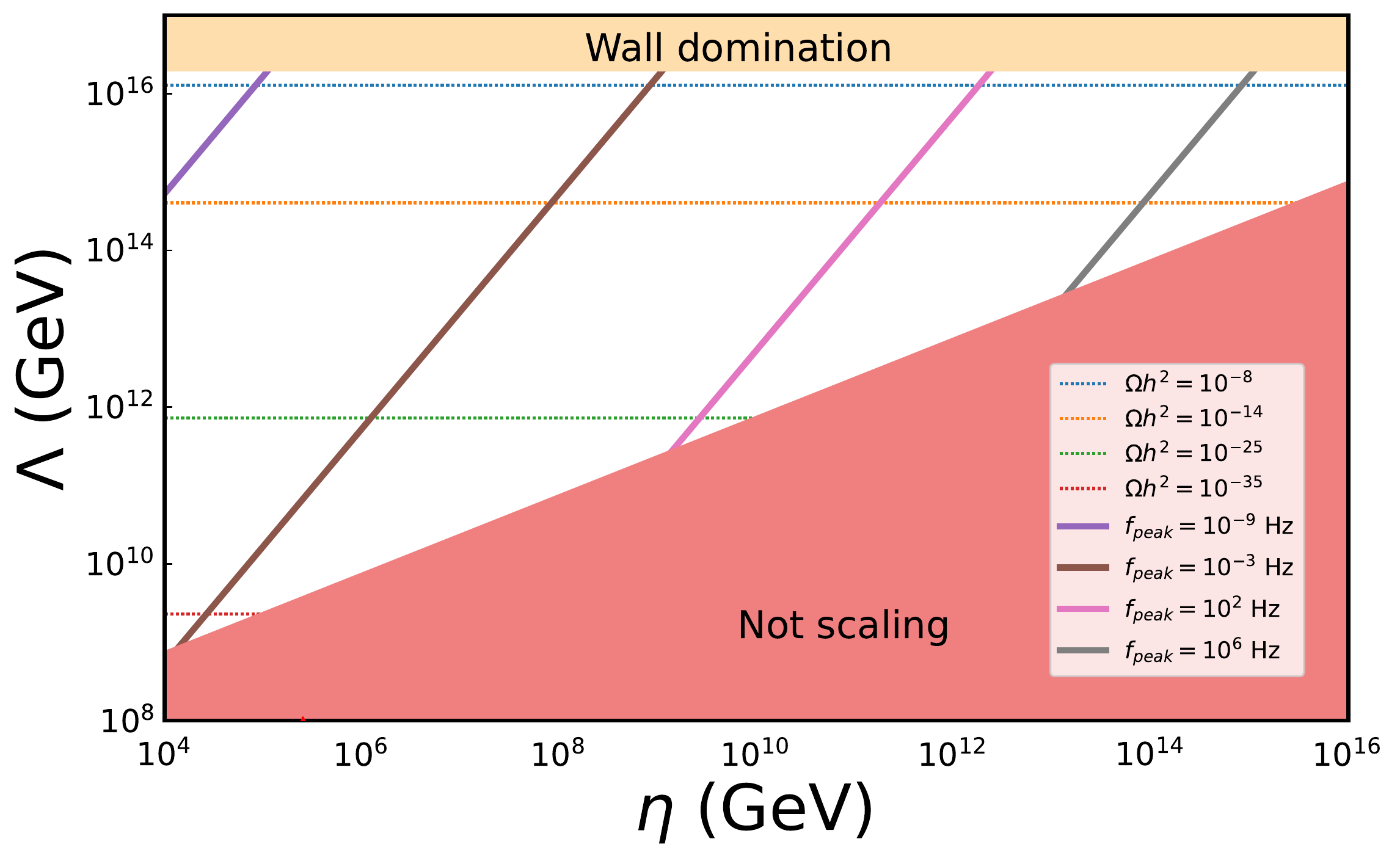}
    \caption{\small Contours of constant peak frequency and amplitude in the parameter space of $\eta$ and $\Lambda$ fixing $\epsilon=f_L-f_R=10^{-5}$. The light red region corresponds to large biases such that domain walls don't reach scaling regime, see Eq.~\eqref{eq:biasscaling}. Light yellow region corresponds to the case where bias is small and domain walls dominate the energy density before they decay, see Eq.~\eqref{eq:biasdomination}. The white region is allowed. Here we have taken $\rho_1=0.4, \rho_3=1.7$ and $g_*=134$.}
    \label{fig:paramScan}
\end{figure}

\begin{figure}[ht]
    \centering
    \includegraphics[width=0.8\linewidth]{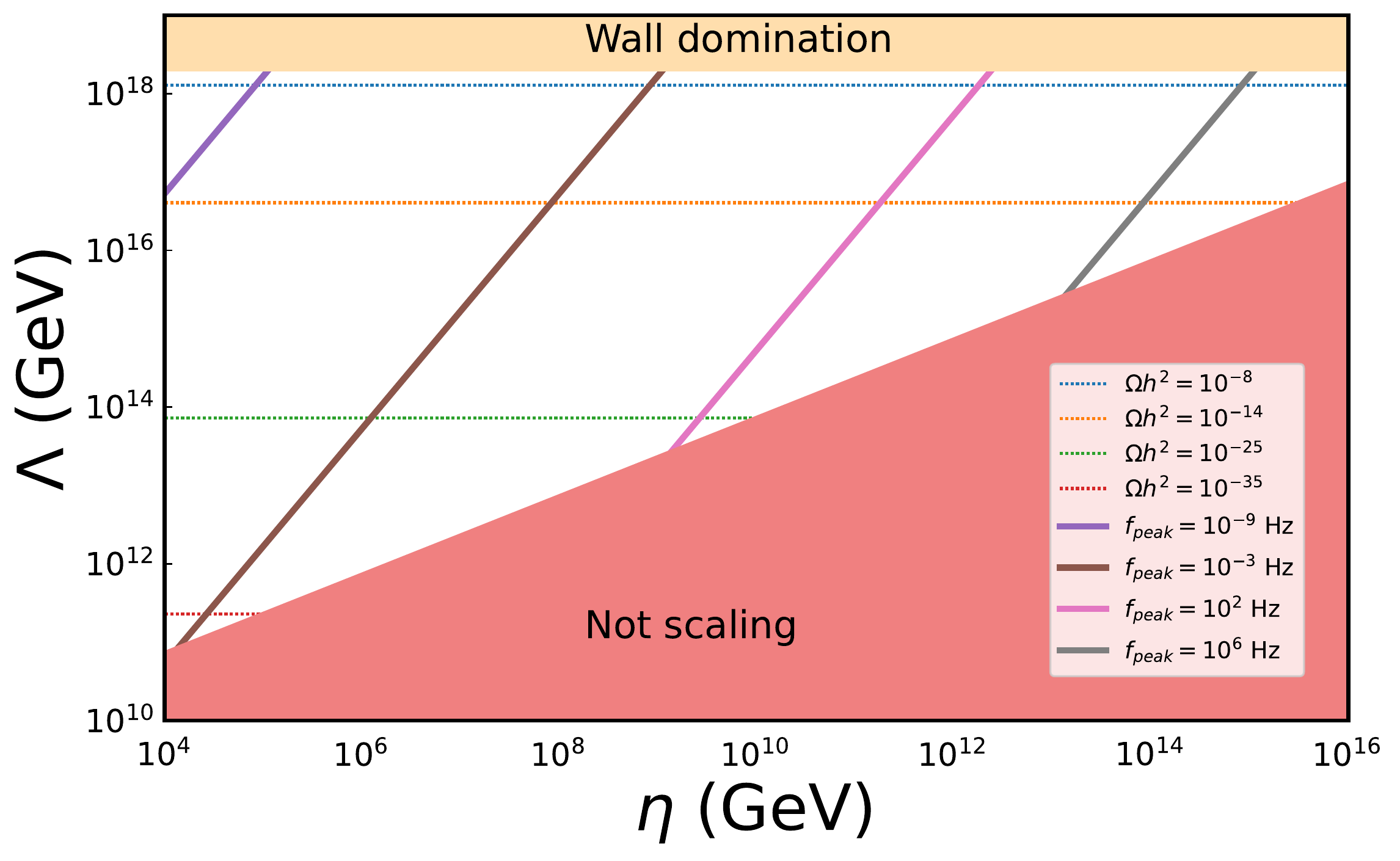}
    \caption{\small Same as Fig. \ref{fig:paramScan} but for $\epsilon=0.1$.}
    \label{fig:paramScan2}
\end{figure}

\section{Dissociating Parity breaking from LR Symmetry Breaking}\label{sec:CMPmodel}
In the minimal model that we discussed so far, parity $\mathcal{P}$ and LR symmetry are broken simultaneously where the right-handed triplet $\Delta_R$ takes a high VEV whereas $\Delta_L$ VEV is assumed to be negligible. This assumption is based on the fact that Standard Model is predominantly left-handed, and what breaks this parity is still unknown. To address this issue, Chang et al \cite{Chang1984NewTheories,Chang1984DecouplingModels} described a way to dissociate parity from LR symmetry by adding a scalar $\chi$ into the particle content, with negative parity, i.e. $\chi\rightarrow-\chi$ under $D$-parity transformation. This $\chi$ adds extra terms to our tree level potential,

\begin{align}
    V_\chi &= -\mu_\chi^2\chi^2+\gamma_{1}\chi^4 \\
    V_{\chi\Delta} &= M\chi\text{ Tr}\left( \Delta_L^\dag\Delta_L - \Delta_R^\dag\Delta_R \right) + \gamma_2\chi^2\text{ Tr}\left( \Delta_L^\dag\Delta_L + \Delta_R^\dag\Delta_R \right)\\
    V_{\chi \phi} &= \gamma_2\chi^2\text{ Tr}\left( \phi^\dag \phi\right) +\gamma_2^1\chi^2\left(\text{Det} \phi+\text{Det} \phi^\dag\right)
\end{align}
In this scenario $\chi$ takes a VEV, breaking parity at $M_\text{parity}$ scale before $SU(2)_R$ is broken at scale $M_R$. Depending on which minima the field $\chi$ takes out of $\pm M_\text{parity}=\mu_\chi/\sqrt{2\gamma_1}$ in a region, the Universe becomes left or right-handed in that region, subject to minimal fine-tuning. Since our Universe is left-handed, we want the right-handed regions/domains to vanish eventually. This can be achieved again by making the domain walls unstable by adding a small bias term into the $\chi$ potential. If $M_\text{parity}\gg M_R$, the contribution from $SU(2)_R$ or $SU(2)_L$ breaking within left-like and right-like domains to the GW background will be negligible compared to the contribution from the $\chi$ domain walls. Thus we can assume that this type of dissociated high-energy parity breaking scheme can also be responsible for detectable GW signals.

\section{Conclusion}
\label{sec:conclusion}

We have emphasized the qualitatively different considerations which arise for phase transitions when dealing with a theory with left-right symmetry. In particular, the SOPT case, usually not expected to leave behind any signature indeed creates a GW signature, which should be accessible to experiments being currently planned. If the parameter space signals an FOPT then in fact we get two separate peaks, one from the usual bubble wall collision dynamics and an additional peak at much lower frequencies arising from the residual DW that arise due to the presence of two or more degenerate fields present in the theory. This case is marginally accessible for low scale L-R models, Fig. \ref{fig:doublepeak}. A high scale case such as in Fig. \ref{fig:doublepeak10} is beyond the reach of currently planned experiments. 

In the present study, we performed numerical simulations of disintegrating domain walls in MLRSM, which is relevant for both SOPT and FOPT, and compared our findings with existing results on GW spectra from $Z_2$ and axionic domain walls. We show that the domain wall structure in MLRSM resembles axionic domain walls for small $\rho_3-\rho_1$ values with $N_{\rm DW}=4$ (see Fig.\ref{fig:axionkink}). For the two-field domain walls of the L-R model we introduced a bias term as in Eq.~\eqref{eq:bias1}. Thus the $M_R$ scale $\eta$ is supplemented by an additional high scale $\Lambda$ needed to introduce the bias between the left vs. right vacua. Note that such a bias is required in all considerations of spontaneously broken left-right symmetries in order to avoid persistent domain walls~\cite{Rai1994GravityProblem}. Fig.~\ref{fig:paramScan} and \ref{fig:paramScan2} provide the acceptable regions in the parameter space of $\Lambda$ and $\eta$. They contain the contours of peak frequency and peak amplitude of gravitational waves spectrum from domain wall decay, fixing $\epsilon=10^{-5}$ in Fig. \ref{fig:paramScan} and $10^{-1}$ in Fig. \ref{fig:paramScan2}. This shows the region of the parameter space where an observable signal is produced for fixed $\epsilon$ value. We find that for $M_R$ scales $10^4-10^6$ GeV, the GW peak from domain wall annihilation will be detectable in PTA experiments such as NANOGrav, IPTA, EPTA etc. while for scales $>10^8-10^{10}$ GeV, LISA, DECIGO, BBO, CE and ET will be able to probe the spectrum.

In the case of an FOPT, it has been determined that the bubbles always remain in the non-runaway scenario. This implies that the major contribution to the gravitational wave (GW) spectrum arises from sound waves propagating through the plasma. For symmetry breaking scales ranging between $10^4$ and $10^6$ GeV, the peak of the GW spectra from FOPT can potentially be detected by several upcoming and planned GW observatories such as LISA, DECIGO, BBO, CE, and ET, as illustrated in Fig.~\ref{fig:spectrumFOPT}. However, at the upper end of this scale, specifically for a symmetry breaking scale of $10^6$ GeV, the detectability of the GW signal becomes significantly more challenging. This is primarily due to the limited range of very small values of the ratio $\beta/H_*$ (where $\beta$ is the inverse time duration of the phase transition and $H_*$ is the Hubble parameter at the time of the phase transition) that falls within the sensitivity range of these detectors. This restriction is depicted in Fig. \ref{fig:albeta} for $10^6$ GeV.

While the high scale $10^{10}$ GeV  as in Fig. \ref{fig:doublepeak10} remains beyond the reach of currently planned experiments, it is nevertheless the lower end of the scales envisioned for grand unification. Thus, in addition to search for  the low energy left-right model as proposed here, it remains interesting to explore similar effects in realistic high scale theories. The methods developed here can  be applied with suitable modification to the case of  SUSY $SO(10)$, as considered for instance in \cite{Banerjee:2018hlp} and \cite{Garg:2018trf}, and could motivate future experiments.

\acknowledgments

Z.A.B. is supported by a DST-INSPIRE fellowship. The research of UAY is partially supported by an Institute Chair Professorship of IIT Bombay.

\newpage
\appendix
\section{Benchmark points for SOPT and FOPT}
\label{app:BP}

\begin{table}[ht]
\centering\setlength\extrarowheight{-1pt}
\begin{tabular*}{\linewidth}{@{\extracolsep{\fill}}ccccccc}
 \hline
 \hline
 & BP1 & BP2 & BP3 & BPF1 & BPF2 & BPF3\\
 \cline{2-7}
$v/$GeV & 246 & 246 & 246 & 246 & 246 & 246 \\
$v_R/$GeV & $10^4$ & $10^5$ & $10^6$ & $10^4$ & $10^5$ & $10^6$ \\
$\tan\beta$ & 0.001 & 0.001 & 0.001 & 0.001 & 0.001 & 0.001\\
$\lambda_1$ & 0.13 & 0.13 & 0.13 & 0.13 & 0.13 & 0.13 \\
$\lambda_2$ & 0 & 0 & 0 & 0 & 0 & 0\\
$\lambda_3$ & 1.95024 & 0.66708 & 1.548816 & 3.452 & 1.3374 & 2.284\\
$\lambda_4$ & 0 & 0 & 0 & 0 & 0 & 0\\
$\rho_1$ & 0.40055 & 0.411194 & 0.350746 & 0.0264 & 0.0386 & 0.0787 \\
$\rho_2$ & 0.88258 & 0.63584 & 0.847746 & 1.0542 & 0.5973 & 0.2495\\
$\rho_3$ & 1.35815 & 1.711196 & 1.14436 & 2.6854 & 2.0801 & 3.3698\\
$\rho_4$ & 0 & 0 & 0 & 0 & 0 & 0\\
$\alpha_1$ & 0 & 0 & 0 & 0 & 0 & 0 \\
$\alpha_2$ & 0.27152 & 0.340437 & 0.2123735 & 0 & 0 & 0 \\
$\alpha_3$ & 0.82883 & 0.603991 & 0.09274 & 0.5796 & 0.4494 & 0.3537 \\
$\beta_{1,2,3}$ & 0 & 0 & 0 & 0 & 0 & 0\\
$g$ & 0.65 & 0.65 & 0.65 & 0.65 & 0.65 & 0.65 \\
$g_{\text{B-L}}$ & 0.4324 & 0.4324 & 0.4324 & 0.4324 & 0.4324 & 0.4324 \\
$y_t$ & 0.95 & 0.95 & 0.95 & 0.95 & 0.95 & 0.95\\
$y_M$ & 1 & 1 & 1 & 0.9362 & 0.7875 & 0.9255\\
$\epsilon=f_L-f_R$ & 0.1 & 0.1 & 0.1 & - & - & - \\
$\Lambda/$GeV & $3\times10^{16}$ & $2\times10^{17}$ & $2\times10^{18}$ & - & - & -\\
 \hline
 \hline
\end{tabular*}
\caption{\small Benchmark points for SOPT denoted by BP1, BP2, BP3 and for FOPT denoted by BPF1, BPF2, BPF3.\label{tab:BPSOPTFOPT}}
\end{table}

\begin{table}[ht]
\centering\setlength\extrarowheight{-5pt}
\begin{tabular*}{0.7\linewidth}{@{\extracolsep{\fill}}cccc}
 \hline
 \hline
 & BPF1 & BPF2 & BPF3 \\
 \cline{2-4}
$T_c/$GeV & 2913 & 28829 & 323004 \\
$T_n/$GeV & 1650 & 21020 & 275040 \\
$T_p/$GeV & 1557.2 & 20426 & 269050 \\
$v_R(T_c)/T_c$ & 4.37146 & 3.9021 & 3.04868 \\
$\alpha$ & 0.395862 & 0.141741 & 0.025492 \\
$\alpha_\infty$ & 0.789406 & 0.334233 & 0.0722176 \\
$\beta/H_*$ & 200.641 & 547.352 & 648.779 \\
$\sqrt{\frac{\Delta V(Tn)}{\alpha \rho_{\text{rad}}}}$ & 0.927674 & 0.82288 & 0.391911 \\
$v_J$ & 0.878237 & 0.802155 & 0.692529 \\
$v_w^\alpha$ & 0.275777 & 0.303166 & 0.325535 \\

 \hline
 \hline
\end{tabular*}
\caption{\small Phase transition parameters calculated for the benchmark points for FOPT given in Table \ref{tab:BPSOPTFOPT}. Here $\sqrt{\frac{\Delta V(Tn)}{\alpha \rho_{\text{rad}}}}>v_J$ implies we take $v_w=1$ for BPF1 and BPF2  while for BPF3 $v_w=0.391911$ as in Eq. \eqref{eq:vw}. \label{tab:BPFOPTres}}
\end{table}

\begin{table}[ht]
\centering\setlength\extrarowheight{-4pt}
\begin{tabular*}{0.7\linewidth}{@{\extracolsep{\fill}}ccccc}
 \hline
 \hline
 & BPD1 & BPD2 & BPD3 & BPD4\\
 \cline{2-5}
$T_n/$GeV & $10^4$ & $10^5$ & $10^6$ & $10^{10}$\\
$\rho_1$ & $0.1$ & $0.1$ & $0.1$ & $0.1$\\
$\rho_3$ & $1.7$ & $1.7$ & $1.7$ & $1.7$\\
$\mu_3^2\sim T_n^2\rho_1$ & $10^7$ & $10^9$ & $10^{11}$ & $10^{19}$\\
$\alpha$ & 0.1 & 0.1 & 0.1 & 0.1\\
$\beta/H_*$ & 1000 & 1000 & 1000 & 1000\\
$\epsilon$ & $0.1$ & $0.1$ & $0.1$ & $0.1$\\
$\Lambda$/GeV & $2\times10^{16}$ & $2\times10^{17}$ & $1.5\times10^{18}$ & $10^{18}$\\
 \hline
 \hline
\end{tabular*}
\caption{\small Example benchmark points for typical strong FOPT in MLRSM to show two peak structure in Fig. \ref{fig:doublepeak} and \ref{fig:doublepeak10}. Here we assumed $T_n\sim\eta$ hence from Eq. \eqref{eq:mu}, $\mu_3^2\sim T_n^2\rho_1$. The residual domain walls annihilate under the influence of a bias $\delta V$, determined by $\epsilon=f_L-f_R$ and the cutoff scale $\Lambda$. Here we fix $\epsilon=0.1$ and vary $\Lambda$.
The resulting doubly peaked spectra are shown in Fig.s \ref{fig:doublepeak} and \ref{fig:doublepeak10}.\label{tab:doublepeak}}
\end{table}

\newpage
\section{More details about numerical simulation}
\label{app:A}

\subsection{Rescaling the Field Equations}
There's only one energy scale in our theory, $\eta$. We can define all other dimensional parameters in terms of $\eta$ to make our equation dimensionless, eg: $t \rightarrow t\eta$, $x \rightarrow x\eta$, $l \rightarrow l\eta^{-1},r \rightarrow r\eta^{-1}$ etc. We assume the evolution takes place entirely in the radiation dominated era such that the scale factor and the Hubble parameter have time dependence
\begin{align}
    \label{hubble1}
    a(t) &= a_0 t^{1/2}\\
    H(t) &= \dot a/a = \frac{1}{2t}
\end{align}
Thus, we make them dimensionless by $a\rightarrow a\eta^{1/2}$, $H\rightarrow H\eta^{-1}$. The rescaled equation is obtained from Eq. \eqref{eomphi} by taking $\eta=1$,
\begin{align}
    \label{rescaled1}
    \nonumber\ddot{l}+3H\dot{l}-\frac{\nabla^2}{a^2}l+\rho_1 l(l^2-1)+\frac{\rho_3}{2}lr^2&=0\\
    \ddot{r}+3H\dot{r}-\frac{\nabla^2}{a^2}r+\rho_1 r(r^2-1)+\frac{\rho_3}{2}l^2r&=0
\end{align}
where all the terms represent their scaled versions now.

\subsection{Simulation Setup}\label{simsetup}

The simulation takes place in a cuboid of size $L_x=L_y=L_z= b=50$ (in units of $\eta^{-1}$), discretized into $N^3=256^3$ lattice points. Our simulation box is a comoving box, which also expands with the expanding Universe. The problem with this setup is that the dynamic range of the simulation is limited. This is because we want our comoving box to be bigger than the correlation length, and also want to resolve the domain wall width, at all times during the simulation. 
Domain wall width remains constant in real space, i.e. decreases as $a(t)^{-1}$ within the simulation box.

Let our physical lattice spacing be $\delta x_{phy} = a(t)\times b/N = a(t)\times50/256$. Assuming initial time of simulation $t_i=1$ and setting $a(t_i)=1$, 
the ratio of domain wall width to lattice spacing is,
\begin{equation}
    \label{widthtospacing}
    \frac{\delta_\text{wall}}{\delta x_{phy}} = \frac{N}{b}\sqrt{\frac{2}{\rho_3-2\rho_1}}\left(\frac{t}{t_i}\right)^{-\gamma}
\end{equation}
where $\gamma$ is from Eq. \eqref{scalefactor}. In our case, the entire process occurs in the radiation dominated universe ($\gamma=1/2$), and we choose $\rho_1=0.4$, $0.1$ and $\rho_3=1.7$, $1$ for the simulations. We define the maximum time of resolution to be $t_\text{res}$ after which the above ratio is $<1$, which means that the walls are not resolvable anymore. Table \ref{tab:tres} shows $t_\text{res}$ for various combinations of $\rho_1$ and $\rho_3$ values. This limits the upper bound on the time integral of the GW calculation in Eq. \eqref{numericalomega1} to be $t=t_\text{res}$, i.e. we consider the source to be active from $t=1$ to $t=t_\text{res}$. For $t>t_{\rm res}$, we will get unphysical results. Still we perform our simulation till $t=151$ to show how the spectrum changes after $t_\text{res}$ and plot them in Figures \ref{fig:nobias},\ref{spectrumgraph} and \ref{fig:2ndpeakspec}. The different plots in each figure correspond to what upper limit we choose in GW calculation in Eq. \eqref{numericalomega1}. Assuming that upper limit to be $t_\text{dec}$ we calculate the required bias $\delta V$ from Eq. \eqref{eq:decaytime}. Ideally $t_\text{dec}$ should be smaller than $t_\text{res}$ in our simulation to obtain realistic results but we also consider larger $t_\text{dec}$ and interpret the results carefully as discussed in Sec. \ref{app:results}.

\begin{table}
\centering
\begin{tabular*}{0.12\linewidth}{ccc}
 \hline
 \hline
  $\rho_3$ & $\rho_1$ & $t_\text{res}$ \\
 \hline
 1.7 & 0.1 & 31 \\
 1.7 & 0.4 & 51 \\
 1 & 0.1 & 61 \\
 1 & 0.4 & 261 \\
 \hline
 \hline
\end{tabular*}
\caption{\small Maximum time of resolution $t_\text{res}$ (in units of $\eta^{-1}$) for different configurations of $\rho_3$ and $\rho_1$ considered here. Beyond $t_\text{res}$, domain walls are not resolvable.\label{tab:tres}}
\end{table}

The correlation length determines the maximal initial size of the domains, provided it's smaller than the causal horizon. Assuming $m_\theta$ and $T_c$ to be of the order of $\eta$, dimensionally we can find $\xi_\text{causal}/\eta^{-1}\sim\left(\frac{M_\text{pl}}{\eta}\right)^{1/3}$, which is around 10 for $\eta=10^{16}$ GeV and 1000 for $\eta=10^{10}$ GeV, i.e. correlation length is larger than causal horizon. So we take causal horizon to be the initial size of the domains which is $\mathcal{O}(1)$ in units of $\eta^{-1}$. But in addition we perform simulations with different initial sizes of domains upto $\mathcal{O}(10)$ just to check the effect of initial domain size on scaling properties of domain walls. Below we list down the steps of calculating the spectrum numerically from field simulations.

\textit{Spectrum calculation procedure:}
\begin{enumerate}
    \item Simulate the fields for a specific time step to obtain $l(\tau,\mathbf{x}),r(\tau,\mathbf{x})$.
    \item Calculate $\left\{\partial_{k} (l,r) \partial_{l} (l,r)\right\}(\tau, \mathbf{k})$ by taking \textbf{Discrete Fourier Transform (DFT)} of $\partial_{k} l(\tau, \mathbf{x}) \partial_{l} l(\tau, \mathbf{x})+\partial_{k} r(\tau, \mathbf{x}) \partial_{l} r(\tau, \mathbf{x})$
    \item Calculate $T_{i j}^{\mathrm{TT}}(\tau, \mathbf{k})$ using Eq. \eqref{energymomentum1}-\eqref{energymomentum3}.
    \item Perform the time integration in Eq. \eqref{numericalomega1} and the sum over $i,j$. We used a simple Simpson's 3/8th rule for the time integration.
    \item Perform the angular integration $\int d\Omega_k$ using \textbf{Monte Carlo method}. Multiply by the appropriate constant as in Eq. \eqref{numericalomega1}.
\end{enumerate}

\subsection{Initial Field Configuration}
We assume that domain walls are already produced at $t_i$ with a characteristic length scale of $\mathcal{O}(1)$ to $\mathcal{O}(10)$, motivated by Eq. \eqref{eq:correlation}. We only consider the positive $l,r$ values so the true and false minima correspond to $\theta=0$ and $\pi/2$ respectively. Our initial field configuration is alternating domains of minima $\theta=0$ and $\pi/2$, separated by domain walls of profile as in \eqref{eq:axionkink}. Simplest case is approximately cuboidal domains. To construct this structure first we take a $sin$ function of spatial coordinates with the wavelength corresponding to twice the required domain length scale at the initial time as our dominant mode, i.e. $k_\text{dominant}=\pi/\xi_\text{domain}$ where $\xi_\text{domain}$ is the required characteristic length scale. One wavelength of the $sin$ function consists of two domains of R and L vacua. We randomize the configuration by adding around 1000 random modes with $k\in(0,k_\text{dominant})$ to it with small coefficients. Then, we take the $tan^{-1}$ of the exponential of this series of Sin functions multiplied by $m_\theta$, as a good model of the domain wall kink, i.e.,
\begin{equation}
    \label{initfieldconfig1}
    \theta(x,y,z)_\text{initial} = \tan^{-1}\,\text{exp}\left[m_\theta\,\mathcal{N}\sum_{n=1}^{999} \frac{C_n}{k_n} \sin{\left(k_n x + \phi_i\right)} \sin{\left(k_n y + \psi_i\right)} \sin{(k_n z + \chi_i)}\right]
\end{equation}
where $C_1=1$ and the rest of the $C_i$ are randomly chosen small numbers of the order $\mathcal{O}(10^{-3})$ to $\mathcal{O}(10^{-1})$. The parameters $k_1=k_\text{dominant}$ and $k_n\in(0,k_\text{dominant})$ for $n\neq1$ represents comoving wavenumbers of the dominant and subdominant modes. The $k_n$ at the denominator is used to keep the domain wall width fixed. Every dimensional parameter is scaled by $\eta$. The parameters $\phi_i,\psi_i,\chi_i$ are random phases and $\mathcal{N}$ is used to normalize the sum of the $sin$ functions. Finally, the initial $l,r$ profiles are obtained from $\theta$ by taking sine and cosine of it. It turns out that the resulting spectrum depends mostly on the dominant fluctuation scales rather than the exact initial configuration.

\subsection{Discretizing the Equation of Motion}
We follow the notation and method of discretization from \cite{langtangen2016finite} adapted to our use case. We start with the equation of motions in Eq. \eqref{rescaled1}, with boundary conditions for $l$ and $r$ as,
\begin{align}
    \label{bc1}
    l,r(x,y,z,0) &= I_{l,r}(x,y,z)\\
    \frac{\partial}{\partial t}l,r(x,y,z,0) &= V(x,y,z) = 0\\
    l,r(0,y,z,t) &= l,r(L_x,y,z,t)\\
    l,r(x,0,z,t) &= l,r(x,L_y,z,t)\\
    l,r(x,y,0,t) &= l,r(x,y,L_z,t)
\end{align}
where $I_{l,r}(x,y,z)$ are the prescribed initial conditions. $L_x=L_y=L_z=b$ is the size of the comoving simulation box. The last three equations represent periodic boundary conditions.

Our 3D mesh contains $N_x\times N_y\times N_z$ mesh points. We also subdivide the time domain $[t_{ini},t_{fin} = T]$ into $N_t$ discrete equidistant points $\left\{t_0 = T_i < t_1 < t_2 <... < T\right\}$. Now a scalar field $\Psi=l$ or $r$ can be defined at each mesh point at a given time by the notation $\Psi(x_i,y_j,z_k,t_n) = \Psi^n_{ijk}$, where superscript $n$ denotes the time coordinate and subscripts $i,j,k$ denote $x,y,z$ coordinates respectively.

To discretize the wave equation, we replace the derivatives with central differences as follows,
\begin{align}
    \label{fd1}
    \frac{\partial}{\partial t}\Psi^n_{ijk} &= [D_{2t}\Psi]^n_{ijk} \approx \frac{\Psi^{n+1}_{ijk} - \Psi^{n-1}_{ijk}}{2 \Delta t}\\
    \frac{\partial^2}{\partial t^2}\Psi^n_{ijk} &= [D_tD_t\Psi]^n_{ijk} \approx \frac{\Psi^{n+1}_{ijk} - 2\Psi^n_{ijk} + \Psi^{n-1}_{ijk}}{\Delta t^2}\\
    \frac{\partial^2}{\partial x^2}\Psi^n_{ijk} &= [D_xD_x\Psi]^n_{ijk} \approx \frac{\Psi^{n}_{i+1jk} - 2\Psi^n_{ijk} + \Psi^{n}_{i-1jk}}{\Delta x^2}
\end{align}
and similarly for $y$ and $z$ coordinates. $D_t, D_x$ represents the derivative operators in a compact notation. In this notation (omitting the space-time indices), Eq. \eqref{rescaled1} becomes,
\begin{align}
    \label{discrete1}
    \nonumber D_t D_t l + 3H D_{2t}l - \frac{1}{a^2}\left[D_xD_x + D_yD_y + D_zD_z\right]l +\rho_1 l(l^2-1) +\frac{\rho_3}{2}lr^2 &=0\\
    D_t D_t r + 3H D_{2t}r - \frac{1}{a^2}\left[D_xD_x + D_yD_y + D_zD_z\right]r +\rho_1 r(r^2-1) +\frac{\rho_3}{2}l^2r &=0
\end{align}
This, in discrete form, can be written as,
\begin{align}
    \label{discrete2}
    \nonumber
    &\frac{ l^{n+1}_{ijk} - 2 l^n_{ijk} +  l^{n-1}_{ijk}}{\Delta t^2} + 3H^n\frac{ l^{n+1}_{ijk} - 2 l^n_{ijk} +  l^{n-1}_{ijk}}{\Delta t^2} - \frac{1}{(a^n)^2}\Bigg[\frac{ l^{n}_{i+1jk} - 2 l^n_{ijk} +  l^{n}_{i-1jk}}{\Delta x^2} \\
    \nonumber
    &+ \frac{ l^{n}_{ij+1k} - 2 l^n_{ijk} +  l^{n}_{ij-1k}}{\Delta y^2} + \frac{ l^{n}_{ijk+1} - 2 l^n_{ijk} +  l^{n}_{ijk-1}}{\Delta z^2}\Bigg] + \rho_1 l^n_{ijk}\left( (l^n_{ijk})^2-1\right)\\
    \nonumber &+ \frac{\rho_3}{2}l^n_{ijk}\left(r^n_{ijk}\right)^2 =0,\quad\text{and}\\
    \nonumber&\frac{ r^{n+1}_{ijk} - 2 r^n_{ijk} +  r^{n-1}_{ijk}}{\Delta t^2} + 3H^n\frac{ r^{n+1}_{ijk} - 2 r^n_{ijk} +  r^{n-1}_{ijk}}{\Delta t^2} - \frac{1}{(a^n)^2}\Bigg[\frac{ r^{n}_{i+1jk} - 2 r^n_{ijk} +  r^{n}_{i-1jk}}{\Delta x^2} \\
    \nonumber
    &+ \frac{ r^{n}_{ij+1k} - 2 r^n_{ijk} +  r^{n}_{ij-1k}}{\Delta y^2} + \frac{ r^{n}_{ijk+1} - 2 r^n_{ijk} +  r^{n}_{ijk-1}}{\Delta z^2}\Bigg] + \rho_1 r^n_{ijk}\left( (r^n_{ijk})^2-1\right)\\
    &+ \frac{\rho_3}{2}\left(l^n_{ijk}\right)^2r^n_{ijk} =0
\end{align}
Provided we know the state $ (l,r)^n$ and $ (l,r)^{n-1}$ at time slices $n$ and $n-1$, we can solve this algebraic equation for $ (l,r)^{n+1}$,
\begin{align}
    \label{recursion1}
    \nonumber
     l^{n+1}_{ijk} &= \left(1+\frac{3}{2}A^n\right)^{-1}\Bigg[\left\{2-\rho_1\Delta t^2\left( (l^n_{ijk})^2-1\right)\right\} l^n_{ijk}-\frac{\rho_3}{2}\Delta t^2l^n_{ijk}\left(r^n_{ijk}\right)^2\\
    &+\left(\frac{3}{2}A^n-1\right) l^{n-1}_{ijk} +B^n_x\left\{ l^n_{i+1jk}-2 l^n_{ijk}+ l^n_{i-1jk}\right\}+B^n_y\left\{...\right\}+B^n_z\left\{...\right\}\Bigg]
\end{align}
and,
\begin{align}
    \label{recursion12}
    \nonumber
     r^{n+1}_{ijk} &= \left(1+\frac{3}{2}A^n\right)^{-1}\Bigg[\left\{2-\rho_1\Delta t^2\left( (r^n_{ijk})^2-1\right)\right\} r^n_{ijk}-\frac{\rho_3}{2}\Delta t^2\left(l^n_{ijk}\right)^2r^n_{ijk}\\
    &+\left(\frac{3}{2}A^n-1\right) r^{n-1}_{ijk} +B^n_x\left\{ r^n_{i+1jk}-2 r^n_{ijk}+ r^n_{i-1jk}\right\}+B^n_y\left\{...\right\}+B^n_z\left\{...\right\}\Bigg]
\end{align}
where $A^n = H^n\Delta t$, and $B^n_x = \frac{\Delta t^2}{(a^n)^2\Delta x^2}$ and similarly for $B^n_y, B^n_z$.

Eq. \eqref{recursion1} is not valid for $n=0$, which is the first time step in the simulation. We note that the initial velocity of the fields are given by,
\begin{equation}
    \label{firststep1}
    V^0_{ijk} = \frac{ \Psi^1_{ijk}- \Psi^{-1}_{ijk}}{2\Delta t} \implies  \Psi^{-1}_{ijk} =  \Psi^1_{ijk} - 2\Delta t V^0_{ijk},\quad\quad(\Psi=l,r)
\end{equation}
Substituting this into Eq. \eqref{recursion1} and \eqref{recursion12}, and solving for $ (l,r)^1_{ijk}$ gives,
\begin{align}
    \label{firststep2}
    \nonumber
     l^1_{ijk} = \frac{1}{2}&\Big[ \left\{2-\rho_1\Delta t^2\left( {l^0}_{ijk}^2-1\right)\right\} l^0_{ijk} -\frac{\rho_3}{2}\Delta t^2l^0_{ijk}\left(r^0_{ijk}\right)^2 - \left(\frac{3}{2}A^0-1\right)2\Delta t V \\
    &+ B^0_x\{...\} + B^0_y\{...\} + B^0_z\{...\} \Big]
\end{align}
and,
\begin{align}
    \label{firststep22}
    \nonumber
     r^1_{ijk} = \frac{1}{2}&\Big[ \left\{2-\rho_1\Delta t^2\left( {r^0}_{ijk}^2-1\right)\right\} r^0_{ijk} -\frac{\rho_3}{2}\Delta t^2\left(l^0_{ijk}\right)^2r^0_{ijk} - \left(\frac{3}{2}A^0-1\right)2\Delta t V \\
    &+ B^0_x\{...\} + B^0_y\{...\} + B^0_z\{...\} \Big]
\end{align}

Eq. \eqref{recursion1} and \eqref{recursion12} along with \eqref{firststep2} and \eqref{firststep22} represent the recursive relations that we will use in our simulation.

\subsection{Periodic Boundary Condition}
Periodic boundary condition is implemented on the 6 sides of the simulation box. It is done to \textit{mimic} the simulation over large distances and long times. The wave exiting through one side of the simulation box enters the box immediately through the opposite side, that way making the wave motion \textit{periodic}. The strategy to implement the periodic boundary condition is as follows: apply \textit{open boundary condition} on side A of the box, and equate the opposite side B with A. That way, any displacement at side A will be instantly reflected at side B. And we have 3 such AB pairs.

An open boundary condition means a wave can pass through the boundary without any resistance. In 3D, the condition we implement is \citep{langtangen2016finite},
\begin{equation}
    \label{periodic1}
    D_t^+ \Psi + c_xD_x^- \Psi + c_yD_y^- \Psi + c_zD_z^- \Psi = 0\big|^n_{N_x,N_y,N_z},\quad(\Psi=l,r)
\end{equation}
where $D$ represents the derivative operators and the superscript $\pm$ represents forward or backward finite difference. In our case, $c_x=c_y=c_z=\Tilde{c}=1/\sqrt{3}$ are wave velocities along $x,y,z$ directions. This equation is implemented at all times (denoted by superscript $n$) at the 3 boundaries of the cube (denoted by subscript $N_x,N_y,N_z$). Solving for $ \Psi^{n+1}_{N_x,j,k}$ at $N_x$ we get,
\begin{align}
    \label{periodic2}
    \nonumber
     \Psi^{n+1}_{N_x,j,k} = & \Psi^n_{N_x,j,k} - \frac{\bar{C}_x}{\sqrt{3}}\left(  \Psi^n_{N_x,j,k}- \Psi^n_{N_x-1,j,k} \right) - \frac{\bar{C}_y}{\sqrt{3}}\left(  \Psi^n_{N_x,j,k}- \Psi^n_{N_x,j-1,k} \right) \\
    &- \frac{\bar{C}_z}{\sqrt{3}}\left(  \Psi^n_{N_x,j,k}- \Psi^n_{N_x,j,k-1} \right)
\end{align}
where $\bar{C}_{x,y,z}=\frac{\Delta t}{\Delta x,y,z}$ are the \textit{Courant numbers}. We get similar recursive relations for $ \Psi^{n+1}_{N_y/N_z}$ at $N_y/N_z$. We use these equations to update the $N_x,N_y,N_z$ boundaries of the box while we equate the opposite sides with them.

\subsection{Stability Condition}
We can consider the coefficients $B_x,B_y,B_z$ from Eq. \eqref{recursion1} to be analogous to \textit{Courant numbers} \citep{langtangen2016finite}, and exert the stability condition for the simulation to be,
\begin{equation}
    \max(B^n_x)+\max(B^n_y)+\max(B^n_z) \leq 1
\end{equation}
As $B^n_x = \frac{\Delta t^2}{a^{n2}\Delta x^2}$, this gives,
\begin{equation}
    \label{stability1}
    \Delta t \leq \min(a^n)\left( \frac{1}{\Delta x^2} + \frac{1}{\Delta y^2} + \frac{1}{\Delta z^2} \right)^{-1/2} = \left( \frac{1}{\Delta x^2} + \frac{1}{\Delta y^2} + \frac{1}{\Delta z^2} \right)^{-1/2}
\end{equation}

\textit{Simulation Procedure}
\begin{enumerate}
    \item Define initial conditions as in Eq. \eqref{initfieldconfig1} by discretizing it.
    \item Perform the first time step using Eq. \eqref{firststep2}.
    \item Update the field using Eq. \eqref{recursion1} in iteration for $n>0$.
\end{enumerate}

\subsection{Calculation of comoving area density}
\label{sec:AreaDensity}

First we need to identify a wall within the simulation box numerically. We define a wall to be the surface where $\tilde{\theta}=\theta-\pi/4$ value is 0, i.e. $\tilde{\theta}$ crosses 0 value. To do this, let us call two lattice points that are nearest neighbours as \textit{links}. These links are oriented in $x,y,z$ directions and we measure the signs $\tilde{\theta}$ at the two ends of each link. If the signs are different at the two ends, we have found a zero crossing. For all such possible links at all the lattice points within the cube, we do the same and assign a variable $\delta_\pm$ to each link, which takes on value $+1$ if we found a zero crossing, otherwise $0$. Once we find such a link that penetrates a wall, we add one unit grid area (area of the square made by nearest points) to the total surface area of walls but divided by a weight factor. The weight factor must be the average number density of points on the surface, based on the orientation of the surface at the location of the link. The orientation can be found through the unit normal vector to the wall (given by the gradient of the field) at that location. Finally, we sum all such contributions from all the links with $\delta_\pm=1$ to obtain the total wall area \citep{Press:1989yh},
\begin{equation}
    \label{wallarea1}
    A = \Delta A\sum_{\text{all links}}\delta_\pm\frac{|\nabla\phi|}{|\partial_x\phi|+|\partial_y\phi|+|\partial_z\phi|}
\end{equation}
where $\Delta A$ is the unit grid area. In our case $\Delta A = \Delta x^2 = \Delta y^2 = \Delta z^2 = (b/N)^2$.

\subsection{Gravitational Waves in FLRW Metric}

Let us look at the theory of GWs in an expanding universe. We use Green's method of \citep{Stauffer:1978kr,Dufaux:2008dn} in this section. We consider GWs to be sourced by tensor perturbations in the spatially flat FLRW metric,
\begin{equation}
    \label{metric}
    ds^2 = -dt^2 + a^2(t)\left(\delta_{ij} + h_{ij}\right)dx^idx^j
\end{equation}
where $a(t)$ is the scale factor, $h_{ij}$ is the tensor perturbation satisfying the transverse-traceless (TT) condition $\partial_ih_{ij} = 0$ and which obeys the linearized Einstein equation,
\begin{equation}
    \label{einstein}
    \ddot{h}_{i j}(t, \mathbf{x})+3 H \dot{h}_{i j}(t, \mathbf{x})-\frac{\nabla^{2}}{a^{2}} h_{i j}(t, \mathbf{x})
    =\frac{16 \pi G}{a^{2}} T_{i j}^{\mathrm{TT}}(t, \mathbf{x})
\end{equation}
where dot denotes derivative with respect to cosmic time $t$ and $T^{TT}$ is the TT part of the energy-momentum tensor. Our treatment is valid for any arbitrarily expanding universe with,
\begin{equation}
    \label{scalefactor}
    a(t)\propto\tau^\alpha\propto t^\gamma
\end{equation}

In the spatial Fourier space and in terms of conformal time $d\tau=dt/a(t)$, this equation takes the form,
\begin{equation}
    \label{eom1}
    h^{\prime\prime}_{ij}(t,\mathbf{k}) + \frac{2\alpha}{\tau}h^\prime_{ij}(\tau,\mathbf{k}) + k^2h_{ij}(\tau,\mathbf{k}) = 16\pi G T^{TT}_{ij}(\tau,\mathbf{k})
\end{equation}
where prime notation denotes derivative with respect to conformal time. In terms of the rescaled metric,
\begin{equation}
    \bar{h}_{ij} = ah_{ij}
\end{equation}
we obtain,
\begin{equation}
\label{bessel}
    \left[ \frac{\partial^2}{\partial x^2} + \left( 1- \frac{4\nu^2-1}{4x^2} \right) \right]\bar{h}_{ij}(\tau,\mathbf{k}) = \frac{16\pi Ga(\tau)}{k^2}T^{TT}_{ij}(\tau,\mathbf{k}) 
\end{equation}
where $x = k\tau$, and $\nu$ is defined as, 
\begin{equation}
    \nu = \alpha - \frac{1}{2} = \frac{3\gamma - 1}{2(1-\gamma)}
\end{equation}

Our source $T^{TT}$ is active in the interval $\tau_i<\tau<\tau_f$. The solution of Eq. \eqref{bessel} with initial condition $\bar{h}_{ij}(\tau_i) = \bar{h}^\prime_{ij}(\tau_i) = 0$ can be found using Green's method (for $\tau\leq\tau_f$),

\begin{equation}
\label{green1}
    \bar{h}_{i j}(\tau, \mathbf{k})=\frac{8 \pi^{2} G}{k^{2}} \int_{x_{i}}^{x} d y(y x)^{1 / 2}\left[N_{\nu}(x) J_{\nu}(y)-J_{\nu}(x) N_{\nu}(y)\right] a(y) T_{i j}^{\mathrm{TT}}(y, \mathbf{k})
\end{equation}
where $J_{\nu}(x)$ and $N_\nu(x)$ are the Bessel and Neumann functions. For $\nu = 1/2$, we get the solution for radiation dominated era.

When $\tau\geq\tau_f$, the source becomes negligible and the homogeneous solution to Eq. \eqref{bessel} is a linear combination of $J_\nu(x)$ and $N_\nu(x)$,
\begin{equation}
\label{bessel2}
\bar{h}_{i j}(\tau, \mathbf{k}) = A_{i j}(\mathbf{k})(k \tau)^{1 / 2} J_{\nu}(k \tau) +B_{i j}(\mathbf{k})(k \tau)^{1 / 2} N_{\nu}(k \tau) \quad\left(\text { for } \tau \geq \tau_{f}\right)
\end{equation}

Coefficients $A_{ij}(\mathbf{k}), B_{ij}(\mathbf{k})$ can be found by equating Eq. \eqref{green1} and \eqref{bessel2} at $\tau = \tau_f$,
\begin{align}
\label{bessel3}
\nonumber
A_{i j}(\mathbf{k}) &= -\frac{8 \pi^{2} G}{k^{2}} \int_{x_{i}}^{x_{f}} d x \sqrt{x} a(x) N_{\nu}(x) T_{i j}^{\mathrm{TT}}(x, \mathbf{k}) \\
B_{i j}(\mathbf{k}) &= \frac{8 \pi^{2} G}{k^{2}} \int_{x_{i}}^{x_{f}} d x \sqrt{x} a(x) J_{\nu}(x) T_{i j}^{\mathrm{TT}}(x, \mathbf{k})
\end{align}

Thus, if we can construct the $T^{TT}(x,\mathbf{k})$, we can calculate the perturbation $h_{ij}(\tau,\mathbf{k})$. The energy density of GWs is given by \cite{maggiore2008gravitational},
\begin{align}
\label{energydensitygw}
\nonumber
\rho_{\mathrm{gw}} &=\frac{1}{32 \pi G}\left\langle\dot{h}_{i j}(t, \mathbf{x}) \dot{h}_{i j}(t, \mathbf{x})\right\rangle \\
& \simeq \frac{1}{32 \pi G a^{4}(\tau)}\left\langle\bar{h}_{i j}^{\prime}(\tau, \mathbf{x}) \bar{h}_{i j}^{\prime}(\tau, \mathbf{x})\right\rangle
\end{align}
where we have ignored higher order terms of $aH$ for the second equality, since we assume that the wavelength of GWs is smaller than the Hubble radius, e.g. $k\tau\gg 1$.

A more useful quantity for observational cosmology is the fraction of energy density of GWs to the total energy density of the universe at time $t$, defined by,
\begin{equation}
    \label{omega}
    \Omega_{gw}(t) = \frac{1}{\rho_c(t)}\frac{d\rho_{gw}(t)}{d\ln{k}}
\end{equation}
where $\rho_c(t)$ is the critical density of the universe at time $t$. Note that $k$ is the magnitude of the \textit{comoving wavevector}. It is related to the frequency observed at time $t$ as,
\begin{equation}
\label{freqwavenumber}
    f = \frac{k}{2\pi a(t)}
\end{equation}
where we set the scale factor at the initial time $t_i$ to be unity, $a(t_i) = 1$.

\subsection{Results from simulation}\label{app:results}

\begin{figure}[htb]
\centering
\begin{subfigure}{0.5\textwidth}
  \includegraphics[width=\linewidth]{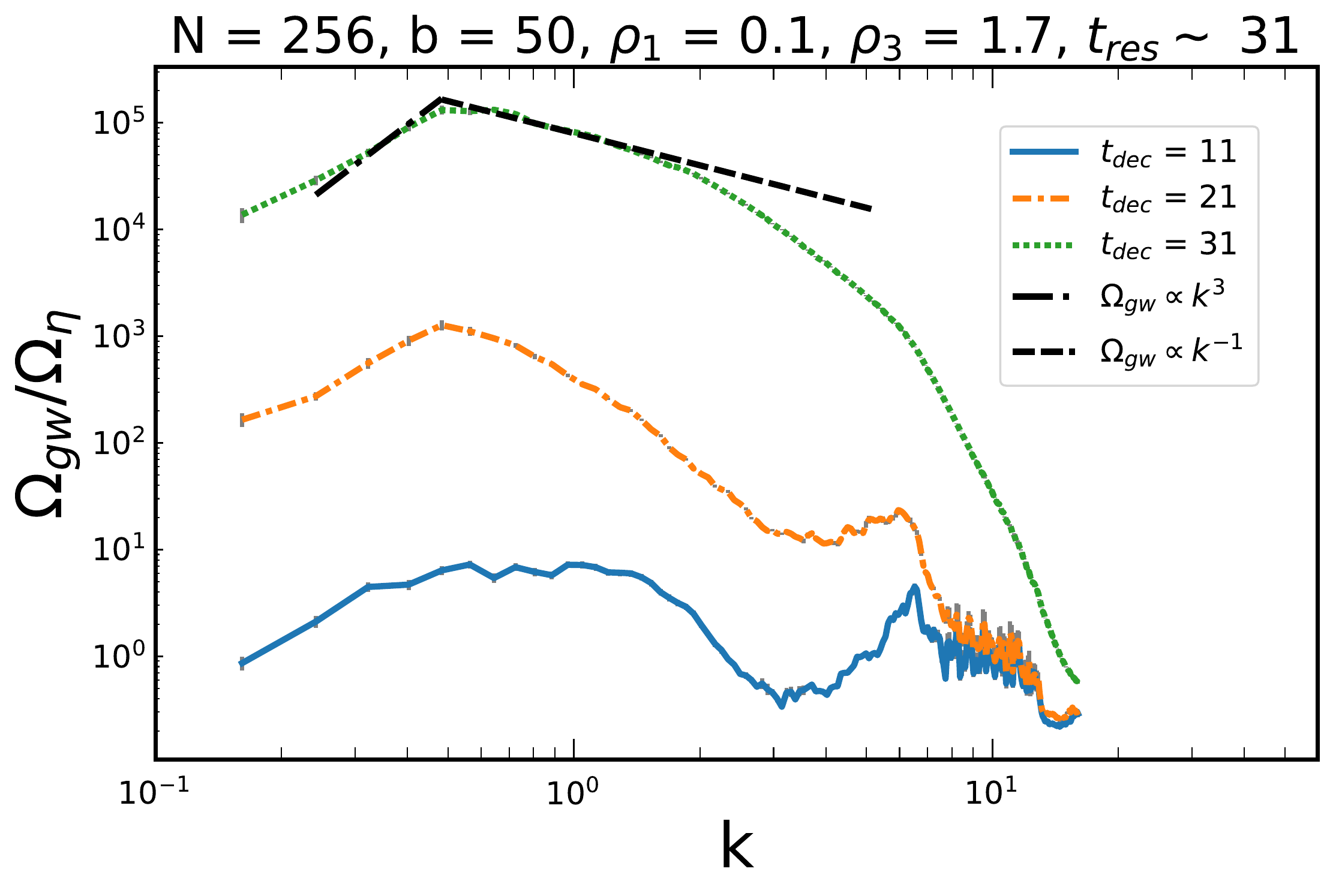}
  \caption{}
  
\end{subfigure}\hfil
\begin{subfigure}{0.5\textwidth}
  \includegraphics[width=\linewidth]{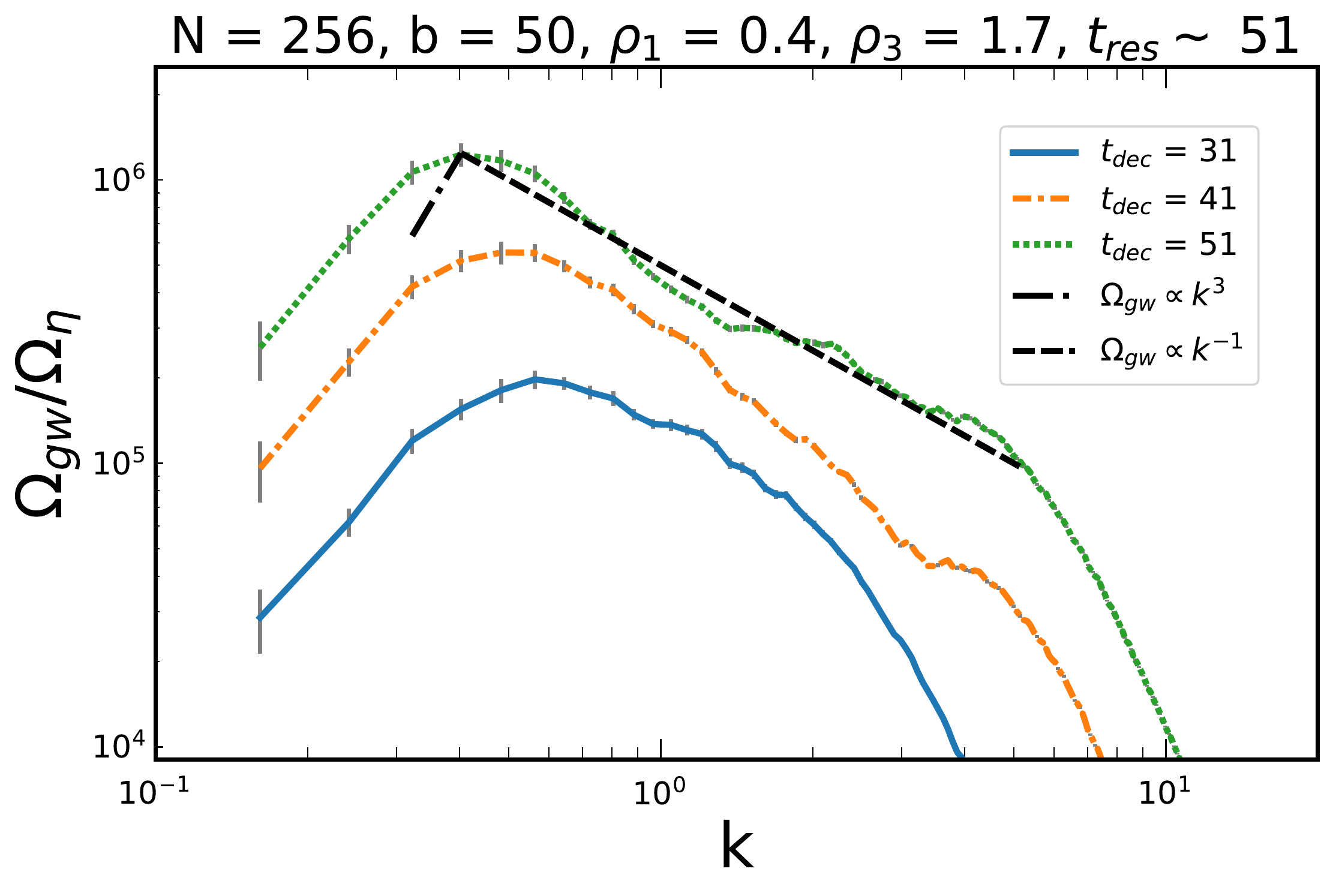}
  \caption{}
  
\end{subfigure}

\medskip
\begin{subfigure}{0.5\textwidth}
  \includegraphics[width=\linewidth]{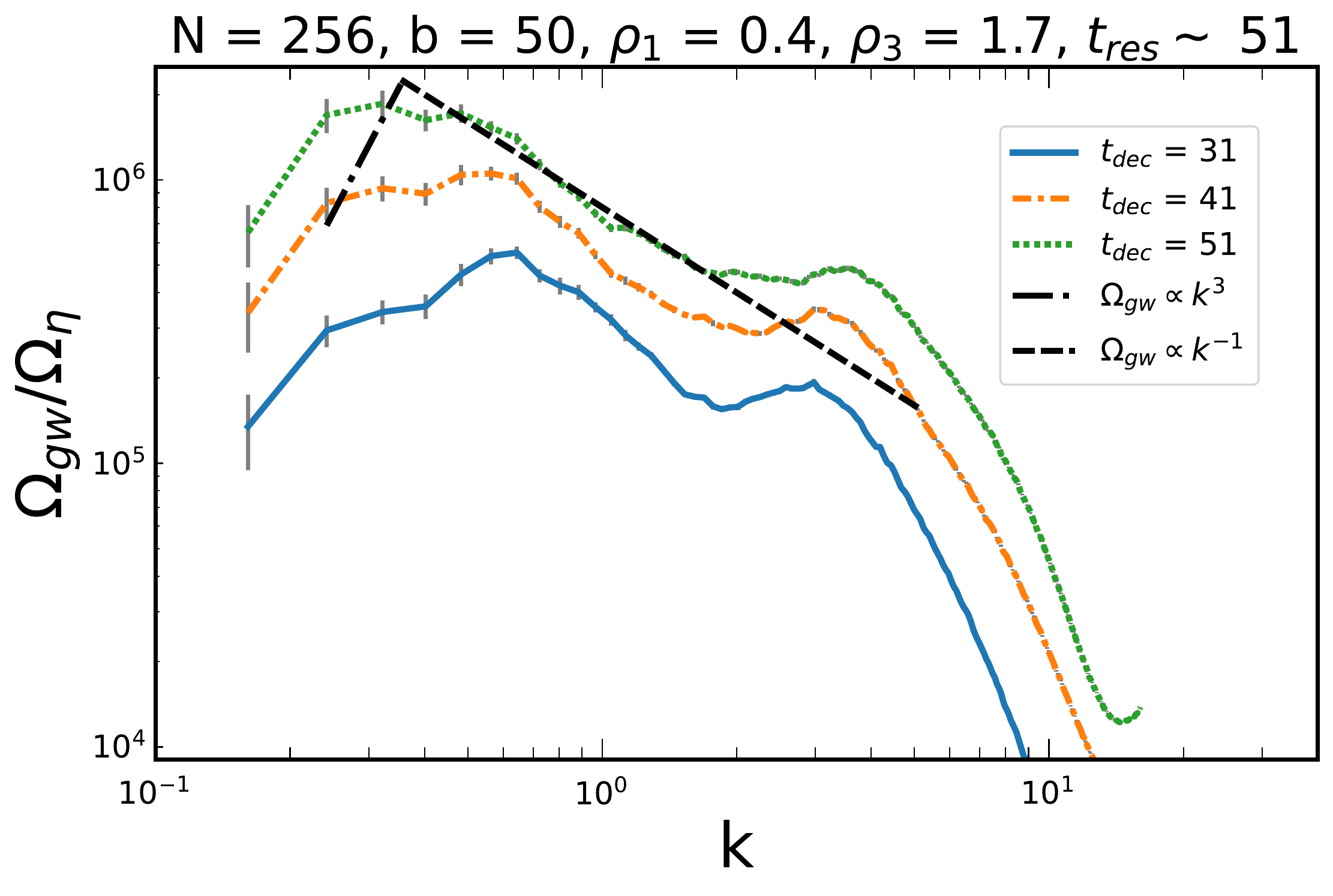}
  \caption{}
  
\end{subfigure}\hfil
\begin{subfigure}{0.5\textwidth}
  \includegraphics[width=\linewidth]{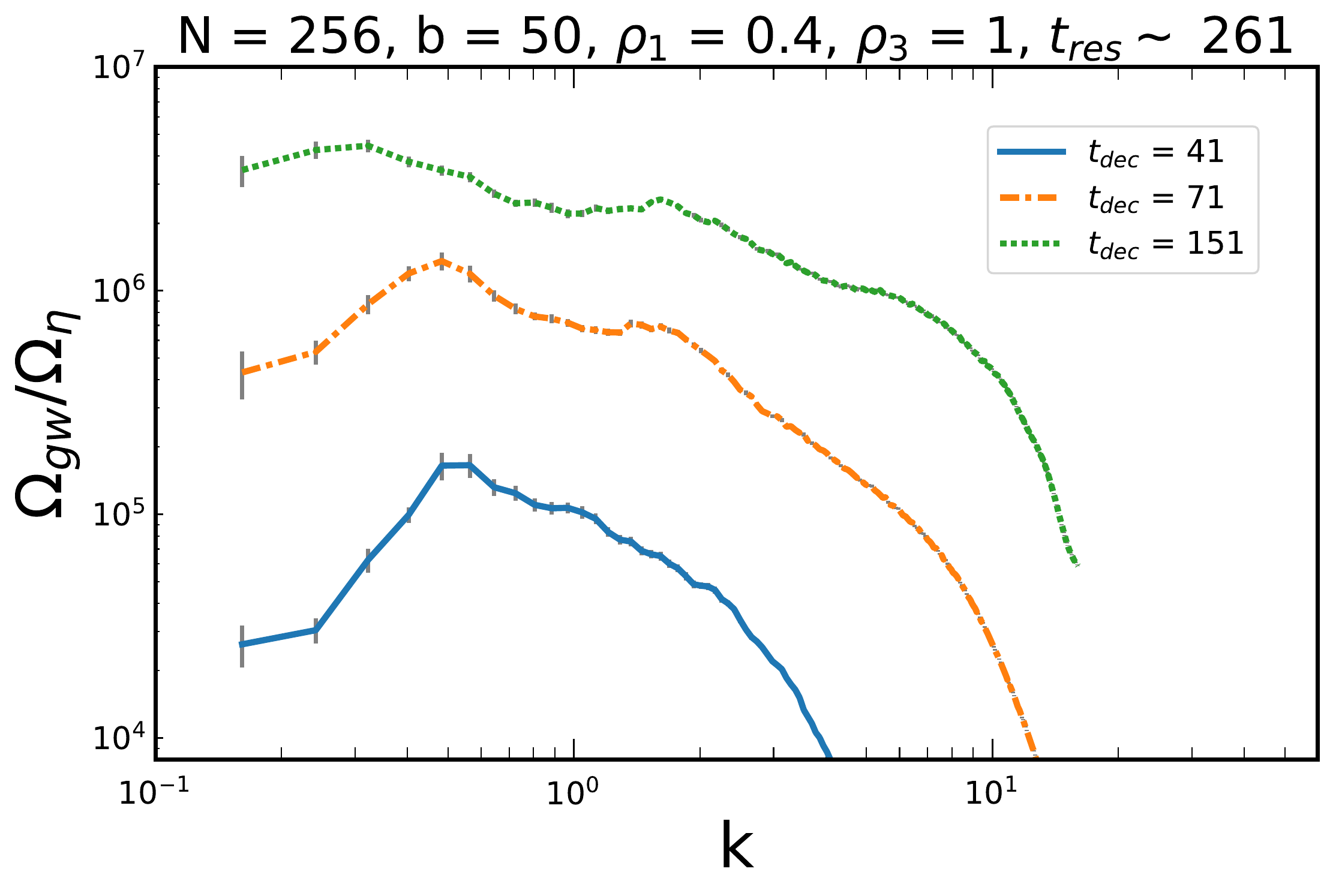}
  \caption{}
  
\end{subfigure}

\caption{\small Same as Fig. \ref{fig:nobias} but for different parameters. All of them start with dominant initial domain sizes comparable to initial Hubble size $\sim1$ (in units of $\eta^{-1}$). The black straight segments are fitted at $t=t_{\rm res}$ in (a)-(c) while in (d) $t_{\rm res}=261$ which is larger than the maximum simulation time we took.}
\label{spectrumgraph}
\end{figure}

\begin{figure}[htb]
\centering
\begin{subfigure}{0.5\textwidth}
  \includegraphics[width=\linewidth]{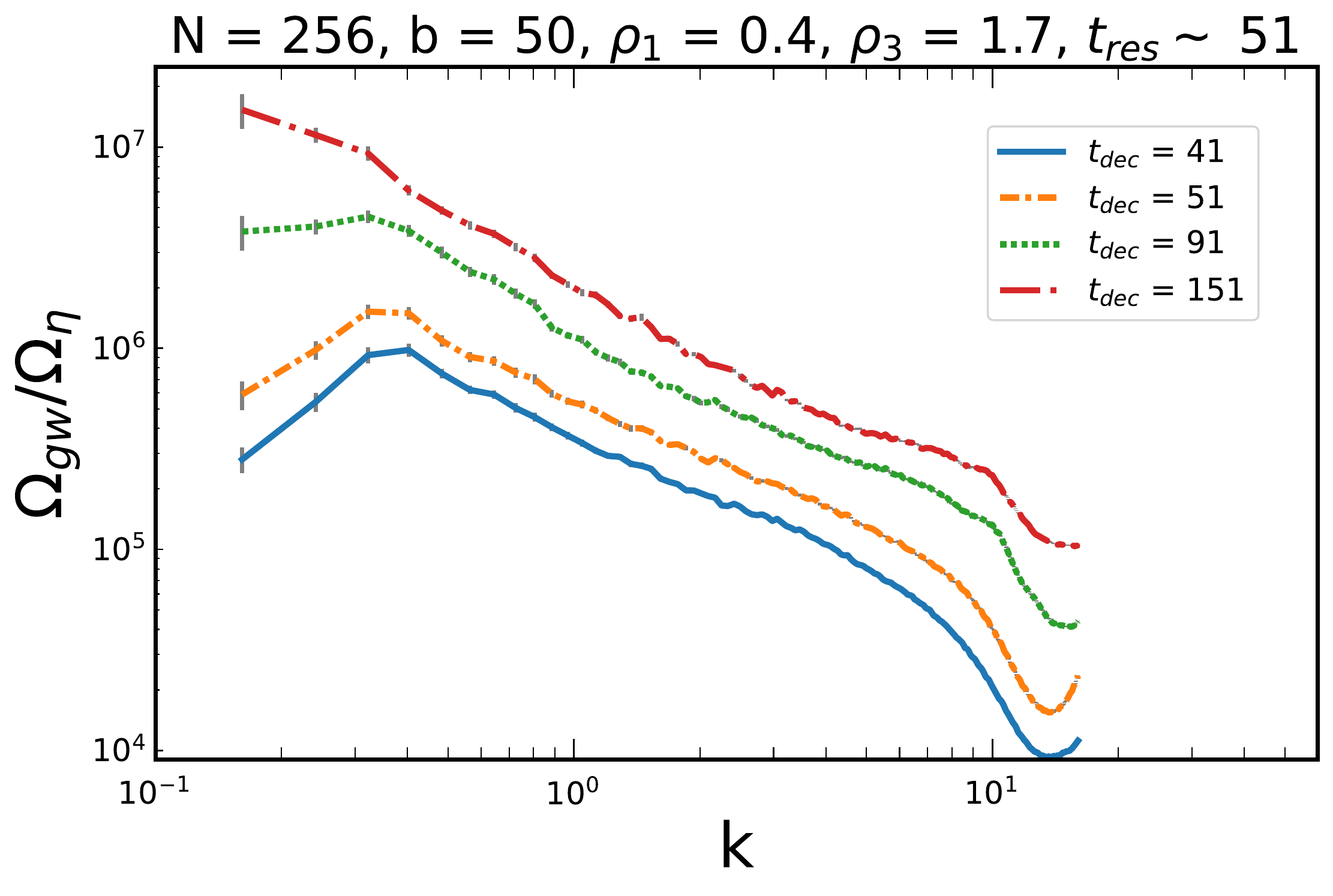}
  \caption{}
  \label{fig:1}
\end{subfigure}\hfil
\begin{subfigure}{0.5\textwidth}
  \includegraphics[width=\linewidth]{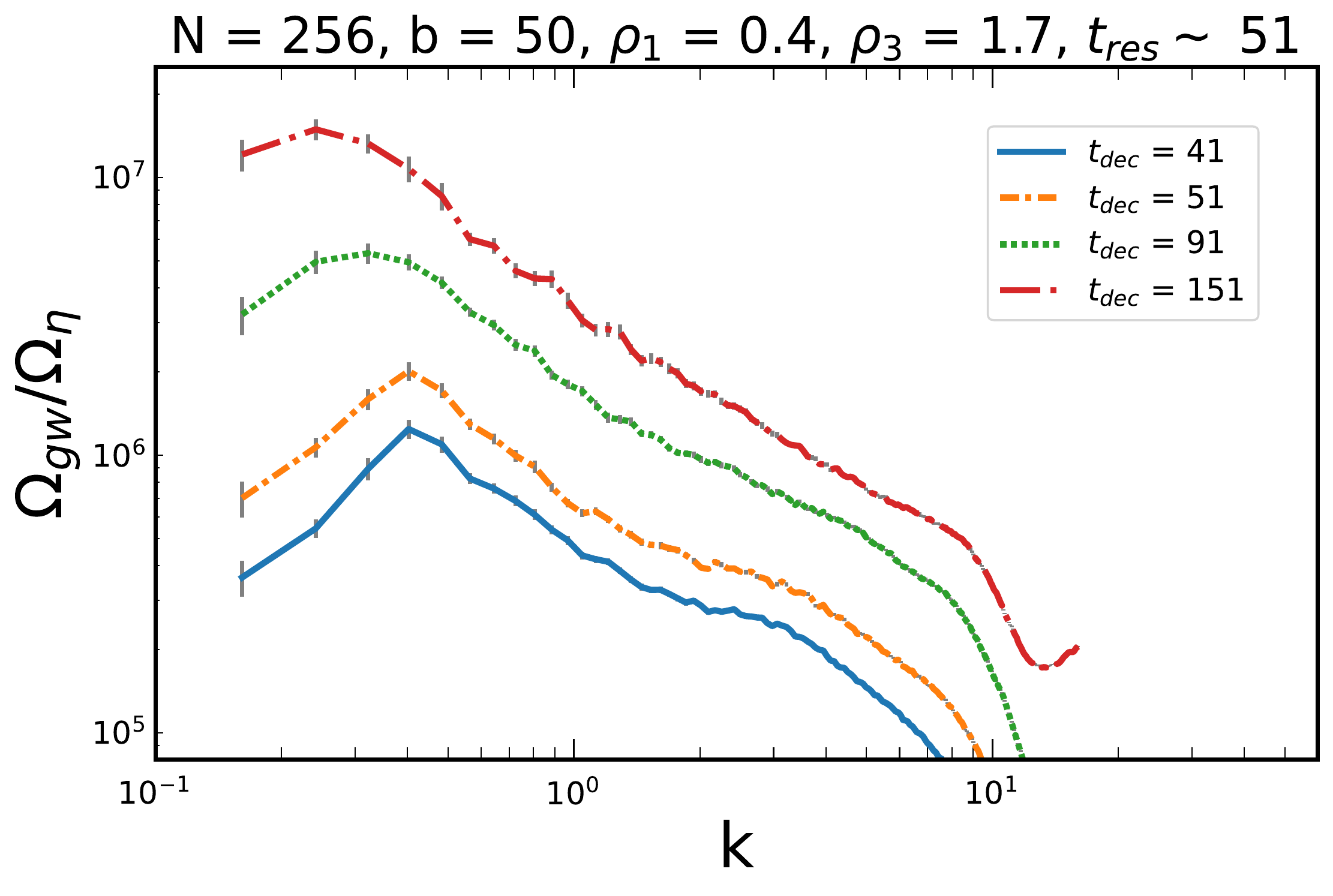}
  \caption{}
  \label{fig:2}
\end{subfigure}

\medskip
\begin{subfigure}{0.5\textwidth}
  \includegraphics[width=\linewidth]{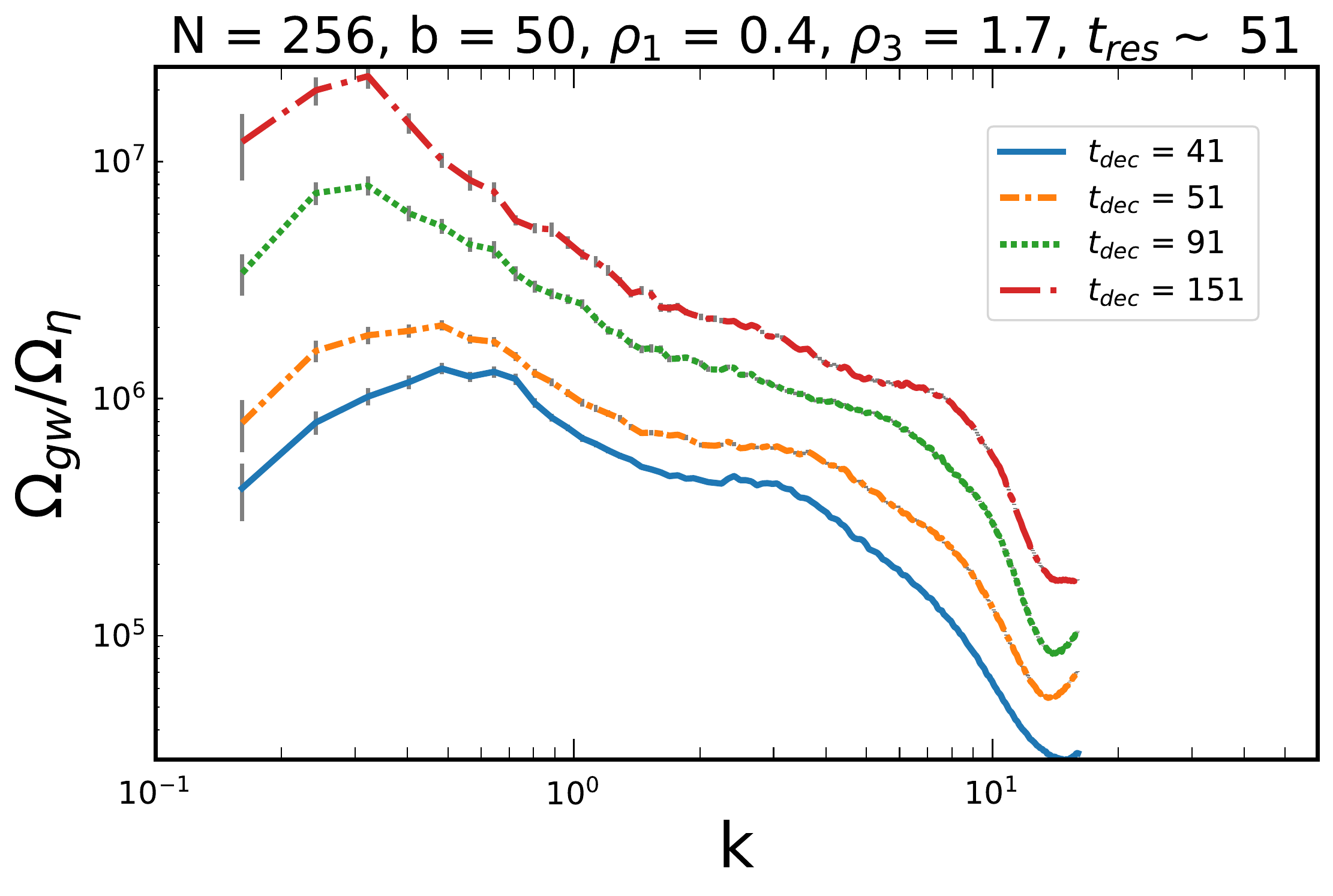}
  \caption{}
  \label{fig:3}
\end{subfigure}\hfil
\begin{subfigure}{0.5\textwidth}
  \includegraphics[width=\linewidth]{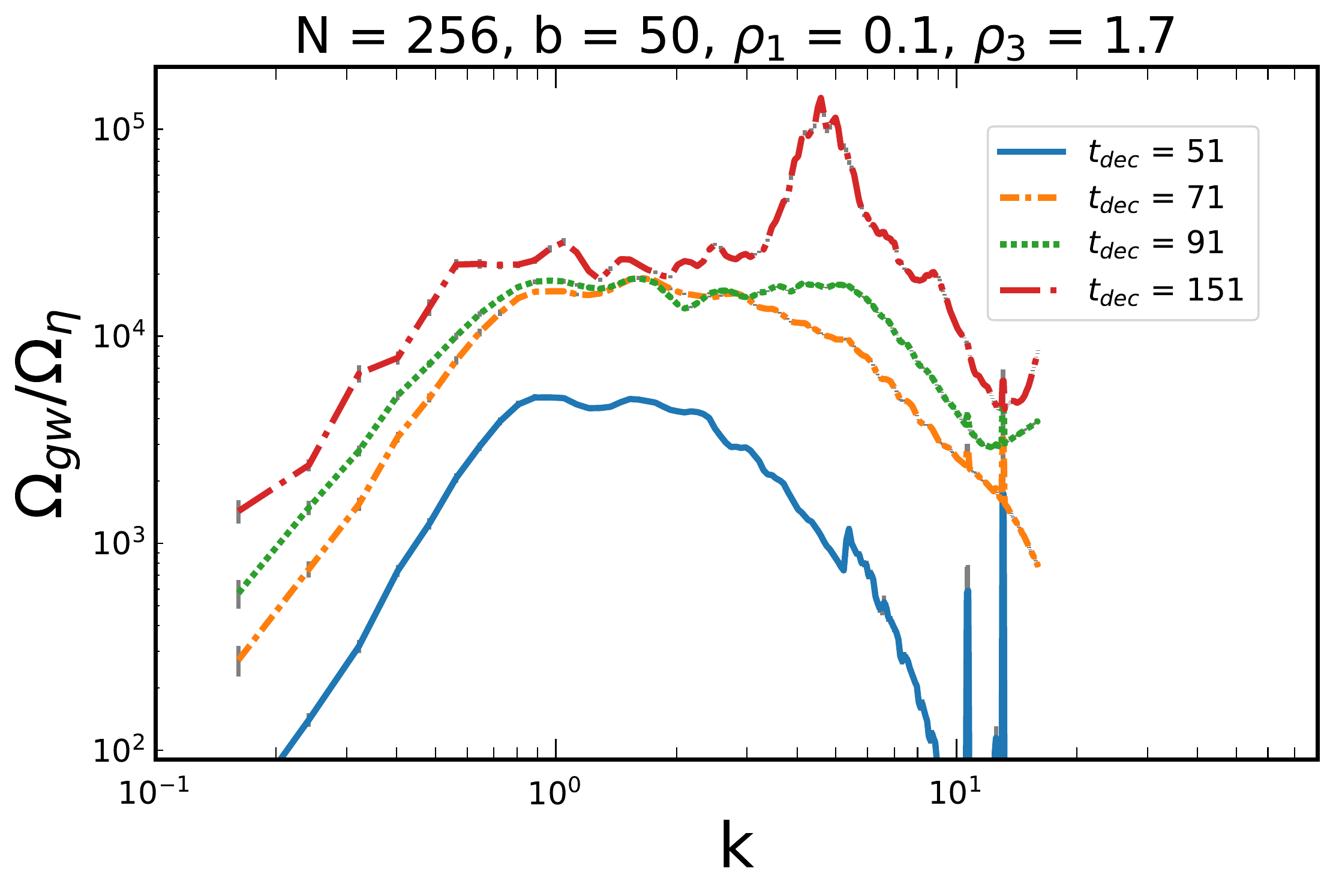}
  \caption{}
  \label{fig:4}
\end{subfigure}

\caption{\small GW spectra for initial domain sizes $\sim$ a) 10, b) 4, c) 2, d)$\ll 1$. In a),b) and c), we see almost no enhancement in high frequencies as in Fig. \ref{fig:nobias} and \ref{spectrumgraph} due to the absence of very small domain walls. This kind of scenario is possible in FOPT case where initially bubble dynamics can create big domains. In d) the initial domain sizes are much smaller than initial Hubble size $\sim 1$, therefore we see a lot of high frequency peaks due to limited resolution of the discrete lattice. These peaks are unphysical and are to be ignored.}
\label{fig:2ndpeakspec}
\end{figure}

Here we show plots of GW spectra from different runs of our simulation and carefully interpret our results. A summary of our interpretation is given in Sec.~\ref{sec:numerical} for plotting the spectrum for general parameter values. We plot the GW spectrum for different $\rho_1$ and $\rho_3$ values with Hubble sized domains in Fig. \ref{fig:nobias} and \ref{spectrumgraph}. The horizontal axis is the amplitude of the comoving wave vector. It is related to the frequency as in Eq. \eqref{freqwavenumber}. Notice that in the Monte Carlo method used to perform the angular integration $d\Omega_k$ as described in Appendix~\ref{app:A}, we define the shell of radius $k$ and average out the values of the fields at the lattice points within that shell. Therefore in the small $k$ region, the number of points within the shell is also small which generates a statistical error in the calculation, i.e. the lower $k$ range of $\mathcal{O}(0.1)$, where the peak occurs, is not properly resolvable. We plot the error bars for the Monte Carlo integration to show the statistical uncertainties in the calculation. Other sources of error include numerical errors from the usage of a simple difference method for simulation and Simpson's rule for time integration. Therefore it is difficult to provide conclusive remarks about the spectrum with our small dynamical range. However, we can deduce some information about the spectrum as follows:
\begin{enumerate}
    \item In Fig.~\ref{fig:nobias} and \ref{spectrumgraph}, we show GW spectra for different runs. Without bias, the walls never vanish. We assume the walls vanish at $t=t_{\rm dec}$ and calculate the bias using Eq. \eqref{eq:decaytime}. Each plot shows spectra at times $t=11,...,151$. Smaller biases lead to longer-lived walls and more gravitational waves, as seen from the increased amplitudes.
    \item It is seen that the spectrum peaks at around $k=k_h$ which corresponds to the Hubble size at the time of GW production, $k_h/a(t)=2\pi H(t)$.
    \item As discussed in Sec. \ref{simsetup}, we fit the curve (solid black line) at \(t \sim t_\text{res}\) in Fig. \ref{fig:nobias} (see Table \ref{tab:tres}). Beyond \(t_\text{res}\), results are unreliable due to unresolvable domain wall width, leading to higher amplitudes and contributions in the high-frequency range. Resonance, seen as a bump around \(k \sim \mathcal{O}(10)\), occurs because the wall width becomes comparable to or smaller than the physical lattice spacing if \(t_{\rm dec}>t_\text{res}\). Thus, the fitted line represents the maximum possible spectrum from the simulation.
    \item For initial domain sizes of \(\mathcal{O}(1)\) or less, we find enhancements above scale \(k \sim 1-2\), corresponding to the initial lattice spacing size that scales as \(t\), i.e. \(k/a(t) = \frac{2\pi N}{b\,t}\), where \(b=50\) and \(N=256\). This enhancement, due to high-frequency modes and limited resolution, changes with domain size, as shown in Fig. \ref{fig:2ndpeakspec}. Therefore, we ignore this enhancement for the actual spectrum. In the case of FOPT, large domains can form due to bubble dynamics, and our simulations of large domains apply, though we won't get scaling solutions. A proper study of deviations from scaling and the spectrum shape due to various initial domain sizes is left for future work.
    \item The spectrum shows a dependence of $k^{-1}$ beyond the peak as seen in Fig. \ref{fig:nobias}. At late times an enhancement towards the high frequency region is seen which can be interpreted as contributions coming from unresolvable wall width as the simulation proceeds. For observation in experiments, such high frequency enhancement is irrelevant since we are interested only in the peak at $k_h$. Thus we conclude that the spectrum varies as $k^{-1}$ for $k>k_h$.
    \item Now due to the high error in the small $k$ range, the behavior of the spectrum in that region is unclear. Therefore we assume a $k^3$ dependence for $k<k_h$ considering requirements for causality \cite{Hindmarsh2014GravitationalTransition}.
\end{enumerate}

\newpage


\bibliography{references,refLRetc,Muon_Anomaly}

\end{document}